%% file: PRE_v1.tex
\documentclass[%
 reprint,
 amsmath,amssymb,
 aps,
]{revtex4-2}

\usepackage{graphicx}
\usepackage{dcolumn}
\usepackage{bm}

\usepackage{bbm}
\usepackage{enumitem}
\usepackage{url} 
\usepackage{amssymb,amsfonts,amsmath,amsthm}
\usepackage{esint}
\usepackage{mathtools} 
\usepackage{graphicx}
\usepackage{MnSymbol}
\usepackage{mathtools} 
\usepackage[bottom]{footmisc}
\usepackage[colorlinks=true, pdfstartview=FitV, linkcolor=blue, citecolor=magenta, urlcolor=blue,pagebackref=false]{hyperref}
\usepackage{microtype}
\usepackage{xcolor}
\usepackage{bigints}
\usepackage{bm}
\usepackage{dsfont} 
\usepackage{notations}  
\usepackage{mmacells}

\newtheorem{conjecture}{Result}

\begin{document}

\preprint{APS/123-QED}

\title{Statistical limits of dictionary learning: \\ random matrix theory and the spectral replica method}

\author{Jean Barbier}
\thanks{Corresponding author}
 \email{jbarbier@ictp.it}
\affiliation{%
 International Center for Theoretical Physics (ICTP), Trieste, Italy.}%

\author{Nicolas Macris}
\email{nicolas.macris@epfl.ch}
\affiliation{Ecole Polytechnique F\'ed\'erale de Lausanne (EPFL), Switzerland.}

\begin{abstract}
We consider increasingly complex models of \emph{matrix denoising} and \emph{dictionary learning} in the Bayes-optimal setting, in the challenging regime where the matrices to infer have a rank growing \emph{linearly} with the system size. This is in contrast with most existing literature concerned with the low-rank (i.e., constant-rank) regime. We first consider a class of rotationally invariant matrix denoising problems whose mutual information and minimum mean-square error are computable using techniques from random matrix theory. Next, we analyze the more challenging models of dictionary learning. To do so we introduce a novel combination of the replica method from statistical mechanics together with random matrix theory, coined \emph{spectral replica method}. This allows us to derive variational formulas for the mutual information between hidden representations and the noisy data of the dictionary learning problem, as well as for the overlaps quantifying the optimal reconstruction error. The proposed method reduces the number of degrees of freedom from $\Theta(N^2)$ matrix entries to $\Theta(N)$ eigenvalues (or singular values), and yields Coulomb gas representations of the mutual information which are reminiscent of matrix models in physics. The main ingredients are a combination of large deviation results for random matrices together with a new replica symmetric decoupling ansatz at the level of the probability distributions of eigenvalues (or singular values) of certain overlap matrices and the use of HarishChandra-Itzykson-Zuber spherical integrals. 
\end{abstract}

\keywords{High-dimensional inference, dictionary learning, random matrix theory, spherical integral, replica method}
\maketitle


\input{sections/Intro}
\input{sections/Denoising-H}
\input{sections/Expansions}

\input{sections/Denoising}      
\input{sections/DictionaryLearning-H}

\input{sections/DictionaryLearning}

\section{Comparison to previous works, conclusion and open problems}

The Bayes-optimal setting of linear-rank dictionary learning has been previously studied in the inspiring works \cite{SK13EPL, SK13ISIT, FML2013, kabashima2016phase, thesis_schmidt} (in the real case $\beta=1$). But these approaches provide only approximations to the exact asymptotic formulas. In \cite{kabashima2016phase} they assume that the entries of $\bS\bT^\intercal$ in \eqref{DiL} are jointly gaussian while they are not, see \cite{maillard2021perturbative} for more explanations. It also corresponds to assume that the overlap matrix $\bQ^{ab}$ is parametrized by a single number. In his thesis \cite{thesis_schmidt} C. Schmidt went further. He proposed instead that the overlap really is a matrix but symmetric (while there is no reason for it to be so) and completely independent of the replica indices.  This assumption that both eigenvalues and eigenvectors are replica independent is physically equivalent to assume that the overlaps $(\bQ^{ab})_{a<b}$ concentrate entrywise. However we expect that only the statistics of eigen/singular values can be self-averaging as is often the case in random matrix theory. The same phenomenon happens in large covariances matrices: only the spectral properties become deterministic while the matrix entries fluctuate even in the large size limit. 

\vspace{0.1cm}
In the most generic version of the method leading to Result~\ref{conj:3}, our new ansatz is only at the level of the distribution of singular values of the overlaps $\bQ^{ab}=\bA^{ab}{\bsig}_Q^{ab}\bB^{ab}$: the singular values of the matrices ${\bsig}_Q^{ab}$ are assumed to decouple and to have identical statistics in the large size limit. Nothing is assumed about the singular vectors. The spectral decoupling assumption \eqref{RS_assump} allows us to carry on the computation while completely capturing the relevant rotational degrees of freedom and invariances. 

The spectral replica method therefore allows us to reduce the challenging task of computing the quenched free entropy (or mutual information), which is an integral over $\Theta(N^2)$ matrix elements (and that additionally should be averaged over the data distribution), into two well defined RMT sub-problems: 
\begin{enumerate}
  \item Obtaining the j.p.d.f. of the singular values (or eigenvalues) of a product of two i.i.d. random matrices drawn from the prior distribution $P_{X,N}$. Or alternatively, to obtain the rate functional controlling the large deviations of the empirical spectral distribution.
  \item Analyzing a Coulomb gas, i.e., an optimization problem over $\Theta(N)$ interacting degrees of freedom representing the singular values (or eigenvalues) of certain matrix order parameters entering the analysis. Or equivalently, solving a functional optimization problem over the associated asymptotic densities.
\end{enumerate}
As already discussed, the first task does not require a-priori the prior $P_{X,N}$ to be factorized over the entries of the matrix $\bX$. So in cases where the j.p.d.f. $p_M$ (or the rate functional $L$) can be evaluated, the spectral replica method yields concrete asympotic formulas for the mutual information and MMSE.


To conclude, we have studied two classes of matrix inference problems in the Bayesian-optimal setting: {\it matrix denoising}, and symmetric and non-symmetric {\it dictionary learning}. Most importantly, they are analysed in the challenging scaling regime of \emph{linear-rank} matrix signals for which very few prior results exist. These problems bear some similarity with matrix models originating in high-energy physics, but with one crucial new aspect when applied to inference, namely the presence of a quenched average over the data.
The analysis of matrix denoising could be carried out using known tools from random matrix theory, in particular the use of spherical integrals and related perturbative expansions. On the other hand for dictionary learning our analysis combines the use of spherical integrals with a novel replica symmetric ansatz, which we call the {\it spectral replica method}. For dictionary learning, the solutions are expressed in terms of non-trivial variational problems reminiscent of Coulomb gases (but with extra complications), which reduce the number of degrees of freedom from $\Theta(N^2)$ down to $\Theta(N)$.

Finding numerical solutions of the variational problems appears to be highly non-trivial. There are two difficulties: one due to the appearance of spherical integrals, and a second due to the determination of the j.p.d.f. of products of i.i.d. random matrices (the distribution $p_M$), or of the rate functional associated with the large deviations of their empirical spectral densities. We have outlined in Section~\ref{sec:Ginibre} how the second problem can sometimes be reduced to the calculation of other spherical integrals. 

We conjecture that our solutions of the denoising and dictionary learning problems are exact in the thermodynamic limit. In Section~\ref{sec:simple-scenario} we have presented a sanity check in a simple case. However this case hardly constitutes a satisfying check. Finding other more interesting checks remains one of the most important problems. A more promising and much less trivial check has been outlined in Section~\ref{sec:Ginibre}. The proposed strategy is based on the method of bi-orthogonal polynomials for exactly computing the integral resulting from our analysis in the case of Hermitian dictionary learning with a Ginibre signal, which will then be comparable against our predictions for matrix denoising using Matytsin's hydrodynamic mapping \cite{matytsin1994large}. While we believe that this is not out of reach, it is certainly not an easy task.

The variational problems can also form the starting point of perturbative expansions in regimes of small and large signal-to-noise ratio, and/or expansions in powers of the inverse rank. It may then be possible to obtain non-trivial comparisons with the ``high-temperature'' expansions derived in \cite{maillard2021perturbative} based on the Plefka-George-Yedidia formalism. As a matter of fact, even if our proposed solutions were not exact, since they are based on a rather weak form of replica symmetric ansatz (the spectral replica symmetry) they could still constitute a good approximation, that would be useful in non-perturbative regimes.

Finally, we hope that eventually our study can inspire rigorous progress on these problems. A first desirable step would be to provide proofs of Conjectures \ref{conjectureHermitianRotInv_simple} and \ref{conjectureRotInv_simple} for the easier denoising problem.

\appendix
\input{sections/AppendixSpherical}

\vspace{7pt}
\input{sections/AppendixMMSEderivative}

\input{sections/AppendixCoulombMoments}
\input{sections/mathematica}

\section*{Acknowledgement}
We thank L. Foini, A. Maillard, M. M\'ezard, F. Krzakala and L. Zdeborov\'a for discussions and pointing out the problematic issues with the replica ansatz in \cite{kabashima2016phase}. J.B. also thanks A. Krajenbrink for interesting discussions on random matrices.

\bibliography{refs}

\end{document}

%% file: sections/Intro.tex
\section{Introduction}
The simplest linear-rank matrix inference task is the problem of recovering the rotationally invariant full-rank matrix $\bS$ from noisy observations $\bY$ generated as
\begin{align*}
\bY=\sqrt{\lambda}\bS +{\bm \xi},
\end{align*}
where $\bxi$ is some Wigner gaussian noise matrix. We refer to this problem as \emph{matrix denoising}. In the random matrix theory (RMT) literature, typical problems are concerned with deriving spectral properties: spectral density and correlation functions of the eigenvalues or singular values of $\bY$, its bulk statistics, the fluctuations of its largest and smallest eigenvalues, potential universality properties, etc. The literature is too large to be exhaustively reviewed here, and relevant references will be cited along the paper. We refer to \cite{TaoRMT,anderson2010introduction} for mathematics books, or \cite{mehta2004random,bouchaudpotters,livan2018introduction} for more physics-oriented presentations. In this paper we instead consider information-theoretic questions such as: \emph{``given a certain signal-to-noise ratio $\lambda$, what is the mutual-information between the hidden matrix signal $\bS$ and the observed noisy data $\bY$?''}, or \emph{``what is the statistically optimal reconstruction error on $\bS$?''}. We are interested in answering these questions in certain asymptotic large size limits. Despite the apparent simplicity of the model, these questions turn out to be highly non-trivial.

In matrix denoising we are ``only'' interested in the reconstruction of the matrix $\bS$. This allows to analyze the model using solely RMT. But there exist models where $\bS$ possesses some additional internal structure other than the (possibly non-trivial) statistics of its spectrum and/or $\bS$ may not be rotationally invariant. This is the case in the model we study next: \emph{dictionary learning}, where a product structure arises. 

Let $M$ noisy $N$-dimensional data points $(\bY_j)_{j\le M}$ be stacked as the columns of $\bY\in\mathbb{K}^{N\times M}$, with $\mathbb{K}=\mathbb{R}$ or $\mathbb{C}$. The unsupervised dictionary learning task we consider is to find a representation of this data $\bY$ in the form
\begin{equation*}
\bS \bT^\dagger +\bZ.
\end{equation*} 
The unknowns are both the ``dictionary'' $\bS\in\mathbb{K}^{N\times K}$ made of $K$ features and the coefficients $\bT \in\mathbb{K}^{M\times K}$ in the decomposition of the clean data $\bS\bT^\dagger$ in feature basis. Here $\bZ$ represents undesired noise. We also analyze a symmetric (or Hermitian) version of the problem where one aims to find a positive-definite first term in the decomposition of $\bY\in \mathbb{K}^{N\times N}$ of the form
\begin{equation*}
\bX\bX^\dagger +\bZ,
\end{equation*} 
where $\bX\in \mathbb{K}^{N\times M}$. The rich internal structure coming from the product between matrices requires new ideas for the analysis: RMT alone does not seem sufficient for analyzing the optimal reconstruction performance on $\bX, \bS,\bT$ themselves (instead of the products $\bS \bT^\dagger$, $\bX\bX^\dagger$ seen a individual matrices). This is where the statistical mechanics of spin glasses, and in particular the novel \emph{spectral replica method} developed in this paper, enter. 

We will consider the Bayes-optimal ``matched setting'' where $\bY$ is truly generated according to one of the models described above, and the statistician has perfect knowledge of $i)$ this data-generating model (i.e., knows the additive nature and statistics of the noise and therefore the likelihood distribution) as well as $ii)$ the prior distributions underlying the hidden random matrix signals $\bX, \bS,\bT$. The statistician can thus exploit this knowledge to write down the correct posterior distribution in order to perform inference. Each model will be analyzed in both cases of real and complex matrices. 

Given its fundamental nature and central role in signal processing and machine learning \cite{mairal2009online,dictionaryLearning}, dictionary learning has generated a large body of work with applications in representation learning \cite{bengio2013representation}, sparse coding \cite{olshausen1996emergence,olshausen1997sparse,kreutz2003dictionary}, robust principal components analysis \cite{candes2011robust,perry2018optimality}, sub-matrix localization \cite{hajek2017information}, blind source separation \cite{belouchrani1997blind}, matrix completion \cite{candes2009exact,candes2010power} and community detection \cite{abbe2017community,2016arXiv161103888L,caltagirone2017recovering}. \emph{Low-rank} (i.e., finite-rank) versions of dictionary learning have been introduced in statistics under the name of ``spike models'' as statistical models for sparse principal components analysis (PCA) \cite{johnstone2001distribution,johnstone2004sparse,zou2006sparse,johnstone2012consistency}. In the low-rank regime $M,N\to+\infty$ proportionally and $K=\Theta(1)$ (or $N\to+\infty$ and $M=\Theta(1)$ for the symmetric case), these models have become paradigms for the study of phase transition phenomena in the recovery of low-rank information hidden in noise. In PCA classical rigorous results
go back to Baik, Ben-Arous and P\'ech\'e \cite{baik2005phase,baik2006eigenvalues} who analyzed the performance of spectral algorithms. More recently, low-dimensional variational formulas for the mutual information and corresponding phase transitions at the level of the Bayes-optimal minimum mean-square error estimator, as well as the algorithmic transitions of message passing and gradient descent-based algorithms and their associated computational-to-statistical gaps, have been derived thanks to the global effort of an highly inter-disciplinary community \cite{koradamacris,lesieur2017constrained,deshpande2017asymptotic,2016arXiv161103888L,miolane2017fundamental,2017arXiv170108010L,XXT,BarbierM17a,BarbierMacris2019,2018arXiv181202537B,2017arXiv170910368B,alaoui2017finite,el2018estimation,mourrat2018hamilton,mourrat2019hamilton,barbier2020all,reeves2020information,alberici2021multi,camilli2021inference,pourkamali2021mismatched,alberici2021solution,mannelli2019afraid,mannelli2019passed,liu2021noisy,maillard2019high,bodin2021rank}. 

In contrast, \emph{much} less is known in the challenging \emph{linear-rank} regime studied here where $N,M,K\to+\infty$ at similar rates, so that ${\rm rank}(\bS\bT^\dagger)$ or ${\rm rank}(\bX\bX^\dagger)$ diverge linearly with $N$. References closely related to our work are  \cite{SK13EPL, SK13ISIT, FML2013, kabashima2016phase, thesis_schmidt} which consider the same Bayesian setting. It has been suspected for some time and eventually confirmed, by the authors of the  very recent new study \cite{maillard2021perturbative}, that the mean field replica assumptions of these works give approximate formulas which are not asymptotically exact. Related problems affect the state evolution analysis \cite{FML2013, kabashima2016phase} of message passing algorithms developed in these papers and in \cite{PSCvI, PSCvII}. This state of affairs for linear-rank dictionary learning is to be contrasted to the low-rank situation where both the replica solutions and state evolution analyses of message passing algorithms are under rigorous mathematical control.
We will discuss the differences between the replica assumptions in the above works and our spectral replica approach in a dedicated section. Another important work is \cite{bun2016rotational} which considers the same models as ours but focuses on a certain class of rotational invariant estimators instead of the information-theoretic performance \footnote{It is now known that the estimator proposed in paper \cite{bun2016rotational} is Bayes-optimal in certain settings, see \cite{maillard2021perturbative}.}.

The main contribution of this paper is a {\it spectral replica ansatz} introduced in Sections \ref{sec:hermitianDL} and \ref{sec:DiL_ST} which seems to us the ``minimal'' possible ansatz to perform a replica symmetric calculation. This ansatz leads to new 
variational expressions for the mutual information between hidden representations and noisy data, whose solutions yield overlaps quantifying the theoretical reconstruction error. Because the replica ansatz is minimal and the setting is Bayes-optimal we believe that our variational expressions are exact. Their solution however is far from trivial and is beyond the scope of this paper. We are still able to check the validity of our formulas analytically in a special case. In Sections \ref{sec:2}, \ref{sec:pertexp}, \ref{sec:non-H_matDnoising} we start by addressing the easier denoising problem for {\it rotation invariant} matrix ensembles. We propose expressions, conjectured to be exact, for the mutual information and mean-square errors. These are again challenging to explicitly compute as a function of the signal-to-noise ratio because they involve non-trivial \emph{spherical integrals} \cite{harish1957differential,itzykson1980planar}. But this time we are able to exploit Bayes-optimality to get rid of the variational problem. Here we limit ourselves to provide perturbative expansions for some examples. We refer to \cite{maillard2021perturbative} for recent work on the denoising problem, also providing expansions and numerical solutions.

Spherical integrals and the associated HarishChandra-Itzykson-Zuber formula \cite{harish1957differential,itzykson1980planar} turn out to be an important tool in our analysis. In physics such integrals appear in multi-matrix models from high-energy physics with applications in string theory, quantum gravity, quantum chromodynamics, fluctuating surfaces and map enumeration \cite{bleher2001random,brezin1993large,mehta1993method,chadha1981method,kazakov2000solvable,kazakov1993induced,kazakov1993d,eynard2015random,zinn2000dilute,stephanov2005random,zvonkin1997matrix,matytsin1997kosterlitz}. In certain aspects multi-matrix models from high-energy physics are similar to matrix \emph{inference} models appearing in denoising and dictionary learning. For example, in contrast to the low-dimensional order parameters of standard low-rank inference problems \cite{barbier2019overlap}, in linear-rank regimes the order parameters are eigenvalues/singular values densities, as in multi-matrix models. Also, at first sight, the variational formulas found in the present contribution look very much like those appearing in these models, see \cite{guionnet2004first}. It may thus be tempting to think that matrix inference models are special cases of known matrix models. This is \emph{not} the case however. The presence of frozen, correlated randomness in inference, namely the data, radically changes the nature of the problem: new tools are needed. This essential difference prevents borrowing various important techniques from this field, but nevertheless certain methods used in the analysis of matrix models will be crucial, in particular the use of spherical integrals in order to integrate rotational degrees of freedom.

Concretely, having frozen data $\bY$ translates, as in spin glasses \cite{MezardParisi87b,MezardMontanari09}, into the need to evaluate the expectation of the \emph{logarithm} of the partition function with respect to it. This yields the behavior of the model for typical realizations of the signals and data. The difference is clear: the canonical two-matrix model from physics reads \cite{mehta1993method,kazakov2000solvable,guionnet2004first} 
\begin{align*}
\ln \mathcal{Z}_{2MM}=\ln \int d\bA \,d\bB \exp\text{Tr} [f(\bA)+g(\bB)+ h(\bA \bB)],
\end{align*}
with $\bA=\bA^\dagger$, $\bB=\bB^\dagger$ and with $f,g,h$ just depending on the spectra. Instead the free energy of matrix inference models (which is essentially the Shannon entropy of the data) looks like 
\begin{align}
\mathbb{E}_\bY\ln \,&\mathcal{Z}_{INFER}(\bY)= \int d\bY \exp  [\text{Tr} f(\bY)] \nonumber\\
&\times\ln \int d\bU \exp \text{Tr}[g(\bU)+h(\bU \bY)]. \label{Z_infer}  
\end{align}
This is a model with ``quenched randomness'' that needs to be averaged, namely the data, while the log-partition function of the two-matrix model is non-random and is a purely ``annealed'' model in physics terminology. This form gets more complicated in non-symmetric multipartite systems such as the $\bS\bT^\dagger$-dictionary learning problem, with an integration over more matrices which are not necessarily symmetric/Hermitian. The presence of a quenched average in \eqref{Z_infer} is far from innocent. It generates a whole new level of difficulty, since the methods in the previous references all rely on a direct saddle-point evaluation and  do not apply \cite{guionnet2004first}. This quenched average is the reason behind the fact that, even if only Hermitian matrices are present in the original inference model, non-Hermitian matrices will appear along the analysis. The role of the spectral replica method combined with RMT is precisely to deal with these new difficulties, at a non-rigorous level. 

Many derivations presented in this paper are based on heuristics, yet they are conjectured to be exact in proper asymptotic limits. We believe that some of our new methodology, in particular the spectral replica method, will pave the way to the analysis of a whole new class of inference and learning problems involving large linear-rank matrices, which remained inaccessible until now. Moreover, given the breadth of applications of such disordered matrix models in information processing systems but also physics, our results may have an impact in a broader context.\\

\noindent \emph{Organization}:\ In Section~\ref{sec:2} we start by analyzing the simplest linear-rank matrix inference model using RMT techniques, namely, denoising of an Hermitian rotationally invariant matrix. Two special cases (one only being non-trivial) are  completely treated, in the sense of deriving explicit enough formulas to draw a phase diagram. In Section~\ref{sec:pertexp} we provide generic systematic expansions of the previously derived formulas in the low signal-to-noise regimes. Section~\ref{sec:non-H_matDnoising} generalizes the analysis to the case of non-Hermitian matrix denoising. Section~\ref{sec:hermitianDL} is devoted to the analysis of Hermitian (i.e., positive-definite) dictionary learning. RMT tools do not suffice anymore, and we introduce the {\it spectral replica method} to go forward. The first part of this section reduces the model to effective Coulomb gases of singular values, while the second one expresses it in terms of eigenvalues. We also discuss the main differences with previous attempts to analyse this model. Finally in Section~\ref{sec:DiL_ST}, we consider the non-symmetric case of dictionary learning. Appendix~\ref{app:sphInt} recalls known facts about full-rank spherical integrals which are of crucial importance in our analyses. Appendix~\ref{app:MMSEderivative} derives a generic formula for the minimum mean-square error  in matrix denoising. The last Appendix~\ref{app:mathematica} provides the necessary \textsc{Mathematica} codes to reproduce our numerical results.\\

\noindent \emph{Notations}: \ Let the field $\mathbb{K}=\mathbb{R}$ if $\beta=1$ or $\mathbb{K}=\mathbb{C}$ if $\beta=2$, where $\beta$ refers to the Dyson index. The symbol $\dagger$ corresponds to the transpose $\intercal$ in the real case $\beta=1$ or to the transpose conjugate when $\beta=2$, the conjugate being $\bar z:={\Re}z-{\rm i}\,{\Im}z$ with ${\Re}z$ and ${\Im}z$ the real and imaginary parts of $z\in\mathbb{C}$ and ${\rm i}=\sqrt{-1}$. Vectors and matrices are in bold. Vectors are columns by default and their transpose (conjugate) $\bx^\dagger$ are row vectors. When no confusion can arise we denote the trace ${\rm Tr}f(\bA)={\rm Tr}[f(\bA)]$ so, e.g., ${\rm Tr}\bA^2={\rm Tr}[\bA^2]$ or ${\rm Tr}\bA\bB = {\rm Tr}[\bA\bB]$. Similarly $\EE X^2 = \EE(X)^2 =\EE[X^2]\ge (\EE X)^2=\EE[X]^2$. The usual inner product between (possibly complex) vectors $\sum_{i} \bar x_i y_i$ is denoted $\bx^\dagger \by$ or $\langle \bar \bx,\by\rangle$ (with $\bar \bx=\bx$ if $\bx$ is real); the matrix inner product is ${\rm Tr}\bX^\dagger \bY$. The usual $L^2$ vector (squared) norm is $\sum_{i} |x_i|^2=\|\bx\|^2$. Every sum or product over $j\le t$ means over $j=1,\ldots,t$. We will often drop parentheses, e.g., $\exp \cdots = \exp(\cdots)$. Depending on the context, the symbol $\propto$ means ``equality up to a normalization'', ``equality up to an irrelevant additive constant'' or ``proportional to''. We denote $[N]:=\{1,\ldots,N\}$. For a diagonal matrix $\bSig$ we will write the diagonal elements with a single index $\Sigma_i:=\Sigma_{ii}$. For a diagonalizable matrix $\bA$ the diagonal matrix of eigenvalues is $\bLam^A=\bLam_A$ and the individual eigenvalues are $\lambda^A_i$; similarly for a generic matrix $\bB$ the matrix of singular values is $\bsig^B=\bsig_B$ (with the singular values on the main diagonal) and the individual singular values are $\sigma^B_i$. We generically denote $\rho_A$ the asymptotic limit of the empirical distribution of eigenvalues or singular values of a matrix $\bA$. Symbol $\mathcal{P}$ refers to the set of probability densities with finite support; $\mathcal{P}_{\ge 0}$ emphasizes that the support is non-negative. A matrix $\bM$ with elements $M_{ij}$ may also be written $[M_{ij}]$ or $[M_{cd}]$. Finally, the symbol $x \sim y$ means equality in distribution for two random variables; $x\sim p$ instead means that $x$ is a sample from $p$ whenever $p$ is a probability distribution or a sample from $p(x)dx$ if $p$ is a probability density function.

%% file: sections/Denoising-H.tex
\section{Denoising of an Hermitian rotationally invariant matrix}\label{sec:2}
We start with the simplest possible model of inference of a large matrix: linear-rank rotationally invariant matrix denoising. It will only require known tools from random matrix theory.

\subsection{The model} 

Let a matrix signal $\bS=\bS^\dagger\in \mathbb{K}^{N\times N}$ with  $\bS\sim P_{S,N}$ for some known prior distribution (which in general does not factorize over the matrix entries), and ${\bm \xi}=\bxi^\dagger\in \mathbb{K}^{N\times N}$ a standard Wigner noise matrix with probability density function (p.d.f.) $$dP_{\xi,N}(\bxi)= C_N\, d\bxi \exp{\rm Tr}\Big[-\frac{\beta N}{4}\bxi^2\Big]$$ with $C_N$ the normalization factor. Consider a matrix denoising problem with data $\bY=\bY^\dagger\in \mathbb{K}^{N\times N}$ generated according to the observation model
\begin{align}\label{model-exactly}
\bY=\sqrt{\lambda}\bS +{\bm \xi}.
\end{align}
The hidden matrix $\bS$ to be recovered from the data is rotationally invariant in the sense that it is drawn from a prior distribution such that 
$$
dP_{S,N}(\bS)=dP_{S,N}(\bO^\dagger \bS\bO)
$$ 
for any orthogonal ($\beta=1$) or unitary ($\beta=2$) matrix $\bO$. It can thus be diagonalized as $\bS=\tilde \bU^\dagger  \bLam^S \tilde \bU$ where $ \tilde \bU\sim \mu_N^{(\beta)}$ with $\mu_N^{(\beta)}$ the normalized Haar measure over the orthogonal group $\mathcal{O}(N)$ if $\beta=1$ or over the unitary group $\mathcal{U}(N)$ if $\beta=2$. Matrix $\bS$ has $O(1/\sqrt{N})$ entries. This scaling for the entries of $\bS$ and of the Wigner matrix are such that the (real) eigenvalues of $\bS,{\bm \xi}$ and therefore $\bY$ remain $O(1)$ in the limit $N\to +\infty$. The joint probability density function (j.p.d.f.) of eigenvalues of the matrix $\bY$ generated according to model \eqref{model-exactly} is rigorously established in the case where $\bS$ has independent entries (which we do \emph{not} necessarily assume), and is obtaind with techniques of a similar flavor as our strategy (i.e., based on the use of spherical integrals) \cite{johansson2001universality}; see also \cite{TaoRMT} for an approach based on Dyson's Brownian motion.

The above model defines a random matrix ensemble for $\bY$ which is linked to the Rosenzweig-Porter random matrix model \cite{rosenzweig1960repulsion} from condensed matter. A generalized version of it has a rich behavior with a localization transition and regions with ``multifractal eigenstates'', see \cite{brezin1996correlations,biroli2021levy,kravtsov2015random,facoetti2016non}. The regime we are interested in, namely with both $\bLam^S$ and the eigenvalues of ${\bm \xi}$ being order $1$, corresponds precisely to the critical scaling regime where a recently discovered transition from non-ergodic extended states to ergodic extended states happens in the model (i.e., the transition towards multifractality; see the discussion on the regime $\gamma=1$ in \cite{kravtsov2015random}). We find the connection with inference particularly intriguing and the results of this paper may thus may be of independent interest in the condensed matter context. In particular, if there is an information-theoretic transition in the inference problem it may happen to be connected to the $\gamma=1$ ergodicity-breaking transition found in \cite{kravtsov2015random}.

We consider a generic j.p.d.f. $p_{S,N}(\bLam^S)$ of eigenvalues which is symmetric \footnote{We require symmetry except in the trivial case $p_{S,N}(\bLam^S)=\delta(\bLam^S-\bLam^S_0)$ for some fixed $\bLam^S_0$.} (i.e., invariant under any permutation of the entries of $\bLam^S$) and whose one-point marginal is assumed to weakly converge as $N\to +\infty$ to a well defined measure $\rho_{S}$ with finite support and without point masses; in particular $\bS$ needs to be full-rank. We discuss at the end of this section how to overcome this latter constraint. Generically, rotational invariance implies that the prior over $\bS=\tilde \bU^\dagger  \bLam^S \tilde \bU$ can be decomposed as
\begin{align}
dP_{S,N}(\bS)=  d\mu_N^{(\beta)}(\tilde\bU) \,dp_{S,N}(\bLam^S). \label{prior_gene}
\end{align}
A (rather generic) special case of rotationally invariant measures for the signal are those of the form
\begin{align}
&dP_{S,N}(\bS)\propto  \nn 
&\quad\mu_N^{(\beta)}(\tilde\bU)\, d\bLam^S\exp{\rm Tr}\Big[-\frac {\beta N}4V(\bLam^S)\Big]\,|{\Delta}_N(\bLam^S)|^\beta \label{priorWithVan}
\end{align}
for a rotation invariant matrix potential ${\rm Tr} V(\bS) = {\rm Tr} V(\bLam^S)$, and where the Vandermonde determinant for a $N\times N$ diagonal matrix $\bA$ with diagonal entries $(A_{i})_{i\le N}$ reads
\begin{align}
\Delta_N(\bA):=\prod_{i<j}^{1,N}(A_i-A_j)={\rm det}[(A_c)^{d-1}].\label{Vander}
\end{align}
In this case the eigenvalues j.p.d.f. has the form
\begin{align}
 p_{S,N}(\bLam^S)\propto \exp{\rm Tr}\Big[-\frac {\beta N}4V(\bLam^S)\Big]\,|{\Delta}_N(\bLam^S)|^\beta.\label{typicalcase}
\end{align}
The case of a standard real symmetric or complex Hermitian Wigner matrices then corresponds to $V(\bLam^S) = \bLam_S^2$. Wishart matrices $\bS = \bX\bX^\dagger$ with $\bX\in \mathbb{K}^{N\times M}$ have a density for $N\leq M$ corresponding to $V(\bLam^S) = 2(1 - {M}/{N} - 1/{N} + {2}/(\beta N)) \ln \bLam^S  +  2(M/N)\bLam^S$ (see more details later in Section~\ref{sec:pertexp}).

The main object of interest is the mutual information between data and signal:
\begin{align}
&I(\bY;\bS)=H(\bY)-H(\bY \mid \bS)\nn
&=H(\bY)-H(\bxi)\nn
&=-\EE_\bY\ln \int dP_{S,N}(\bs)\,C_N\exp{\rm Tr}\Big[-\frac{\beta N}{4}\big(\bY-\sqrt{\lambda}\bs\big)^2\Big]\nn
&\quad+\EE\ln C_N\exp{\rm Tr}\Big[-\frac{\beta N}{4}\bxi^2\Big]	\nn
&=-\EE_\bY\ln\int dP_{S,N}(\bs)\exp\frac{\beta N}{2}{\rm Tr}\Big[\sqrt{\lambda}\bs\bY-\frac{\lambda}{2}\bs^2\Big] \nn
&\quad+ \frac{\beta \lambda N}{4}\EE{\rm Tr} \bS^2.\label{mut_inf_conj1}
\end{align}

\subsection{Free entropy and mutual information through random matrix theory}

We define the free entropy $f_N=f_N(\bY)$ as minus the first term in \eqref{mut_inf_conj1} divided by $N^2$, without the expectation. The mutual information will directly be deduced from the free entropy, and using that the later is expected to concentrate onto its $\bY$-average. Using the eigen-decomposition $\bs=\bU^\dagger\bLam^s\bU$ the free entropy reads 
\begin{align}
f_N&:= \frac1{N^2}\ln \int dP_{S,N}(\bs)\exp \frac{\beta N}{2} {\rm Tr}\Big[\sqrt{\lambda}\bs\bY-\frac\lambda 2\bs^2\Big]\nn
&= \frac{1}{N^2}\ln \int dp_{S,N}(\bLam^s) \exp\Big[-\frac{\beta\lambda N} 4{\rm Tr}\bLam_s^2\Big]\nn
&\quad\times \int d\mu_N^{(\beta)}(\bU)\exp\frac{\beta \sqrt{\lambda}}{2}N{\rm Tr}\big[\bU^\dagger\bLam^s\bU\bY\big]\nn
&=\frac{1}{N^2}\ln \int d\bLam^s \exp {N^2}\Big(\frac{1}{N^2}\ln p_{S,N}(\bLam^s)\nn
&\quad -\frac{\beta\lambda }{4N}{\rm Tr}\bLam_s^2+I_N^{(\beta)}(\bLam^s,\bLam^Y,\sqrt{\lambda})\Big). \label{f_N1}
\end{align}
There appears the HarishChandra-Itzykson-Zuber (HCIZ) spherical integral \cite{itzykson1980planar,guionnet2002large}, which is only a function of the eigenspectra of its arguments. For $N\times N$ symmetric/Hermitian matrices $\bA$ and $\bB$,
\begin{align}\label{IZhermitian}
&I_N^{(\beta)}(\bA,\bB,\gamma)=I_N^{(\beta)}(\bLam^A,\bLam^B,\gamma)\nn
&\quad:=\frac1{N^2}\ln\int d\mu_N^{(\beta)}(\bU) \exp \frac{\beta \gamma}2 N{\rm Tr}\, \big[{\bm U}^{\dagger} \bLam^A{\bm U}\bLam^B\big],
\end{align}
where the integration is over $\mathcal{O}(N)$ when $\beta=1$ or $\mathcal{U}(N)$ when $\beta=2$. We recall known facts about it in Appendix~\ref{app:sphInt}. It has a well-defined limit \cite{matytsin1994large,guionnet2002large}:
\begin{align}
I^{(\beta)}[ \rho_{{A}},\rho_{B},\gamma]:=\lim_{N\to+\infty}I_N^{(\beta)}(\bLam^A,\bLam^B,\gamma),	\label{sphInt_lim}
\end{align}
where $\rho_{A},\rho_{B}$ are the asymptotic \emph{densities} of eigenvalues (i.e., one-point correlation functions) of $\bA$ and $\bB$, respectively. In the present case, the eigenvalues $\bLam^Y$ of the data matrix $\bY$ and associated asymptotic density $\rho_{{Y}}$ are fixed by the model; $\bLam^Y$ can be simulated and $\rho_Y$ can be obtained using free probability, see, e.g., \cite{mingo2017free,pielaszkiewicz2015closed}. Thus, a standard saddle-point argument leads to the following conjecture for the free entropy $f_N=f_N(\bY)$ as $N\to +\infty$.
\begin{conjecture}[Free entropy of Hermitian rotationally invariant matrix denoising]\label{conjectureHermitianRotInv}
 The free entropy of model \eqref{model-exactly} is
\begin{align}
f_N= \sup_{\bLam^s\in\mathbb{R}^N} &\Big\{\frac{1}{N^2}\ln p_{S,N}(\bLam^s)\nn
&\quad-\frac{\beta\lambda }{4N}{\rm Tr}\bLam_s^2+I_N^{(\beta)}(\bLam^s,\bLam^Y,\sqrt{\lambda})\Big\}+\tau_N.
\label{formula:conj1}
\end{align}
The constant $\tau_N$ fixes the constraint $f_N({\lambda=0}) = 0$ (the spherical integral \eqref{IZhermitian} vanishes when $\lambda=0$):
\begin{align*}
\tau_N&:= -\sup_{\bLam^s\in\mathbb{R}^N} \frac1{N^2}{\ln p_{S,N}(\bLam^s)} + o_N(1) 
\end{align*}
\end{conjecture}

Because we do not rigorously control the saddle-point estimation we state the result as a conjecture, but it should not be out of reach to turn it into a theorem using techniques as in \cite{guionnet2004first}. This free entropy was not averaged with respect to $\bY$. But it is expected that additionally it is self-averaging as it depends on $\bY$ only through its spectrum $\bLam^Y$:
$$\EE f_N=f_N+o_N(1).$$

Note that from this conjecture, the minimum mean-square error (MMSE) can be deduced using the I-MMSE relation for gaussian channels \cite{GuoShamaiVerdu} \footnote{The factor $4$ that differs from the $2$ in the usual I-MMSE relation \cite{GuoShamaiVerdu} comes from the fact that the Wigner matrix to denoise has only a fraction $N(N+1)/(2N^2)= 1/2+O(1/N)$ of independent entries. The $O(1/N)$ correction comes from the diagonal terms in matrix $\bS$ for which the signal-to-noise ratio is different than the one of the off-diagonal entries. The complex noise case of the I-MMSE relation is discussed in Section V.D of \cite{GuoShamaiVerdu}.} 
\begin{align}
&\frac1{N^2}\mathbb{E}\|\bS-\mathbb{E}[\bS\mid \bY]\|^2=\frac{4}{\beta N^2}\frac{d}{d\lambda}I(\bY;\bS)+O(1/N)\nn
&\qquad= \frac{1}{N}\EE{\rm Tr} \bLam_S^2-\frac 4\beta\frac{d}{d\lambda}\EE f_N+O(1/N) .\label{IMMSE}  
\end{align}

\noindent \emph{Regularity of eigenvalues and singular values distributions.} \ All along the paper we assume that all eigenvalues (and later singular values) distributions are such that empirical distributions of eigen/singular values converge weakly to well defined asymptotic probability densities with $i)$ (possibly disconnected) finite support, and $ii)$ without any point masses. Cases of distributions with point masses (such as a matrix $\bS$ of rank lower than $N$ with a point mass $\delta_0$ in its eigenvalues distribution; for example a rank-deficient Wishart matrix with $N>M$) can be approximated by considering regularizations. If the original signal matrix, say $\bS\in\mathbb{R}^{N\times N}$, has a $\text{rank}(\bS)<N$ with a finite fraction of eigen/singular values strictly null, one may instead consider from the beginning the same inference model but with full-rank signal $\bS_\eps:=\bS+\bZ_\eps$ where $\bZ_\eps$ is an independent rotationally invariant regularization with norm smaller than $\eps$, such as a Wigner matrix with sufficiently small variance. In certain cases it should then be possible to obtain the j.p.d.f. of the resulting matrix ensemble. The asymptotic formulas for the free entropies and mutual informations are expected to be continuous in $\eps$. Thus assuming that the convergence to the asymptotic value is uniform in $\eps$, we can permute the $N\to+\infty$ and $\eps\to 0_+$ limits to obtain the formulas for ``non full-rank'' cases and densities with point masses. For the rest of the paper, we will thus restrict all theoretical arguments to full-rank cases without point masses. 

Let us also mention that despite we focus on full-rank square models of matrix denoising \eqref{model-exactly} with $\bS$ a $N\times N$ matrix, we believe that by combining our approach together with the idea of ``quadratization of rectangular matrices'' found in \cite{fischmann2012induced}, and exploited, e.g., in \cite{akemann2013products,ipsen2014weak}, then it should not require too much work to generalize the results on matrix denoising of Sections~\ref{sec:2} and \ref{sec:pertexp} to the rectangular setting $\bS\in\mathbb{K}^{N\times M}$, $N\neq M$.

\subsection{Expressing the result using a density order parameter thanks to a large deviation principle}\label{sec:LDP}
For the typical form of eigenvalues density \eqref{typicalcase} we have 
\begin{align}
\frac{\ln p_{S,N}(\bLam^s)}{N^2}=\frac\beta{2N^2}\sum_{i\neq j}^{1,N}\ln|\lambda_i^s-\lambda_j^s|-\frac {\beta }{4N}{\rm Tr}V(\bLam^s).\label{typform}  
\end{align}
When potential $V$ acts entry-wise this can be expressed using the empirical spectral distribution (ESD)
\begin{align}
{\hat{\rho}}^{\boldsymbol{\lambda}}_N(x)&:=\frac1N\sum_{i\le N}\delta(\lambda_{i}^s -x)\label{ESD_denoising}
\end{align}
as follows:
\begin{align*}
\frac{1}{N^2}\ln p_{S,N}(\bLam^s)&=\frac\beta{2}\int d{\hat{\rho}}^{\boldsymbol{\lambda}}_N(x) d {\hat{\rho}}^{\boldsymbol{\lambda}}_N(y)\ln|x-y|\nn
&\qquad-\frac {\beta }{4}\int d {\hat{\rho}}^{\boldsymbol{\lambda}}_N(x) V(x)+c_N.  
\end{align*}
The ``constant'' $c_N$ is formally infinity and takes care of removing the divergence coming from the doule integral when $x=y$; it simplifies with the same constant appearing in $\tau_N$. This combined with \eqref{sphInt_lim} suggests that the formula \eqref{formula:conj1} can be expressed as an optimization over a density of eigenvalues. To see that, note that the integral  \eqref{f_N1} can be written as
\begin{align}
 \frac1{N^2}\ln \int D\hat p_{S,N}[\hat \rho] &\exp N^2\Big(-\frac{\beta\lambda }{4}\int d\hat \rho(x) x^2\nn
 &+I^{(\beta)}[\hat \rho ,\rho_Y,\sqrt \lambda]+o_N(1)\Big)\label{int_denoi_density}
\end{align}
where the density of the ESD induced by $p_{S,N}(\bLam^s)$ can be formally written as
\begin{align}
 D\hat p_{S,N}[\hat \rho]=D[\hat \rho]\int dp_{S,N}(\bLam^s)\delta({\hat{\rho}}^{\boldsymbol{\lambda}}_N-\hat \rho).\label{densityESD}
\end{align}
For a large class of potentials $V$ a large deviation principle (LDP) at scale $N^2$ with rate functional $L$ holds for the random ESD \cite{hiai2000large,arous1997large,arous1998large}:
\begin{align}
 D\hat p_{S,N}[\hat{\rho}]\sim D[\hat{\rho}]\exp\big(-N^2 L[\hat{\rho}]\big)  .\label{LDP}
 \end{align} 
If $L$ is known, plugging this expression in \eqref{int_denoi_density} and evaluating the integral by saddle point approximation over the density provides the desired expression.

Let us discuss how to obtain the rate function $L$. Given a potential $V$ defining a rotationally invariant ensemble of symmetric/Hermitian matrices given by \eqref{prior_gene}, \eqref{priorWithVan}, the associated rate function $L$ of the ESD $\hat \rho_N^{\boldsymbol{\lambda}}$ is rather generically given, up to an irrelevant additive constant, by \cite{hiai2000large,arous1997large,arous1998large} 
\begin{align}
L[\hat \rho]&=\frac\beta 2\int d\hat \rho(x)\Big( \frac{V(x)}2-\int d\hat \rho(y)\ln|x-y|\Big).\label{rateFunc}
\end{align}
A direct way to see that starting from \eqref{densityESD} is by using \eqref{typform} to get
\begin{align}
&\int dp_{S,N}(\bLam^s)\delta({\hat{\rho}}^{\boldsymbol{\lambda}}_N-\hat \rho)  \nn
&\quad= \exp {N^2}\Big(\frac\beta 2\int d\hat \rho(x)\, d\hat \rho(y)\ln|x-y|\nn
&\quad\quad-\frac{\beta} 4\int d\hat \rho(x)V(x)+\frac1{N^{2}}\ln \int d\bLam^s \delta({\hat{\rho}}^{\boldsymbol{\lambda}}_N-\hat \rho)\Big).
\end{align}
Now, the entropic contribution $H[\hat \rho]:=\ln \int d\bLam^s \delta({\hat{\rho}}^{\boldsymbol{\lambda}}_N-\hat \rho)$ can be evaluated by introducing a Fourier representation of the Dirac delta:
\begin{align*}
H[\hat\rho]&=    \ln \int d\bLam^s\int D[g] \nn
&\quad\times\exp\Big({\rm i}\int dx \,g(x)\big(N\hat\rho(x) -\sum_{i\le N}\delta(\lambda_i^s-x)\big)\Big)\nn
&\propto -N\int d\hat\rho(x)\ln \hat\rho(x) = O(N).
\end{align*}
We recognize $N$ times the Shannon entropy of $\hat\rho$ (see Section 4.2 in Chapter 4, equations (4.14)--(4.18) in \cite{livan2018introduction}, or Appendix C of \cite{fyodorov2007replica}). Therefore, at leading order $\exp\Theta(N^2)$ this entropic contribution is negligible and we identify from \eqref{LDP} the rate functional \eqref{rateFunc} from the LDP approach.

Now imagine that the potential $V$ underlying the RMT ensemble \eqref{priorWithVan} is not known but, instead, its asymptotic spectral density is. The generalized Tricomi formula (see, e.g., Chapter 5 of \cite{livan2018introduction}) states that for a rotationally invariant ensemble of symmetric/Hermitian random matrices whose j.p.d.f. of the matrix entries is \eqref{prior_gene}, the equilibrium spectral density $\rho$ describing its $O(1)$ eigenvalues verifies
\begin{align}
 \mathcal{H}_{\rho}(x):={\rm Pr}\int d\rho(t) \frac{1}{x-t}= \frac\beta 4 V'(x) 
\end{align}
where $\mathcal{H}_{\rho}$ is the Hilbert transform of density $\rho$ whose domain is $\mathbb{R}$ and $\rm Pr$ is the principal value. When unknown, it allows to reconstruct the potential $V$ given $\rho$ by integration:
\begin{align}
 V(x) = \frac 4\beta \int^x \mathcal{H}_{\rho}(t)dt + C\label{Vfromrho}
\end{align}
for a certain irrelevant constant $C$ that can be set to $0$ without loss of generality. Therefore at the level of rigor of this paper, a RMT ensemble of the form \eqref{priorWithVan} can equivalently be defined from the knowledge of the potential or its asymptotic spectral distribution. 

With all these results in hand we can now re-state the previous replica symmetric formula using a density order parameter:
\begin{conjecture}[Free entropy of Hermitian rotationally invariant matrix denoising]\label{conjectureHermitianRotInv_density}
 The asymptotic free entropy of model \eqref{model-exactly} when the signal is drawn from the random matrix ensemble \eqref{priorWithVan} is
\begin{align*}
f_N\to\! &\sup_{\rho_s\in \mathcal{P}}\! \Big\{\frac\beta2\int d \rho_s(x)d \rho_s(y)\ln|x-y|- \frac{\beta\lambda }4\int d\rho_s(x)\,x^2 \nn
&\qquad-\frac\beta 4\int d \rho_s(x) V(x)+I^{(\beta)}[ \rho_s,\rho_{Y},\sqrt{\lambda}]\Big\}+\tau.
\end{align*}
The optimization is over the set $\mathcal{P}$ of probablity densities with finite support. The constant $\tau$ fixes the constraint $f_N({\lambda=0}) = 0$ and is given by
\begin{align*}
-\sup_{\rho_s\in \mathcal{P}} \Big\{\frac\beta2\int d \rho_s(x)d \rho_s(y)\ln|x-y|-\frac\beta 4\int d \rho_s(x) V(x)\Big\}.
\end{align*}
\end{conjecture}

\subsection{Simplifications in the Bayes-optimal setting using the Nishimori identity}  

The above conjectures have already reduced the computation of an integral over $\Theta(N^2)$ degrees of freedom (the matrix elements) onto an optimization problem over $\Theta(N)$ eigenvalues or a functional optimization over a density. But we claim that because we are in the Bayesian optimal setting the formulas can be further simplified in the form of Result~\ref{conjectureHermitianRotInv_simple} below. In the matched setting the posterior is the ``correct'' one, and as a consequence, a fundamental property known as the Nishimori identity holds. This identity states that for any well-behaved function $g:\mathbb{R}^{N\times N}\mapsto \mathbb{R}$ we have (here we state a restricted form of the most general identity found in \cite{barbier2019overlap})
\begin{align}
  \mathbb{E}\langle g(\bs)\rangle  =\mathbb{E}\,g(\bS) \label{Nishi}
\end{align}
where the signal $\bS\sim P_{S,N}$, while $\bs$ is a sample from the Bayes-optimal posterior j.p.d.f.
\begin{align*}
&dP_{S\mid Y,N}(\bs\mid \bY)\nn
&\quad=\frac1{\mathcal{Z}(\bY)} dP_{S,N}(\bs)\exp{\rm Tr}\Big[-\frac{\beta N}{4}\big(\bY-\sqrt{\lambda}\bs\big)^2\Big],   
\end{align*}
and the Gibbs-bracket $\langle \,\cdot\,\rangle$ is the associated expectation. In particular we have
\begin{align}\label{Nishim}
\mathbb{E}\Big\langle \frac{{\rm Tr} \bLam_s^k}{N} \Big\rangle  = \mathbb{E}\frac{{\rm Tr} \bLam_S^k}{N},
\end{align}
for the $k$th moment of the empirical density of eigenvalues of the signal. 

We now give an heuristic argument based on four steps and leading to Result~\ref{conjectureHermitianRotInv_simple} below, and believe that this may be the starting point of a rigorous proof strategy. Recall that $\hat\rho_{N}^{\bLam}$ given by \eqref{ESD_denoising} is the ESD associated with the eigenvalues $\bLam^s$ of the posterior sample. Define its moments 
$$m_{k, N} := \int dx\, x^k \hat\rho_{N}^{\bLam}(x) = \frac{{\rm Tr}\bLam_s^k}N .$$
\begin{enumerate}
\item
First, note that in expression \eqref{f_N1} or \eqref{int_denoi_density} the ESD plays the role of an order parameter for a ``mean-field'' free entropy functional given by the logarithm of the integrand in \eqref{int_denoi_density}, or equivalently, by the functional to be extremized in Result~\ref{conjectureHermitianRotInv_density}. Concretely, one can express the integrand in \eqref{f_N1} or \eqref{int_denoi_density} entirely in terms of the moments $(m_{k,N})_{k\ge 1}$. 
\\ 
\item
Second, we assume that the ESD $\hat \rho_{N}^*$ corresponding to the extremizer in Result~\ref{conjectureHermitianRotInv} is such that its moments $m_{k,N}^*$ are close to the Gibbs averages $\langle m_{k,N}\rangle=N^{-1}\langle {\rm Tr} \bLam_s^k\rangle$. In other words $$m_{k,N}^* = \langle m_{k,N}\rangle+o_N(1).$$ This is a natural self-consistency hypothesis for any ``replica-symmetric'' mean-field theory, where the optimal value of the order parameter generally coincides with the Gibbs average (the reader may recall the solution of the Curie-Weiss model for the prime example of this mechanism). Replica symmetry, namely the self-averaging/concentration of the order parameters (or the moments $(m_{k,N})_k$), is generically rigorously valid in Bayes-optimal inference of low-rank models \cite{barbier2019overlap,barbier2020strong} and we think that this property extends to linear-rank regimes.\\
\item
Third, we assume that the Gibbs expectation of the moments concentrates with respect to the data $\bY$: $\langle m_{k,N}\rangle = \mathbb{E}\langle m_{k,N}\rangle + o_N(1)$. This translates to 
\begin{align*}
 \langle m_{k,N}\rangle &= N^{-1}\mathbb{E}\langle {\rm Tr} \bLam_s^k \rangle+o_N(1)\nn
 &=\mathbb{E}\langle (\lambda^s_1)^k \rangle+o_N(1). 
\end{align*}
This is again true in low-rank Bayes-optimal inference \cite{barbier2019overlap,barbier2020strong}.\\
\item
Finally, from the two previous points and the Nishimori identity \eqref{Nishim} we conclude $$m_{k,N}^* = \mathbb{E}(\lambda^S_1)^k+o_N(1).$$
\end{enumerate}

We have thus found that, somewhat remarkably, the extremizer in Result~\ref{conjectureHermitianRotInv} has a corresponding ESD matching the one of the 
signal $\hat\rho_{N}^* = \hat \rho_{S,N}$. Taking $N\to+\infty$, the argument becomes exact: the empirical density $\hat\rho_{N}^*\to \rho^*= \rho_{S}$ the asymptotic density of eigenvalues of the signal. This yields the formula \eqref{conj2_dens} below. In particular the supremum in both the non trivial term where the spherical integral appears in Result~\ref{conjectureHermitianRotInv} and the constant term $\tau_N$ (or $\tau$ in the case of infinite $N$) are the same. Consequently we obtain a greatly simplified expression for the mutual information using relation \eqref{mut_inf_conj1}, the fact that $N^{-1}{{\rm Tr}\bS^2}=N^{-1}{{\rm Tr}\bLam_S^2}$ concentrates onto $\mathbb{E}(S_1)^2$ when $N\to+\infty$, and the concentration assumption for the free entropy $\EE f_N=f_N+o_N(1)$. We also obtain a formula for the MMSE using the I-MMSE relation \eqref{IMMSE}.

\begin{conjecture}[Mutual information of Hermitian rotationally invariant matrix denoising]\label{conjectureHermitianRotInv_simple} 
 Let $\bLam^s\in\mathbb{R}^N$ be the eigenvalues of a random matrix $\bs\sim P_{S,N}$, i.e., $\bLam^s\sim p_{S,N}$. The mutual information of model \eqref{model-exactly} verifies
\begin{align*}
\frac{1}{N^2}I(\bY;\bS)=\frac{\beta \lambda}{2N}{{\rm Tr}\bLam_s^2}- I_N^{(\beta)}(\bLam^s,\bLam^Y,\sqrt{\lambda})+o_N(1).
\end{align*}
Introducing $\rho_s=\rho_S$ the asymptotic spectral density of $\bs\sim P_{S,N}$ we get for $N\to +\infty$
\begin{align}
\frac{1}{N^2}I(\bY;\bS)\to   \frac{\beta\lambda}2\int d\rho_s(x)\, x^2-I^{(\beta)}[ \rho_s,\rho_{Y},\sqrt{\lambda}].\label{conj2_dens}
\end{align}

We deduce from \eqref{IMMSE} and a convexity argument (as $(I(\bY;\bS))_N$ is a sequence of concave functions in $\lambda$) that the minimum mean-square error verifies
\begin{align}
&\frac{1}{N^2}\mathbb{E}\|\bS-\mathbb{E}[\bS\mid \bY]\|^2\nn
&\quad=\frac{2}N \EE{\rm Tr} \bLam_s^2-\frac{4}\beta\frac{d}{d\lambda }I_N^{(\beta)}(\bLam^s,\bLam^Y,\sqrt{\lambda})+o_N(1),\label{MMSE_conj2}
\end{align}
or, working with the eigenvalue densities, for $N\to +\infty$
\begin{align*}
&\frac{1}{N^2}\mathbb{E}\|\bS-\mathbb{E}[\bS\mid \bY]\|^2\nn
&\quad\to 2\int d\rho_s(x) \,x^2-\frac4\beta\frac{d}{d\lambda }I^{(\beta)}[ \rho_s,\rho_{Y},\sqrt{\lambda}].
\end{align*}
\end{conjecture}

\noindent\emph{Remark 1:} \ The spherical integral $I^{(\beta)}$ or $I^{(\beta)}_N$ is difficult to compute. One route is to try using the HCIZ formula, but it is known that the ratio of determinants involved in the formula (see Appendix~\ref{app:sphInt}) is notoriously difficult to evaluate analytically or even numerically. Another one is to employ its asymptotic hydrodynamic description \cite{matytsin1994large,guionnet2002large} but this is challenging too. The HCIZ formula can be evaluated exactly in very special cases (e.g., the uniform and Wigner cases below) or perturbatively (see Section~\ref{sec:pertexp}), or approximated by using sampling techniques \cite{leake2021sampling}. We wish to point out that an easy and nice application of the HCIZ formula is to check that the asymptotic mutual information obtained in Result~\ref{conjectureHermitianRotInv_simple} is the same for a signal $\bS$ or its centered (trace-less) version 
$\bS -  {\rm I}_{\rm d,N}\,N^{-1}{\rm Tr} \bS $ (where ${\rm I}_{\rm d,N}$ is the identity of size $N$); this can be checked using basic properties of determinants. We know a-priori that this should be so because in the Bayesian-optimal case the statistician knows the asymptotic value of $N^{-1}{\rm Tr} \bS\to \EE\lambda_1^S$ (which is nothing else than the first moment of distribution of the signal) and can subtract it from the data matrix, so information-theoretically it has no influence. \\

\noindent \emph{Remark 2:} \ Result~\ref{conjectureHermitianRotInv_simple} was first obtained in the thesis of C. Schmidt \cite{thesis_schmidt} in the real case $\beta=1$ (see Appendix 7). But what we believe are crucial steps and justifications were omitted in his derivation, and it is not obvious to us how the final (correct) result was obtained. In particular, \cite{thesis_schmidt} jumps from equation \eqref{f_N1} to the final Result~\ref{conjectureHermitianRotInv_simple} without justification (see the transition from equation (A.79) to (A.83) in Appendix 7 of \cite{thesis_schmidt}). \\

\noindent  \emph{Remark 3:} \ The formula for the MMSE involves the derivative of the HCIZ formula with respect to $\lambda$. This can be computed in cases where some expression for the asymptotic value $I^{(\beta)}[\rho_S, \rho_Y, \sqrt\lambda]$ is known. This is for example the case in the sanity checks of the next paragraphs, and also in terms of perturbative expansions presented in Section~\ref{sec:pertexp} for small and large signal-to-noise ratio. It is possible to deduce from the HCIZ formula an expression for the derivative directly in terms of the eigenvalues and eigenvectors of $\bY$. While this is not directly used in the present paper it could be of interest in numerical approaches and for the analysis of the various variational problems in this paper. For this reason we include it in Appendix~\ref{app:MMSEderivative}.\\

Let us comment on an a-priori quite surprising observation. Consider three scenarios for the prior over the eigenvalues $\bLam^S$ of the signal:
\begin{enumerate}[label=(\Alph*)]
  \item The prior over the eigenvalues is of the form \eqref{typicalcase}. When $N\to+\infty$ (which can be thought of as a vanishing temperature limit), strongly coupled eigenvalues $\bLam^S$ drawn according to $p_{S,N}(\bLam^S)$ ``freeze'' into configurations of low energy (which includes the external potential $V$ plus the long range Coulomb repulsion due to the Vandermonde). The resulting one-point marginal is a non-trivial density $\rho_S$.\\

  \item The prior is factorized as $p_{S,N}(\bLam^S)=\prod_{i\le N}\rho_S(\lambda^S_i)$, where $\rho_S$ corresponds to the asymptotic marginal from the prior in case (A). In this case the prior does not induce any sort of interaction among eigenvalues and fluctuations survive even in the limit $N\to+\infty$: the ``temperature remains finite'' and no ``freezing'' occurs.\\

  \item The eigenvalues are deterministic and given to the statistician, i.e., $p_{S,N}(\bLam^S)=\delta(\bLam^S-\bLam^S_0)$, where the fixed configuration $\bLam^S_0$ has an empirical density weakly converging to $\rho_S$.\\
\end{enumerate}
By construction these three priors have the same one-point marginals (in the large size limit). For example, in the Wigner case, (A) would correspond to \eqref{typicalcase} with $V(\bLam^S) = \bLam_S^2$, and case (B) to $p_{S,N}(\bLam^S)=\prod_{i\le N}(4-(\lambda^S_i)^2)^{1/2}/(2\pi)$ a product of semicircle laws. For case (C) one can generate a typical sample from priors (A) or (B) and fix it. For the Wishart ensemble it would correspond to $V(\bLam^S) = 2(1 - {M}/{N} - 1/{N} + {2}/(\beta N)) \ln \bLam^S  +  2(M/N)\bLam^S$ in \eqref{typicalcase} for case (A), and $p_{S,N}(\bLam^S)=\prod_{i\le N}\rho_{\rm MP}(\lambda^S_i)$ a product of Marcenko-Pastur laws \eqref{rhoMP}.

Now, we claim that in all three cases Result~\ref{conjectureHermitianRotInv_simple} holds without any difference apart from possible $o_N(1)$ corrections. Indeed, in case (C) the integration over $\bLam^s$ in \eqref{f_N1} is trivial and it gives directly Result~\ref{conjectureHermitianRotInv_simple}. In case (B) if one plugs $p_{S,N}(\bLam^S)=\prod_{i\le N}\rho_S(\lambda^S_i)$ in formula~\eqref{formula:conj1} the term $N^{-2}\ln p_{S,N}(\bLam^s)=o_N(1)$, so one may think that the prior has no influence on the formula. But this is not true, because the influence of this prior manifests itself through the data $\bY$ in the spherical integral which strongly depends on it. Going again through the four points above leading to the simplified Result~\ref{conjectureHermitianRotInv_simple}, one can see that they all remain valid. And because the moments $\EE(\lambda_1^S)^k$ are the same in scenarios (A) and (B) (and (C) as well), the last point based on the Nishimori identity \eqref{Nishim} identifies the same maximizing $(m^*_{k,N})$, i.e., the same optimal density $\rho^*= \rho_S$. 

Let us provide an alternative information-theoretic counting argument in order to obtain Result~\ref{conjectureHermitianRotInv_simple} ``directly'' without going through all the previous steps, and that justifies a-posteriori the equivalence of these seemingly very different situations at the level of the mutual information, which is thus insensitive to possible strong correlations between the eigenvalues of $\bS$ and only depends on their density. By the chain rule for mutual information it can be decomposed as (recall $\bS=\tilde \bU^\dagger  \bLam^S \tilde \bU$)
\begin{align*}
\frac{1}{N^2}I(\bY;\bS)&=\frac{1}{N^2}I\big(\bY;(\tilde \bU,\bLam^S)\big)\nn
&=\frac{1}{N^2}I(\bY;\tilde \bU\mid \bLam^S)+\frac{1}{N^2}I(\bY;\bLam^S).
\end{align*}
Now, because there are only $N$ eigenvalues while there are $N^2$ (resp. $N(N-1)/2$) angles defining the eigenbasis $\tilde \bU \in \mathcal{U}(N)$ (resp. $\tilde \bU \in \mathcal{O}(N)$), the second term in the right-hand side in the above decomposition is $O(1/N)$. Thus, at leading order $\Theta(N^2)$, the mutual information $I(\bY;\bS)$ and the one given the eigenvalues $I(\bY; \tilde \bU \mid \bLam^S)$ are equal. Said differently, there are so much fewer eigenvalues $\bLam^S$ than angular degrees of freedom and data points that their inference has comparably negligible cost. In particular in $I(\bY; \tilde \bU \mid \bLam^S)$ the set of eigenvalues is given, so that their correlations does not matter and the mutual information can only depend on their density $\rho_S$. Since the priors (A)--(C) above have the same density the corresponding mutual informations are identical. Finally, note that by the same arguments we also have that $I(\bY; \tilde \bU \mid \bLam^S)$ (and thus $I(\bY;\bS)$ too) is equal, at leading order in $N$, to $$I(\bY; \tilde \bU)=I(\sqrt{\lambda}\,\tilde \bU^\dagger\bLam^S\tilde \bU +{\bm \xi}; \tilde \bU).$$

\subsection{A sanity check: the case of a Wigner signal} Consider the problem of denoising a Wigner matrix: $\bS$ is itself a standard Wigner with same distribution as the noise $\bxi$. So $V(\bS)=\bS^2$ in \eqref{priorWithVan}. The data $\bY$ is therefore also a centered Wigner matrix with law 
$$
P(\bY)\propto \exp{\rm Tr}\Big[-\frac {\beta N}{4(1+\lambda)}\bY^2\Big]
$$ 
whose asymptotic spectral density is a semicircle of width $\sigma_Y:=\sqrt{1+\lambda}$. This case is completely decoupled in the sense that each i.i.d. entry of the matrix $\bS$ is corrupted independently by an i.i.d. gaussian noise, so we should recover the known formulas for scalar decoupled gaussian channels \cite{GuoShamaiVerdu}. This can be verified as follows: in this case the supremum over $\rho_s$ in Result~\ref{conjectureHermitianRotInv} is attained for $\rho_s$ being itself a semicircle of width $\sigma_s=1$. Note that in this particular case, this can be deduced without making use of the Nishimori identity by realizing that whenever $\lambda\to +\infty$ or $\lambda\to 0_+$ it has to be so. Indeed in the noiseless limit $\lambda\to +\infty$ the posterior is peaked on the ground-truth signal $\bS$ and thus a sample $\bs$ will match it and have the same spectrum $\bLam^s=\bLam^S$ whose density is a semicircle of width $1$. In the opposite completely noisy limit limit $\lambda=0$, a sample from the posterior is simply drawn according to the prior $P_{S,N}$ which is the law of a standard Wigner matrix. Therefore in both cases the density $\rho_s$ is a semicircle of width $1$, but only in the second case the actual eigenvalues will match those of $\bS$. For any intermediate value of $\lambda$ the eigenvalues $\bLam^s$ will be in a mixture that polarize more towards $\bLam^S$ as $\lambda$ increases, but which maintains the same asymptotic \emph{density}. In the complex case $\beta=2$, the asymptotic spherical integral $I^{(2)}[ \rho_s,\rho_{Y},\sqrt{\lambda}]$ has a closed expression when evaluated for two semicircle laws \cite{bun2014instanton}:
\begin{align}
I^{(2)}[ \rho_s,\rho_{Y},\sqrt{\lambda}]&=\frac12\Big(\sqrt{4\sigma(\lambda)^4+1}-1\nn
&\quad-\ln\big(1+\sqrt{4\sigma(\lambda)^4+1}\big)+\ln 2\Big),
\end{align}
where $\sigma(\lambda)^2:=\sqrt{\lambda}\sigma_Y\sigma_s=\sqrt{\lambda(1+\lambda)}$. Moreover, according to ``Zuber's $\frac12$-rule'' \cite{zuber2008large} we can simply relate the real case $\beta=1$ to the complex one $\beta=2$:
\begin{align}
I^{(1)}[ \rho_s,\rho_{Y},\sqrt{\lambda}]=\frac12 I^{(2)}[ \rho_s,\rho_{Y},\sqrt{\lambda}].
\end{align}
Using that the second moment of the semicircle law $\mathbb{E}(\lambda^S_1)^2=\int_{-2}^2 dx \,x^2\,\sqrt{4-x^2}/(2\pi) =1$ we reach from Result~\ref{conjectureHermitianRotInv_simple} the expected expression:
\begin{align}\label{MIgauss}
 &\frac1{N^2} I(\bY;\bS)\to \frac{\beta}{4}\Big(2\lambda+1-\sqrt{4\lambda(1+\lambda)+1}\nn
 &\quad+\ln\big(1+\sqrt{4\lambda(1+\lambda)+1}\big)-\ln 2\Big)=\frac{\beta}{4}\ln(1+\lambda).
\end{align}
The minimum mean-square error is thus
\begin{align}
\frac1{N^2}\mathbb{E}\|\bS-\mathbb{E}[\bS\mid \bY]\|^2  \to \frac{1}{1+\lambda}.  \label{MMSEgauss}
\end{align}
So we recover the formulas of \cite{GuoShamaiVerdu}. Note that in the present case, the convergence $\to$ in the above identities are actually equalities for any $N$ (but our derivation here is asymptotic in nature).

\subsection{An explicit model with uniform spectral distribution}
We consider model \eqref{model-exactly} with $\bLam^S$ being a uniform permutation of equally spaced eigenvalues in $[-\sqrt3,\sqrt3)$: 
\begin{align}
p_{S,N}(\bLam^S)&=\frac{1}{N!}\boldsymbol{1}\Big(\bLam^S\in \Pi\Big(\sqrt\gamma\Big(-\frac12,\frac1N-\frac12,\nn
&\quad\quad\frac2N-\frac12,\ldots,\frac12-\frac1N\Big)\Big)\Big)\label{priorUni}  
\end{align}
where $\Pi(v)$ is the set of all $N!$ permutations of $v\in \mathbb{R}^N$, $\boldsymbol{1}(\cdot)$ is the indicator function, and $\gamma=\gamma_N\to12$ enforces ${\rm Tr}\bLam_S^2=N$. The advantage of this model is that the HCIZ integral appearing in Result~\ref{conjectureHermitianRotInv_simple} is explicit when $\beta=2$. Let $\bLam^s\sim p_{S,N}$. The HCIZ integral (see Appendix~\ref{app:sphInt}) does not depend on the ordering of the eigenvalues, therefore we can consider the increasing ordering $\lambda_i^s=\sqrt\gamma(i-1)/N-\sqrt\gamma /2$. Denote $\sigma:=\gamma \lambda$. The HCIZ formula then gives (because the ratio of determinants is non-negative we can insert an absolute value)
\begin{align*}
N^2 I_N^{(2)}(\bLam^s,\sqrt\lambda\bLam^Y,1)
&=\ln \frac{\prod_{k\leq N-1} k!}{ N^{N(N-1)/2}}\nn
&\hspace{-2cm}+\ln\Big|
\frac{{\rm det} [(\exp\sqrt{\sigma} \lambda_{j}^Y)^{i-1}\exp( -N\sqrt{\sigma} \lambda_{j}^Y/2)]}{\Delta_N(\bLam^S)\Delta_N(\sqrt\lambda\bLam^Y)}\Big|\\
&=\ln \frac{\prod_{k\leq N-1} k!}{ N^{N(N-1)/2}}\nn
&\hspace{-2cm}+\ln\Big|
\frac{{\rm det} [(\exp\sqrt{\sigma}  \lambda_{j}^Y)^{i-1}]}{\Delta_N(\bLam^S)\Delta_N(\sqrt\lambda\bLam^Y)}\Big|-\frac N2\sqrt{\sigma}\,  {\rm Tr}\bLam^Y.
\end{align*}
The matrix $[(\exp\sqrt\sigma \lambda_{j}^Y)^{i-1}]$ is a generalized Vandermonde, and thus 
\begin{align*}
 {\rm det}[(\exp\sqrt\sigma \lambda_{j}^Y)^{i-1}]&=\prod_{i<j}^{1,N}(\exp\sqrt\sigma \lambda_{i}^Y-\exp\sqrt\sigma \lambda_{j}^Y).
\end{align*}
The mutual information from Result~\ref{conjectureHermitianRotInv_simple} then reads:
\begin{align}
&\frac{1}{N^2}I(\bY;\bS)=\lambda -\frac1{N^2}\sum_{i< j}^{1,N}\ln|\exp\sqrt\sigma \lambda_{i}^Y-\exp\sqrt\sigma \lambda_{j}^Y|\nn
&\quad+\frac1{N^2}\sum_{i< j}^{1,N}\ln(\sqrt\lambda|\lambda^Y_i-\lambda^Y_j|)+\frac1{N^2}\sum_{i< j}^{1,N}\ln\Big(\sqrt\gamma\frac{|i-j|}{N}\Big) \nonumber\\
&\quad+\frac{\sqrt\sigma}{2N} {\rm Tr}\bLam^Y-\frac1{N^2}\ln \frac{\prod_{k\leq N-1} k!}{ N^{N(N-1)/2}} +o_N(1).\label{MI_uni_finiteN}
\end{align}
The MMSE can then be obtained using the I-MMSE relation \eqref{IMMSE}:
\begin{align*}
 &\frac\beta{N^2}\mathbb{E}\|\bS-\mathbb{E}[\bS\mid \bY]\|^2\nn
 &\quad=4-\frac4{N^2}\sum_{i< j}^{1,N}\frac{e^{\sqrt\sigma \lambda_{i}^Y}\frac d{d\lambda}(\sqrt\sigma \lambda_{i}^Y)-e^{\sqrt\sigma \lambda_{j}^Y}\frac d{d\lambda}(\sqrt\sigma \lambda_{j}^Y)}{e^{\sqrt\sigma \lambda_{i}^Y}-e^{\sqrt\sigma \lambda_{j}^Y}}\\
 &\quad\quad+\frac1\lambda + \frac4{N^2}\sum_{i< j}^{1,N}\frac{\frac d {d\lambda}(\lambda^Y_i-\lambda^Y_j) }{\lambda^Y_i-\lambda^Y_j}+\nn
 &\quad\quad \sqrt{\frac{\gamma}{\lambda }}\frac1N{\rm Tr}\bLam^Y+\frac{2\sqrt\sigma}{N}\sum_{i\le N}\frac d{d\lambda}\lambda_i^Y+o_N(1).
\end{align*}
Introducing the $\bY$-eigenvectors $\bY \boldsymbol{\psi}_i^Y = \lambda_i^Y   \boldsymbol{\psi}_i^Y$, the Hellmann-Feynman theorem implies 
\begin{align*}
 \frac{d }{d \lambda}\lambda_i^Y = \frac{1}{2\sqrt\lambda} ( \boldsymbol{\psi}_i^Y)^\dagger  \bS  \boldsymbol{\psi}_i^Y =:  \frac{1}{2\sqrt\lambda}p_i 
\end{align*}
where $\bS$ is the ground-truth in \eqref{model-exactly} (not to be confused with $\bs$, another independent sample from $P_{S,N}$). As a consequence we finally obtain the explicit expression
\begin{align}
&\frac\beta{N^2}\mathbb{E}\|\bS-\mathbb{E}[\bS\mid \bY]\|^2\nn
&\quad=4-\frac{2\sqrt\gamma}{N^2}\sum_{i< j}^{1,N}\frac{e^{\sqrt\sigma \lambda_{i}^Y} (\frac1{\sqrt\lambda} \lambda_{i}^Y+p_i)-e^{\sqrt\sigma \lambda_{j}^Y}(\frac1{\sqrt\lambda} \lambda_{j}^Y+p_j)}{e^{\sqrt\sigma \lambda_{i}^Y}-e^{\sqrt\sigma \lambda_{j}^Y}}\nonumber \\
 &\quad\quad+\frac1\lambda+\frac2{\sqrt\lambda N^2}\sum_{i< j}^{1,N}\frac{p_i-p_j}{\lambda^Y_i-\lambda^Y_j}\nn
 &\quad\quad+\sqrt\frac\gamma{\lambda }\frac1N{\rm Tr}\bLam^Y+\frac{\sqrt\gamma}{N}\sum_{i\le N}p_i+o_N(1).\label{MMSE_unif}
\end{align}

Let us introduce the asymptotic spectral densities $\rho_s$ and $\rho_Y$ associated with the matrices $\bs$ and $\bY$. Then the above expression reads, in the large size limit $N\to+\infty$,
\begin{align}
&\frac{1}{N^2}I(\bY;\bS)\to \lambda+\frac{\ln \lambda\gamma}{4} \nn
&+\frac1{2}\int d\rho_Y(x)\,d\rho_Y(y)\ln\Big|\frac{x-y}{\exp x\sqrt{\lambda\gamma} -\exp y\sqrt{\lambda\gamma} }\Big|. \label{MI_uni_asympt}
\end{align}
We used that
\begin{align*}
&\frac1{N^2}\sum_{i< j}^{1,N}\ln\frac{|i-j|}{N}\to\frac1{2}\int_{[0,1]\times [0,1]}dx \,dy\ln|x-y|\nn
&\quad=-\frac34=\lim_{N\to+\infty} \frac1{N^2}\ln \frac{\prod_{k\leq N-1} k!}{ N^{N(N-1)/2}}
\end{align*}
so these two terms asymptotically cancel each others. We also used that $\bY$, as a sum of asymptotically trace-less matrices, is asymptotically trace-less too and therefore the term $\int d\rho_Y(x)\,x=0$. 

Note that, as explained below Result~\ref{conjectureHermitianRotInv_simple}, we could have fixed from the beginning one arbitrary permutation of the eigenvalues: $p_{S,N}(\bLam^S)=\delta(\bLam^S-\sqrt\gamma(-1/2,1/N-1/2,2/N-1/2,\ldots,1/2-1/N))$, instead of considering the uniform measure \eqref{priorUni} over permutations. This would have lead to the same calculations as can be easily seen. What is less trivial to see (because in that case we cannot simplify anymore the HCIZ formula using the generalized Vandermonde form) is that the result would be asymptotically the same if the prior was instead uniform but not necessarily equally spaced, i.e., $p_{S,N}=\mathcal{U}[-\sqrt3, \sqrt 3)^{\otimes N}$.

\begin{figure}[t!]
\centering
\includegraphics[width=8.5cm]{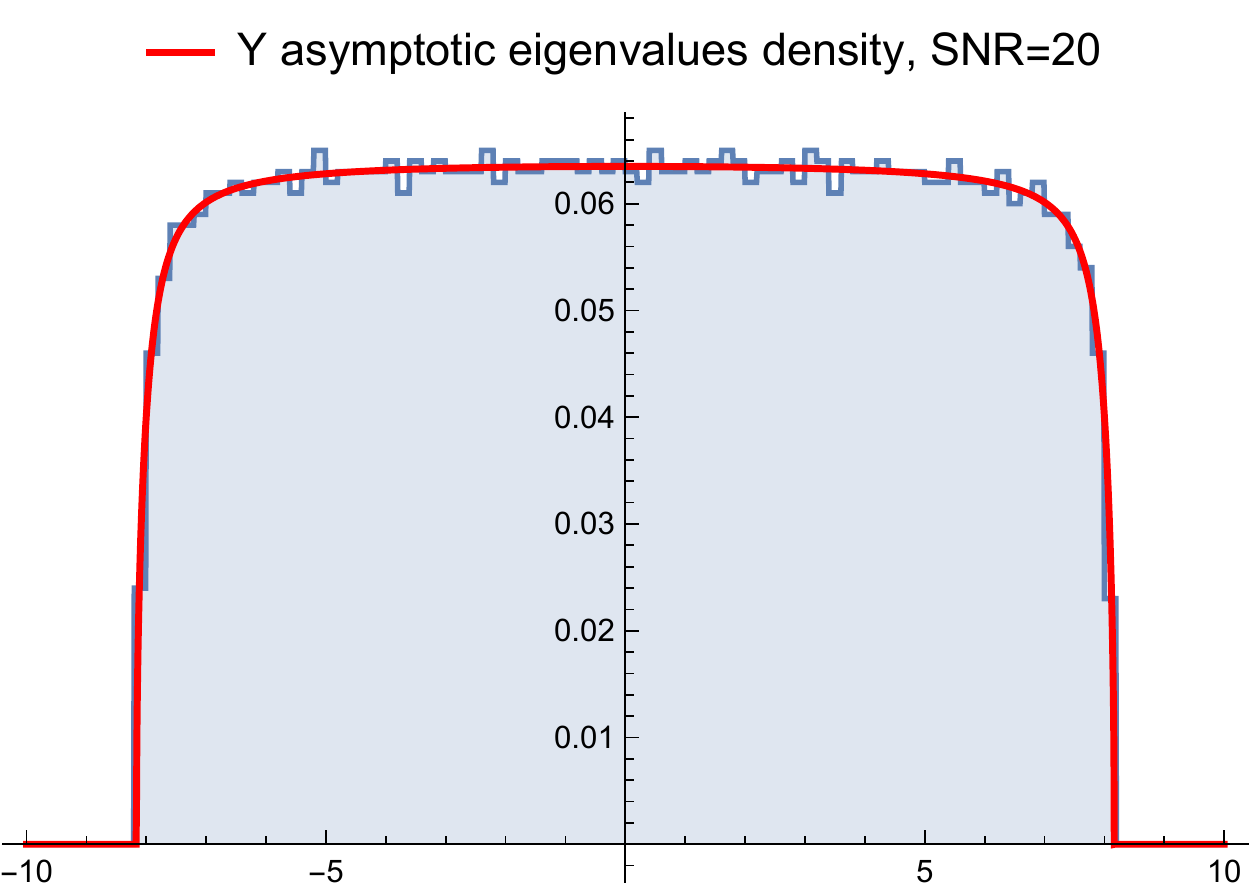}
\includegraphics[width=8.5cm]{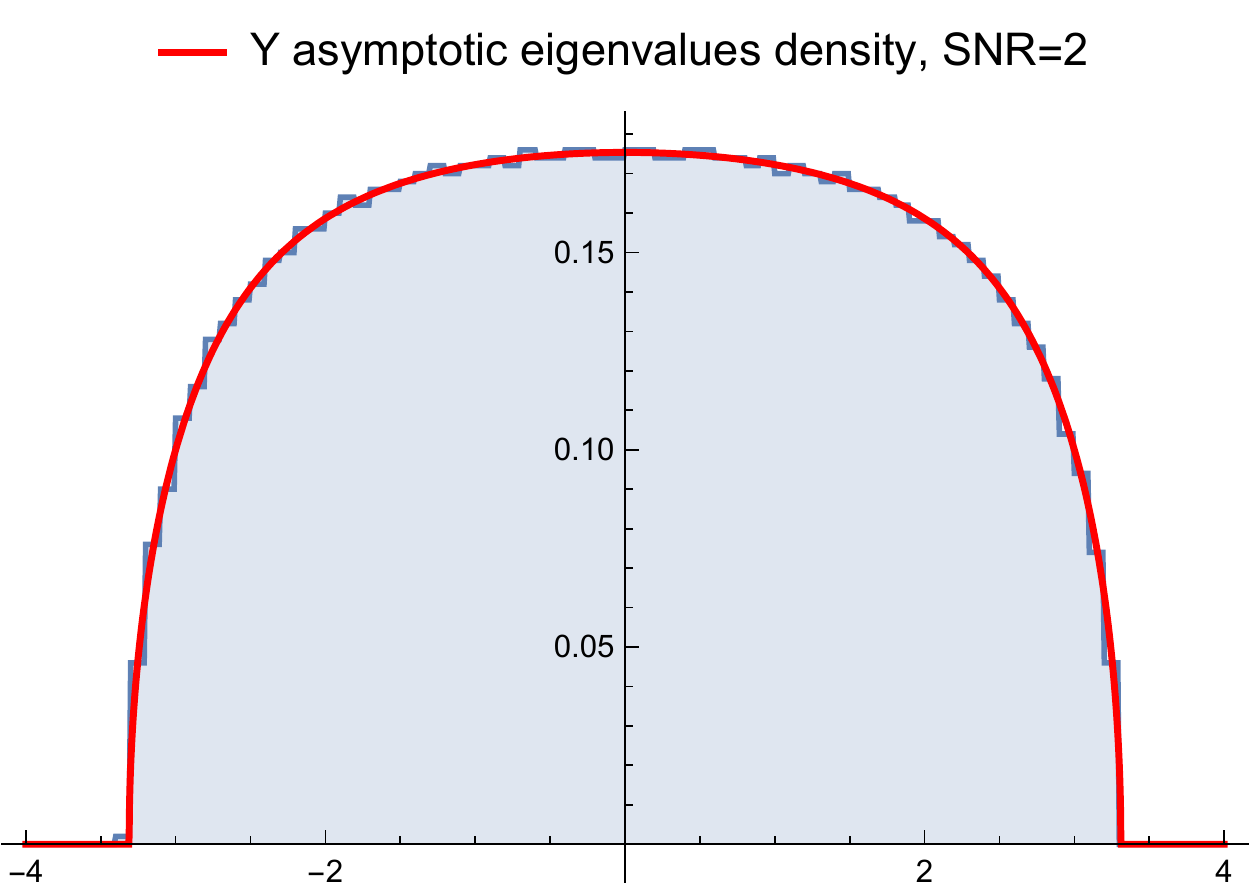}
\caption{\emph{Up:} Asymptotic $N\to+\infty$ spectral density $\rho_Y(x)$ (red) for the denoising model \eqref{model-exactly} with $\lambda=20$ and a signal $\bS$ with uniform eigenvalues in $[-\sqrt3,\sqrt3)$. It is compared to the empirical spectral density of $\bY$ for a realization of size $N=5000$ (blue). \emph{Down:} The same for a smaller signal-to-noise ratio $\lambda=2$. As expected, the spectrum ressembles more the semicircle law in that case. The density does approach Wigner's semicircle law of radius $2$ as $\lambda\to0_+$.}\label{Fig1}
\end{figure} 

The $\lambda$-dependent data spectral distribution $\rho_Y$ can be obtained from free probability as follows; we refer to \cite{potters2020first,mingo2017free} for clean definitions, domains of definitions and properties of the functions we are going to use now. The complex-valued Green function (or minus Stieljes transform) associated with $\rho$, whose domain is the complex plane minus the support of $\rho$, is
\begin{align*}
G_{\rho}(z):=\int d\rho(x) \frac{1}{z-x}.
\end{align*}
The Blue function is its functional inverse verifying $B_{\rho}(G_{\rho}(z))=G_{\rho}(B_{\rho}(z))=z$. Then the complex valued R-transform is defined as
\begin{align*}
 R_{\rho}(z)=B_\rho(z) -\frac1 z=\sum_{i\ge 1}k_{i}\,z^{i-1},
\end{align*}
where the coefficients $(k_{i})_{i\ge 1}$ in its series expansion are the so-called free cumulants associated with density $\rho$. Asymptotically, the matrix $\sqrt{\lambda}\bS$ has eigenvalue density $\rho_{\sqrt\lambda S}$ which is the uniform distribution in $[-\sqrt{3\lambda},\sqrt{3\lambda})$. The associated Green function is 
\begin{align*}
 G_{\rho_{\sqrt\lambda S}}(z)=\frac1{2\sqrt{3\lambda}}\ln\frac{z+\sqrt{3\lambda}}{z-\sqrt{3\lambda}}, 
\end{align*}
thus 
\begin{align*}
\mathcal{R}_{\rho_{\sqrt\lambda S}}(z)=\sqrt {3\lambda}\,{\rm coth}(z\sqrt {3\lambda})-\frac1z.  
\end{align*}
The R-transform of the standard Wigner semicircle law is the identity: $R_{\rho_Z}(z)=z$. Finally, by additivity of the R-transform for asymptotically free random matrices, the R-transform of the spectral density of the data matrix is $${R}_{\rho_Y}(z)={R}_{\rho_{\sqrt\lambda S}}(z)+{R}_{\rho_Z}(z)=\sqrt {3\lambda}\,{\rm coth}(z\sqrt{3\lambda})-\frac1z +z.$$ Its Blue function is thus ${B}_{\rho_Y}(z)=\sqrt {3\lambda}\,{\rm coth}(z\sqrt{3\lambda})+z$ from which we get a transcendental equation for its Green function: 
\begin{align}
z=\sqrt {3\lambda}\,{\rm coth}({G}_{\rho_Y}(z)\sqrt {3\lambda})+{G}_{\rho_Y}(z). \label{eq_for_Green}  
\end{align}
This equation can be solved numerically using a complex non-linear solver. A \textsc{Mathematica} code to do so is provided in Appendix~\ref{app:mathematica}. From its solution we can access the spectral density thanks to
\begin{align}
 \rho_Y(x) =\frac{1}{\pi}\lim_{\varepsilon\to 0_+}|\Im {G}_{\rho_Y}(x-{\rm i} \varepsilon)|. \label{rho_from_Green}  
\end{align}
Figure~\ref{Fig1} shows in red the asymptotic prediction from the spectrum extracted from the numerical solution of \eqref{eq_for_Green} and \eqref{rho_from_Green}. It almsot perfectly matches the empirical density of eigenvalues of $\bY$ for realisations of the model for large sizes, see the blue histograms.

\begin{figure}[t!]
\centering
\includegraphics[trim=0 6cm 0 0, clip, width=8.7cm]{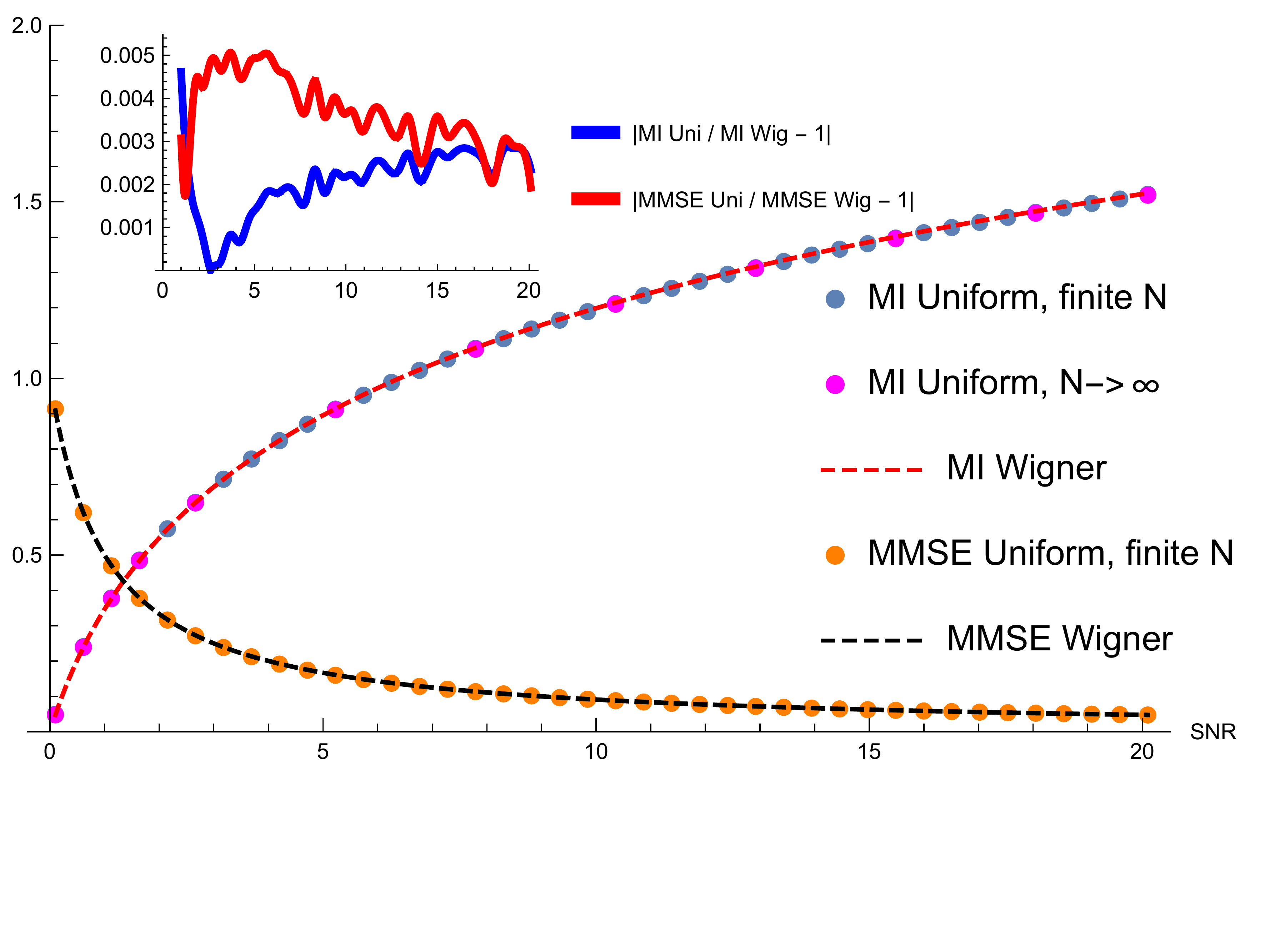}
\caption{\emph{Main:} The abcissa corresponds to the signal-to-noise ratio $\lambda$ in model~\eqref{model-exactly}. The blue dots correspond to the mutual information (MI) for the uniform spectrum case evaluated from \eqref{MI_uni_finiteN} for $N=1000$ averaged over $100$ independent realizations; the orange dots are the MMSE in the same monte-carlo experiment, evaluated from~\eqref{MMSE_unif}. The pink dots correspond to the asymptotic $N\to+\infty$ mutual information for the uniform spectrum case evaluated from~\eqref{MI_uni_asympt}. The finite $N$ and asymptotic $N\to+\infty$ values of the mutual information match very closely as can be seen from the superposition of the pink and blue dots. The red dashed line is the mutual information for the Wigner signal case~\eqref{MIgauss} and the black one the MMSE~\eqref{MMSEgauss}. All is for $\beta=2$. The curves for the uniform and semicircle laws match surprinsingly well but are actually different. \emph{Inset:} These curves quantify the relative difference between the empirical curves (the blue or orange dots) for the uniform case and the (dashed) curves for the Wigner case. The relative difference is typically of order $O(10^{-3})$. When comparing instead the $N\to+\infty$ curve for the uniform case (pink dots) to the Wigner mutual information so that any finite size effects are removed, a difference of the same order survives (which is much higher than the expected numerical precision for these computations). This confirms that the curves are \emph{not} exactly the same.}\label{Fig2}
\end{figure}

Given a signal-to-noise ratio $\lambda$, we can compute the mutual information as $N\to+\infty$ using \eqref{MI_uni_asympt}, \eqref{eq_for_Green} and \eqref{rho_from_Green}. This yields the pink dots of Figure~\ref{Fig2}. The blue dots are instead for the mutual information computed for large realizations of the model \eqref{model-exactly} for a $\bS$ with a uniform spectrum, see formula~\eqref{MI_uni_finiteN}. The orange dots are for the MMSE for that case, see formula~\eqref{MMSE_unif}. These curves are compared to the case of Wigner signal and match surprisingly well up to relative differences of $O(10^{-3})$. But our perturbative expansions of the next section as well as the comparison with the asymptotic predictions from \eqref{MI_uni_asympt} show that this difference, even if small, is not just due to numerical imprecisions: the curves really are different even in the large size limit. Yet, it is interesting to observe that the simple (decoupled) case of matrix denoising with $\bS$ a Wigner matrix allows us to very accurately approximate the information-theoretic quantities of the much less trivial setting where $\bS$ has a uniform spectrum (and therefore the matrix elements of $\bS$ are dependent, as opposed to the Wigner case). Further investigation around this fact is needed and left for future work.

%% file: sections/Expansions.tex
\section{Perturbative expansions for Hermitian matrix denoising}\label{sec:pertexp}

We exploit perturbative expansions of the HCIZ integral for small $\lambda$ to discuss the corresponding expansions of the mutual information and MMSE as predicted by Result~\ref{conjectureHermitianRotInv_simple}. In this section we only consider complex cases $\beta=2$, for which expansions of the HCIZ integral have been worked out \cite{collins2003moments,zinn2003some}.

Let $\bA$ and $\bB$ two hermitian $N\times N$ matrices.
We use an expansion of the HCIZ integral \eqref{IZhermitian} in the complex case $I_N^{(2)}(\bA, \bB, \sqrt\lambda)$ in terms of moments 
\begin{align*}
\theta_p &:= \lim_{N\to +\infty}\frac{1}{N} {\rm Tr} \bA^p = \lim_{N\to +\infty}\frac{1}{N}\sum_{i\le N} (\lambda_i^A)^p ,\nn
 \bar\theta_p &:= \lim_{N\to +\infty}\frac{1}{N} {\rm Tr} \bB^p = \lim_{N\to +\infty}\frac{1}{N}\sum_{i\le N} (\lambda_i^B)^p  
\end{align*}
for integer $p\geq 1$. Note that by concentration $\theta_p$ is also equal to $\lim_{N\to +\infty}N^{-1} \EE{\rm Tr} \bA^p$ and similarly for $\bar \theta_p$. We assume that $\bA$ and $\bB$ are {\it trace-less}, i.e, $\theta_1= \bar\theta_1 =0$. Then, according to \cite{zinn2003some},
\begin{align}\label{smallLambdaExp}
\lim_{N\to+\infty} I_N^{(2)}(\bA, \bB, \sqrt\lambda) &= I^{(2)}[\rho_A,\rho_B,\sqrt\lambda]\nn
&= \sum_{n\ge 2} \lambda^{\frac{n}{2}} F_n(\bA, \bB)
\end{align}
with (terms up to $n=8$ are explicitly derived in \cite{zinn2003some} and diagrammatic rules are given for higher orders; see also Appendix~\ref{app:mathematica} for their complete expressions)
\begin{align}
\begin{cases}
  F_2 &= \frac{1}{2}\theta_2 \bar\theta_2, \\
 F_3 &=\frac{1}{3}\theta_3 \bar\theta_3, \\ 
 F_4 &= \frac{3}{4}\theta_2\bar\theta_2 - \frac{1}{2}\theta_2^2\bar\theta_4 - \frac{1}{2}\theta_4 \bar\theta_2^2 + \frac{1}{4}\theta_4 \bar\theta_4.  
\end{cases}
\label{F123}
\end{align}
%
%
We will also make use of the derivatives with respect to the moments $\bar\theta_p$. These read
\begin{align*}
\frac{\partial}{\partial\bar\theta_p} I^{(2)}[\rho_A, \rho_B, \sqrt\lambda] = \frac{\bar{D}_{p}}{p} \quad \mbox{for} \quad  p\geq 2, 
\end{align*}
with
\begin{align}
\begin{cases}
 \bar{D}_2&= \lambda\theta_2 + \lambda^2\frac12( 3\theta_2 - 4\theta_4\bar\theta_2) + O(\lambda^{5/2}), \\ 
\bar{D}_3&= \lambda^{3/2} \theta_3 + O(\lambda^{5/2}), \\
\bar{D}_4&= -\lambda^2 (2\theta_2^2 - \theta_4) + O(\lambda^{3}).  
\end{cases}  
\end{align}
%
 %
%
where the higher order corrections come from the structure of $F_5$ and $F_6$ and can be worked out from \cite{zinn2003some}.

These formulas are applied for $\bA =\bS$ and $\bB = \bY = \sqrt\lambda \bS + \boldsymbol{\xi}$ with $N^{-1}{\rm Tr} \bS =o_N(1)$ and $N^{-1}{\rm Tr} \bY =o_N(1)$. 
As explained in Remark 1 after Result~\ref{conjectureHermitianRotInv_simple} the mutual information remains the same if we center the signal to make $\bS$ and $\bY$ trace-less. Although this is not necessary, and one can work out the expansion for a non-centered signal and data, this turns out to be a major simplification in the ensuing calculations. 
For the mutual information, according to Result~\ref{conjectureHermitianRotInv_simple} we find the expansion when $\beta=2$:
\begin{align}\label{exp:MI}
\lim_{N\to +\infty}\frac{1}{N^2}I(\bY;\bS) 
=   \lambda\theta_2 - \sum_{n\ge 2} \lambda^{\frac{n}{2}} F_n
\end{align}
and for the MMSE
\begin{align}
&\lim_{N\to +\infty}\frac{\beta}{N^2}\mathbb{E}\|\bS-\mathbb{E}[\bS\mid \bY]\|^2 \nn
&\quad =4\theta_2 - 4\frac{\partial}{\partial\lambda}I^{(2)}[\rho_S,\rho_Y,\sqrt{\lambda}] \nn
&\quad\quad- 4 \sum_{p\ge 2} 
\frac{\partial}{\partial\bar\theta_p}
I^{(2)}[\rho_S,\rho_Y,\sqrt{\lambda}] \frac{d\bar\theta_p}{d\lambda}
\nonumber  \\ 
&\quad= 4\theta_2  
- 4\sum_{n\ge 2} \frac{n}{2} \lambda^{\frac{n}{2}-1} F_n 
- 4 \sum_{p\ge 2} \frac{\bar{D}_{p}}{p} \frac{d\bar\theta_p}{d \lambda}.
\end{align}
In these expressions $F_n$ and $D_p$ are given by their expansions in terms of the moments 
\begin{align*}
\begin{cases}
\theta_p &= \lim_{N\to +\infty}N^{-1}{\rm Tr}\bS^p=\lim_{N\to +\infty}N^{-1}\EE{\rm Tr}\bS^p,\\
\bar\theta_p &= \lim_{N\to +\infty}N^{-1}{\rm Tr}\bY^p=\lim_{N\to +\infty}N^{-1}\EE{\rm Tr}\bY^p ,    
  \end{cases}  
\end{align*}
which themselves are polynomials in $\sqrt\lambda$. 
To go further we must fix a specific model of interest.\\

\noindent \emph{Example 1 (Wigner signal):} \  Let $\bxi=\bxi^\dagger \in \mathbb{C}^{N\times N}$, $\bxi\sim \exp{\rm Tr}[-\frac { N}2\bxi^2]$ a standard Hermitian Wigner matrix; this corresponds to a potential $V(x)=x^2$ in \eqref{priorWithVan}. Take an i.i.d. copy $\boldsymbol\xi^\prime$ and set $\bS = \boldsymbol{\xi}^\prime$ and 
$\bY = \sqrt\lambda \boldsymbol{\xi}^\prime + \boldsymbol{\xi}$. We note that $\bY \sim \sqrt{1+\lambda}\, \boldsymbol\xi$.
Wigner's semicircle law 
$$
\rho_\xi(x) = \boldsymbol{1}(\vert x\vert\leq 2)\frac{\sqrt{4 - x^2}}{2\pi} 
$$ 
implies for even moments (odd moments vanish)
\begin{align*}
  \theta_{2p}&=\lim_{N\to +\infty} \frac{1}{N} \mathbb{E} {\rm Tr} \boldsymbol{\xi}^{2p} = \frac{1}{p+1}\binom{2p}{p}, \nn
 \bar\theta_{2p}&=\lim_{N\to +\infty} \frac{1}{N} \mathbb{E} {\rm Tr} \bY^{2p} = \frac{(1+\lambda)^p}{p+1}\binom{2p}{p}.
\end{align*}
From $\theta_2 = 1$, $\theta_4 = 2$, $\theta_6 = 5$ and $\bar\theta_2 = 1+\lambda$, $\bar\theta_4 = 2 (1+\lambda)^2$, $\bar\theta_6 = 5(1+\lambda)^3$ we find  
\begin{align*}
 F_2 = \frac{1}{2}(1+\lambda), \quad F_4 = -\frac{1}{4}(1+\lambda)^2 \quad \mbox{and}\quad  F_6 = \frac{1}{3}(1+\lambda)^3.     
\end{align*}
The expansion \eqref{exp:MI} for the mutual information
yields 
\begin{align}
\lim_{N\to +\infty} \frac{1}{N^2}I(\bY;\bS) = \frac{\lambda}{2} - \frac{\lambda^2}{4} + \frac{\lambda^3}{6} + O(\lambda^4).
\end{align}
We recognize the expansion of $\frac{1}{2}\ln(1+\lambda)$ and the result is consistent  with \eqref{MIgauss}. \\

\noindent \emph{Example 2 (Wishart signal):} \ 
Consider $\bS = \bX \bX^\dagger$ and $\bY = \sqrt\lambda \bX \bX^\dagger + \boldsymbol{\xi}$, where the noise $\bxi\in\mathbb{C}^{N\times N}$ is an Hermitian Wigner matrix normalized as in the previous example and $\bX\in \mathbb{C}^{N\times M}$ is drawn from $P_{X,M}(\bX)\propto \exp{\rm Tr}[-M\bX\bX^\dagger]$. Let $\varphi:=N/M$. For $\varphi\leq 1$ the eigenvalue j.p.d.f. is well defined, and corresponds to the potential $V(x) = 2(1 - 1/\varphi) \ln\vert x\vert  +  2 x/\varphi$ in \eqref{priorWithVan} (see, e.g., \cite{potters2020first}).  When $\varphi >1$ the matrix is rank deficient and there is no well defined j.p.d.f. for the eigenvalues but the model can be regularized as explained in Section \ref{sec:2} and the final conjectures apply. In particular the Marcenko-Pastur distribution for the eigenvalues of $\bS = \bX \bX^\dagger$ is well defined for all $\varphi > 0$:
\begin{align}
\rho_{\rm MP}(x)&= \max(1-1/\varphi, 0)\delta(x) \nn
&\quad+ \frac{\boldsymbol{1}(c\leq x\leq d)}{2\pi\varphi x}\sqrt{(x-c)(d-x)}\label{rhoMP}
\end{align}
where $c:=(\sqrt\varphi -1)^2$ and $d:= (\sqrt\varphi +1)^2$.
The spectral moments are deduced by standard integration methods and we find 
\begin{align*}
&\lim_{N\to+\infty}\frac{1}{N} {\rm Tr} (\bX \bX^\dagger)^p\nn
&\quad=\lim_{N\to+\infty}\frac{1}{N} \mathbb{E} {\rm Tr} (\bX \bX^\dagger)^p = \frac{1}{p}\sum_{k\le p} \varphi^{k - 1} \binom{p}{k} \binom{p}{k-1}.  
\end{align*}
In particular $\lim_{N\to +\infty}N^{-1} \mathbb{E} {\rm Tr} \bX \bX^\dagger = 1$ for the first moment.

Now, as explained before, in order to compute the mutual information it is convenient to center the signal so that it becomes trace-less. In other words we replace 
$\bX \bX^\dagger$ by $\bS = \bX \bX^\dagger -  {\rm I}_{{\rm d},N}$ so that $N^{-1}{\rm Tr} \bS \to 0$. This also implies $\bY = \sqrt\lambda(\bX \bX^\dagger -  {\rm I}_{{\rm d},N}) + \boldsymbol{\xi}$ and $N^{-1}{\rm Tr} \bY \to 0$.
The first moments of the spectral density of this $\bS$ are in the asymptotic limit
$\theta_2 = \varphi$, $\theta_3 = \varphi^2$, $\theta_4 = \varphi^3 +2\varphi^2$, $\theta_5 = \varphi^4 + 5\varphi^3$, $\theta_6 = 5\varphi^3+9\varphi^4+\varphi^5$ and those of $\bY$ are $\bar\theta_1 = 0$, $\bar\theta_2 = 1+\lambda \varphi$, $\bar\theta_3 = \lambda^{3/2}\varphi^2$, $\bar\theta_4 = \lambda^2(\varphi^3+2\varphi^2) + 4\lambda\varphi +2$, $\bar\theta_5 = O(\lambda^{3/2})$, $\bar\theta_6 = 5 + O(\lambda)$.
This yields
\begin{align}
\begin{cases}
F_2 &= \frac{\varphi}{2} + \lambda\frac{\varphi^2}{2},\\ 
 F_3 &= \lambda^{3/2} \,\frac{\varphi^4}{3}, \\ 
F_4 &= -\frac{\varphi^4}{4} - \lambda\frac{\varphi^3}{2} + \lambda^2 \frac14(\varphi^6-\varphi^4), \\
F_5 &= O(\lambda^{3/2}), \\ 
F_6 &= \frac{\varphi^3}{3} - \frac{\varphi^4}{6} + O(\lambda).
\end{cases}
\end{align}
For the mutual information we find 
\begin{align}
&\lim_{N\to +\infty} \frac{1}{N^2}I(\bY;\bS) \nn
&\quad = \lambda\frac{ \varphi}{2} - \lambda^2\frac{\varphi^2}{4} + \lambda^3\frac{\varphi^3 -\varphi^4}{6} +O(\lambda^4).
\end{align}
We note that the contribution of the order $O(\lambda^3)$ only comes from the order $O(\lambda)$ in $F_4$ and the constant term in $F_6$.
We also remark that for $\varphi=1$ the first two orders are the same than the pure Wigner case of Example 1. It is possible to show that this is a universal feature for all matrices $\bS$ such that $N^{-1}{\Tr} \bS \to 0$ and $N^{-1}{\rm Tr} \bS^2 \to 1$, see the next example. \\

\noindent \emph{Example 3 (general case):} \ As before $(\theta_p)_{p\ge 1}$ correspond to the asymptotic spectral moments of the signal $\bS$. It easy to skim through the above calculations and obtain the first terms of an expansion for general signals with even spectral density such that  
$\theta_2=1$ and $\theta_{2p+1} =0$, $p\geq 0$ (note that for the trace-less Marchenko-Pastur distribution this is true only if $p=0$). At third order the resulting expansion can be read off by removing the contribution of $F_3$ and the term $-\varphi^4/6$ from $F_6$ and we find
\begin{align}
\lim_{N\to +\infty} \frac{1}{N^2}I(\bY;\bS) = \frac{\lambda}{2} - \frac{\lambda^2}{4} +\frac{ \lambda^3}{6} +O(\lambda^4).
\end{align}

The generic case until fourth order is heavy to handle by hand. We provide in Appendix~\ref{app:mathematica} a \textsc{Mathematica} code to get the following expansions. Let $(k_p)_{p\ge 2}$ be the free cumulants associated with the asymptotic spectral density $\rho_S$ of $\bS$ (see, e.g., \cite{bouchaudpotters,mingo2017free} to know about free cumulants). For $N\to+\infty$ followed by $\lambda\to 0_+$ and a $\rho_S$ such that $\theta_1=k_1=0$ and $\theta_2=k_2=1$,
\begin{align}
&\lim_{N\to +\infty}\frac{1}{N^2}I(\bY;\bS) \nn
&\quad = \frac\lambda 2 - \frac{\lambda^2}4+ \lambda^3\frac{1-k_3^2}6  - \lambda^4\frac{1+4 k_3^2+k_4^2}8 +o(\lambda^4). \label{expansion_generic_freeCumul}
\end{align}
Or expressed in terms of the moments (the mapping between moments and free cumulants can be obtained using the routines in Appendix~\ref{app:mathematica}),
\begin{align}
&\lim_{N\to +\infty}\frac{1}{N^2}I(\bY;\bS) \nn
&= \frac\lambda 2 - \frac{\lambda^2}4+ \lambda^3\frac{1-\theta_3^2}6  -  \lambda^4\frac{5\!+\!4 \theta_3^2\!+\!\theta_4^2\!-\!4\theta_4}8+o(\lambda^4).  \label{expansion_generic_mom}
\end{align}
Interestingly, for any even spectral density this matches the pure Wigner case of Example 1 up to third order. However this breaks down at fourth order as soon as the fourth moment $\theta_4$ of $\rho_S$ is different from $2$. This is another indication that the curves for the uniform spectrum and Wigner cases of Figure~\ref{Fig2} are different. Indeed, it can be checked that the $(F_n)$ in the expansion \eqref{exp:MI} for these two cases are very close but different. Or that their respective expansions \eqref{expansion_generic_mom} are the same up to order three, but the order four for the Wigner case is $-\frac18\lambda^4$ while it is $-\frac18\frac{26}{25}\lambda^4$ for the case of uniform spectrum in $[-\sqrt3,\sqrt3)$, and is thus very close.

If one is instead interested in obtaining expansions in the opposite large signal-to-noise ratio regime, it is possible to do so using the expressions found in \cite{bun2014instanton}.

%% file: sections/Denoising.tex
\section{Denoising of a rotationally invariant matrix: non-Hermitian case}\label{sec:non-H_matDnoising}

\subsection{The model}

We consider again a model of the form \eqref{model-exactly} but we now relax the hypothesis that $\bS$ is Hermitian. This time we consider that ${\bm \xi}$ is a standard (non-Hermitian) Ginibre matrix with law 
$$dP_{\xi,N}(\bxi)= C_N \,d\bxi \exp{\rm Tr}\Big[-\frac{\beta N}2\bxi\bxi^\dagger\Big].$$ Its entries are typically of order $O(1/\sqrt N)$ and singular values $O(1)$. The planted full-rank matrix signal $\bS\in \mathbb{K}^{N\times N}$ is no longer Hermitian but is still rotationally invariant in the sense that $$dP_{S,N}(\bS)=dP_{S,N}(\bO \bS\tilde \bO)$$ for any orthogonal/unitary $\bO,\tilde \bO$. It has $O(1/\sqrt{N})$ entries and $O(1)$ singular values. Its singular values decomposition (SVD) reads $\bS=\tilde \bU \bsig^S \tilde \bV$. Its singular values $\bsig^S$ have a generic empirical distribution converging as $N\to +\infty$ to $\rho_S$ with finite support. Left and right rotational invariance implies that its measure is decomposed as
\begin{align*}
dP_{S,N}(\bS)\propto  d\mu_N^{(\beta)}(\tilde \bU)\,d\mu_N^{(\beta)}(\tilde \bV)\, dp_{S,N}(\bsig^S),
\end{align*}
where the Vandermonde determinant and other terms inherent to the change of variable are included in the generic symmetric j.p.d.f. $p_{S,N}(\bsig^S)$ of the singular values. For example, in the case of a measure defined by a rotationally invariant potential it reads \cite{edelman2005random}
\begin{align}\label{prior_S_nonH_new}
dP_{S,N}(\bS)&\propto  d\mu_N^{(\beta)}(\tilde \bU)\,d\mu_N^{(\beta)}(\tilde \bV) \,d\bsig^S \big(\prod_{i\le N}\sigma^S_i\big)^{\beta-1}\nn
&\quad\quad\times\exp{\rm Tr}\Big[-\frac {\beta N}2V(\bsig^S)\Big]\,|{\Delta_N}(\bsig_S^2)|^\beta .
\end{align}

\subsection{Free entropy through random matrix analysis} 

Using the SVD $\bs=\bU\bsig^s\bV$ the free entropy reads 
\begin{align*}
f_N&:= \frac1{N^2}\ln \int dP_{S,N}(\bs)\exp \frac{\beta N}2 {\rm Tr}\Big[\sqrt{\lambda}\bY^\dagger\bs\nn
&\quad\quad\quad\quad+\sqrt{\lambda}\bY\bs^\dagger-\lambda\bs^\dagger\bs\Big]\nn
&\propto \frac{1}{N^2}\ln \int dp_{S,N}(\bsig^s) \exp\Big(-\frac{\beta\lambda N} 2{\rm Tr}\bsig_s^2\nn
&\quad\quad\quad\quad +(\beta-1)\sum_{i\le N}\ln \sigma_i^s\Big)\nn
&\times \int d\mu_N^{(\beta)}(\bU) \,d\mu_N^{(\beta)}(\bV)\exp\beta \sqrt{\lambda}N\Re {\rm Tr}\big[\bY^\dagger\bU{\bsig^s}\bV\big]\nn
&=\frac{1}{N^2}\ln \int d\bsig^s \exp {N^2}\Big(\frac{1}{N^2}\ln p_{S,N}(\bsig^s)-\frac{\beta\lambda}{2N}{\rm Tr}\bsig_s^2\\
&\quad\quad\quad\quad+J_N^{(\beta)}\big( {\bsig^s},\bsig^Y,2\sqrt{\lambda}\big)\Big)+O(1/N).
\end{align*}
Here and everywhere integrals over indivudual singular values are restricted to $\mathbb{R}_{\ge 0}$. The expression of the rectangular log-spherical integral density is \cite{schlittgen2003generalizations,ghaderipoor2008generalization}
\begin{align}\label{geneIZ}
J_N^{(\beta)}( \bA,\bB,\gamma)&=J_N^{(\beta)}( \bsig^A,\bsig^B,\gamma)\nn
&:=\frac1{N^2}\ln\int d\mu_N^{(\beta)}(\bU)\, d\mu_N^{(\beta)}(\bV)\nn 
&\quad\quad\times\exp \frac{\beta \gamma}2  N{\Re}{\rm Tr}\, \big[\bsig^A{\bm U} \bsig^B{\bm V}\big]
\end{align}
for generic $N\times N$ matrices $\bA$, $\bB$ with respective singular values $\bsig^A$ and $\bsig^B$. It has a well-defined limit \cite{forrester2016hydrodynamical,guionnet2021large}:
\begin{align}
J^{(\beta)}[ \rho_A,\rho_B,\gamma]:=\lim_{N\to+\infty}J_N^{(\beta)}( \bsig^A,\bsig^B,\gamma),	\label{lim_RectSphInt}
\end{align}
where $\rho_A,\rho_B$ are the asymptotic normalized densities of singular values associated with $\bA$ and $\bB$, respectively. Let $\rho_Y$ be the the asymptotic singular values density associated with the data $\bY$; again both $\bsig^Y$ and $\rho_Y$ are obtainable. In the large size limit $N\to +\infty$ we obtain by saddle-point the following conjecture for $f_N(\bY)$:
\begin{align*}
f_N= \sup_{\bsig^s\in\mathbb{R}_{\ge 0}^N} &\Big\{\frac{1}{N^2}\ln p_{S,N}(\bsig^s)-\frac{\beta \lambda}{2N} {\rm Tr}\bsig_s^2\nn
&\quad+J_N^{(\beta)}(\bsig^s,\bsig^Y,2{\sqrt{\lambda}})\Big\}+\tau_N.
\end{align*}
Focusing on the case of a prior of the form \eqref{prior_S_nonH_new}, 
\begin{align*}
f_N&= \sup_{\bsig^s\in\mathbb{R}_{\ge 0}^N} \Big\{\frac\beta2\sum_{i\neq j}^{1,N}\frac{\ln|(\sigma^s_i)^2-(\sigma^s_j)^2|}{N^2}\nn
&-\frac\beta 2\sum_{i\le N}\frac{\lambda(\sigma^s_i)^2+V({\sigma}^s_i)}{N}+J_N^{(\beta)}(\bsig^s,\bsig^Y,2{\sqrt{\lambda}})\Big\}+\tau_N.
\end{align*}
Introducing asymptotic densities of singular values and by the same type of LDP as in Section~\ref{sec:LDP}, the previous variational formula can be re-expressed as
\begin{align*}
f_N&\to \sup_{\rho_s\in\mathcal{P}_{\ge 0}} \Big\{\frac{\beta}{2}\int d\rho_s(x)\,d\rho_s(y)\ln|x^2-y^2| \\
&- \frac{\beta}2\int d\rho_{s}(x)\big(\lambda x^2+V(x)\big)+J^{(\beta)}[ \rho_{s},\rho_{Y},2\sqrt{\lambda}]\Big\}+\tau.
\end{align*}
The optimization is over a p.d.f. with bounded non-negative support. The constants $\tau_N$ and $\tau$ are fixed by the constraint $f_N({\lambda=0})=0$:
\begin{align*}
 \tau_N:=-\sup_{\bsig^s\in\mathbb{R}_{\ge 0}^N} &\Big\{\frac\beta2\sum_{i\neq j}^{1,N}\frac{\ln|(\sigma^s_i)^2-(\sigma^s_j)^2|}{N^2}\nn
 &\quad-\frac\beta 2\sum_{i\le N}\frac{V({\sigma}^s_i)}{N}\Big\} + o_N(1), \nn 
 \tau :=-\sup_{\rho_s\in\mathcal{P}_{\ge 0}} &\Big\{\frac{\beta}{2}\int d\rho_s(x)\,d\rho_s(y)\ln|x^2-y^2| \nn
 &\quad- \frac{\beta}2\int d\rho_{s}(x)\,V(x)\Big\} . 
 \end{align*} 
Again, as $f_N$ ends up being solely a function of the singular values of the data matrix, it is expected to be self-averaging with respect to $\bY$: $\EE f_N=f_N+o_N(1)$.

The free entropy is linked to the mutual information through
\begin{align*}
   I(\bY;\bS)=-\EE f_N +\frac{\beta \lambda N}{2}\mathbb{E} {\rm Tr}\bS\bS^\dagger.
\end{align*}
With the Ginibre noise instead of Wigner and for a non-Hermitian signal $\bS$ the I-MMSE relation reads
\begin{align}
\frac1{N^2}\mathbb{E}\|\bS-\mathbb{E}[\bS\mid \bY]\|^2=\frac{2}{\beta N^2}\frac{d}{d\lambda}I(\bY;\bS).\label{IMMSE_2}  
\end{align}
Like in the Hermitian case, the Nishimori identities combined with the concentration of the moments of the density of singular values of the posterior samples imply together that the supremum is attained for the density of singular values of the planted signal $\bS$.
\begin{conjecture}[Mutual information of rotationally invariant matrix denoising]\label{conjectureRotInv_simple}
Let the singular values $\bsig^s\sim p_{S,N}$ of a random matrix drawn according to the prior $P_{S,N}$. The mutual information of model \eqref{model-exactly} in the case where $\bS$ is not necessarily Hermitian and the noise $\bxi$ is a standard Ginibre matrix verifies
\begin{align*}
\frac1{N^2}I(\bY;\bS)=\frac{\beta \lambda}N {{\rm Tr} \bsig_s^2}-J_N^{(\beta)}(\bsig^s,\bsig^Y,2{\sqrt{\lambda}})+o_N(1).
\end{align*}
Introducing asymptotic densities of singular values it reads as $N\to+\infty$
\begin{align*}
\frac1{N^2}I(\bY;\bS)\to \beta\lambda\int d\rho_{s}(x)\, x^2-J^{(\beta)}[ \rho_{s},\rho_{Y},2\sqrt{\lambda}]
\end{align*}    
where $\rho_s=\rho_S$ is the asymptotic density of singular values of $\bs\sim P_{S,N}$.

We deduce from \eqref{IMMSE_2} that the minimum mean-square error verifies
\begin{align*}
&\frac{1}{N^2}\mathbb{E}\|\bS-\mathbb{E}[\bS\mid \bY]\|^2\nn
&\quad=\frac{2}N \EE{{\rm Tr} \bsig_s^2}-\frac2\beta\frac{d}{d\lambda }J_N^{(\beta)}(\bsig^s,\bsig^Y,\sqrt{\lambda})+o_N(1),
\end{align*}
or, working with asymptotic densities of singular values,
\begin{align*}
&\frac{1}{N^2}\mathbb{E}\|\bS-\mathbb{E}[\bS\mid \bY]\|^2\nn
&\quad\to 2\int d\rho_s(x) \,x^2-\frac2\beta\frac{d}{d\lambda }J^{(\beta)}[ \rho_s,\rho_{Y},\sqrt{\lambda}].
\end{align*}
\end{conjecture}

%% file: sections/DictionaryLearning-H.tex
\section{Hermitian positive definite dictionary learning} \label{sec:hermitianDL}

We now move to the more challenging problem of dictionary learning, first in the positive-definite case. Its analysis will require the introduction of the main methodological novelty of this paper: the spectral replica method.
 
\subsection{The model}

Consider a ground-truth matrix signal ${\bm X}=[X_{ik}]\in\mathbb{K}^{N\times M}$, with $$M=\alpha N+o(N)$$ with fixed $\alpha>0$, and with prior distribution $\bX\sim P_{X,N}$ which is centered $\mathbb{E}X_{ik}=0$ and such that typically $X_{ik}=O(1)$. This prior is \emph{not} necessarily rotationally invariant nor factorized over the matrix entries, but we require that it induces a permutation symmetric j.p.d.f. over the singular values of $\bX$. Let $\bZ=\bZ^\dagger\in\mathbb{K}^{N\times N}$ a noise Wigner matrix with p.d.f. 
$$
P_{Z,N}(\bZ)\propto \exp{\rm Tr}\Big[-\frac{\beta}4\bZ^2\Big].
$$ 
With this scaling the eigenvalues of $\bZ$ are $O(\sqrt{N})$. Consider having access to an Hermitian data matrix $\bY=[Y_{ij}]\in\mathbb{K}^{N\times N}$ with entries generated through the following observation channel:
\begin{align}\label{model}
\bY=\sqrt{\frac\lambda N}\,\bX\bX^\dagger + {\bm Z}.
\end{align}
Matrix $\sqrt{\lambda/N}\,\bX\bX^\dagger$ has  $O(\sqrt N)$ eigenvalues like the noise (and generically $O(1)$ entries), hence the scaling 
$\sqrt{\lambda/N}$ of the signal-to-noise ratio. The Bayesian posterior reads 
\begin{align*}
&dP_{X\mid Y, N}(\bx\mid \bY)\nn
&\ \ =\frac{1}{\mathcal{Z}(\bY)}dP_{X,N}(\bx)\exp\frac{\beta}{2}{\rm Tr}\Big[\sqrt{\frac{\lambda}{N}}\bY\bx\bx^\dagger-\frac{\lambda}{2N}(\bx\bx^\dagger)^2\Big].
\end{align*}
Note the invariance of the model under ${\bm X}\to {\bm X}{\bm U}$ for any $M\times M$ orthogonal/unitary ${\bm U}$ such that $P_{X,N}(\bX\bU)=P_{X,N}(\bX)$. 

The mutual information $I(\bY;\bX)$, which we aim at computing, is obtained by similar manipulations as in the previous sections:
\begin{align*}
\frac{1}{MN}I(\bY;\bX)&=-\frac1{MN}\mathbb{E}_\bY\ln \int dP_{X,N}(\bx)\nn
&\quad\times\exp\frac{\beta}{2}{\rm Tr}\Big[\sqrt{\frac{\lambda}{N}}\bY\bx\bx^\dagger-\frac{\lambda}{2N}(\bx\bx^\dagger)^2\Big]\nn
&\quad+\frac{\beta\lambda}{4MN^2}\EE{\rm Tr}(\bX\bX^\dagger)^2	
\end{align*}
where the first term is minus the expected free entropy $$\EE f_N:=\frac{1}{MN}\mathbb{E}_\bY\ln \mathcal{Z}(\bY).$$ 
Note that compared with our analysis on matrix denoising, we now do consider the expectation over the data in the free entropy.

In the case where $\bX$ is rotationally invariant, and therefore $\bX\bX^\dagger$ too, the results of the previous section on denoising can be applied. But it is important to notice right away that even in this case, \emph{the previous conjectures do not give any information about one of the main quantities of interest in order to evaluate the inference performance, namely, the scalar overlap between the ground-truth $\bX$ and a sample $\bx$ from the posterior $P_{X\mid Y,N}$}: 
\begin{align}
q:=\lim_{N\to+\infty}\frac1{N^2}\EE\big\langle \big|{\rm Tr}\bx\bX^\dagger\big|\big\rangle.\label{q}  
\end{align}
The absolute value is needed because $\bY$ contains no information about the sign of $\bX$, so $\bx$ and $-\bx$ have same posterior weight. Only the MMSE on the product $\bX\bX^\dagger$ is obtainable through this approach, through the I-MMSE identity. But this quantity is much less interesting than $q$ as it does not carry information about the reconstruction of the internal structure of $\bX\bX^\dagger$. In particular, as noted in \cite{thesis_schmidt}, in the present linear-rank regime of Hermitian dictionary learning with a factorized prior $P_{X,N}=g^{\otimes N(N+1)/2}$ over the matrix entries, the MMSE on $\bX\bX^\dagger$ is expected to be a universal quantity independent of the specific distribution $g$ of the individual entries of $\bX$ (as long as the first few moments exist). This is because the entries of $\bX\bX^\dagger$ are sums of many independent random contributions and thus the resulting matrix should behave at the level of the mutual information and MMSE on $\bX\bX^\dagger$ as a random matrix from the Wishart ensemble (i.e., as if $\bX$ had i.i.d. standard normal entries) due to strong universality properties \cite{tao2012random}. Therefore, it is crucial to:
\begin{itemize}
  \item access the (a-priori) non-universal $P_{X,N}$-dependent scalar overlap $q$, both in rotationally invariant and non rotationally invariant models;
\item go beyond models with factorized distributions over the components of the hidden matrices.\\
\end{itemize}

Concerning the second point: as it will become clear, the spectral replica method presented below does not \emph{a-priori} require the hidden matrix $\bX$ (or $\bS,\bT$ in the non-symmetric case) to have independent entries. But the possibility to concretely evaluate expressions in the ensuing conjectures depends on the solution of a classical (but in general highly non-trivial) RMT sub-problem, namely that of evaluating the j.p.d.f. of the matrix product between two i.i.d. samples from the prior $P_{X,N}$. As a consequence, in situations where this task can be solved (despite the lack of independence of the signal matrix entries) then quantitative predictions may be reachable. Advancing on the above two points is the main role of the spectral replica method as compared with the pure RMT approaches.

To fix ideas let us consider at the moment the complex case $\beta=2$. Model \eqref{model} is equivalent to three coupled real models:
\begin{align*}
\begin{cases}
\Re Y_{ij}\sim\mathcal{N}\big(\sqrt{\frac{\lambda}{N}}\frac{\langle\bX_{i},\bar \bX_{j}\rangle+\langle\bar \bX_{i},\bX_{j}\rangle}{2},\frac12\Big) \quad &\text{for} \quad i<j \in[N]^2,\nn
\Im Y_{ij}\sim\mathcal{N}\big(\sqrt{\frac{\lambda}{N}}\frac{\langle\bX_{i},\bar \bX_{j}\rangle-\langle\bar \bX_{i}, \bX_{j}\rangle}{2{\rm i}},\frac12\big)\quad &\text{for} \quad i<j \in[N]^2,\nn
Y_{ii}\sim\mathcal{N}\big(\sqrt{\frac{\lambda}{N}}\|\bX_{i}\|^2,1\big) \quad &\text{for} \quad i \in[N].
\end{cases}
\end{align*}
The average free entropy then concretely reads
\begin{widetext}
\begin{align}
\EE f_N&=\frac{1}{NM}\EE \ln \int dP_{X,N}(\bx)\exp\sum_{i\le N}\Big(\sqrt{\frac{\lambda}{N}}Y_{ii}\|\bx_{i}\|^2-\frac{\lambda}{2N}\|\bx_{i}\|^4\Big)\nn
&\qquad\qquad\qquad\qquad\times \exp2\sum_{i<j}^{1,N}\Big(\sqrt{\frac{\lambda}{N}}\Re  Y_{ij}\frac{\langle\bx_{i}, \bar \bx_{j}\rangle+\langle\bar \bx_{i}, \bx_{j}\rangle}{2}-\frac{\lambda}{2N}\Big(\frac{\langle\bx_{i}, \bar \bx_{j}\rangle+\langle\bar \bx_{i}, \bx_{j}\rangle}{2}\Big)^2\Big)\nn
&\qquad\qquad\qquad\qquad\times \exp2\sum_{i<j}^{1,N}\Big(\sqrt{\frac{\lambda}{N}}\Im  Y_{ij}\frac{\langle\bx_i,\bar \bx_j\rangle-\langle\bar \bx_i,\bx_j\rangle}{2{\rm i}}-\frac{\lambda}{2N}\Big(\frac{\langle\bx_i,\bar \bx_j\rangle-\langle\bar \bx_i,\bx_j\rangle}{2{\rm i}}\Big)^2\Big).\label{eq3}
\end{align}
\end{widetext}

\subsection{Replica trick and spectral replica symmetry}

The new important difficulty is that the potential lack of rotational invriance of $\bX$ and therefore of $\bS=\bX\bX^\dagger$ prevents the direct use of spherical integration of the rotational degrees of freedom. Moreover, we want to access the scalar overlap $q$, which random matrix theory alone seems not abble alone to reach. But combining random matrix theory with the replica method will allow to overcome these difficulties. The approach starts from the replica trick:
\begin{align}
\lim_{N\to+\infty}\EE f_N &=\lim_{N\to+\infty}\lim_{u\to0_+}\frac{1}{NMu}\ln\EE \mathcal{Z}(\bY)^u\nn
&=\lim_{u\to0_+}\lim_{N\to+\infty}\frac{1}{NMu}\ln\EE \mathcal{Z}(\bY)^u.\label{replicaTrick}
\end{align}
As in standard replica theory, we assume that the $u$ and $N$ limits commute, and later on that the formulas we will derive for integer $u$ can be analytically continued to real $u\to 0_+$. We therefore need to evaluate the expectation of the replicated partition function. We direclty integrate the quenched gaussian observations in \eqref{eq3} using the following useful formula: 
\begin{align}
Y\sim\mathcal{N}\Big(\sqrt{\frac\lambda N}f_0,\gamma\Big)\Rightarrow \ &\EE_{Y|f_0} \prod_{a\le u}\exp\gamma\Big(\sqrt{\frac{\lambda}N}Yf_a-\frac{\lambda}{2N}f_a^2\Big)\nn
&\quad=\prod_{a< b}^{0,u}\exp\frac{\gamma\lambda}{N}f_af_b.\label{useful_form}
\end{align}
In the Bayes-optimal setting the ground-truth $\bX$ plays a totally similar role as one additional replica, so we rename it $\bx^0:= \bX$. We set $\bx_i^a=(x_{ik}^a)_{k\le M}$ and introduce the notation $$\int dP_{X,N}(\{\bx\}_0^u)\,\cdots = \int_{\mathbb{R}^{MNu}}\prod_{a=0}^u dP_{X,N}(\bx^a)\,\cdots\,.$$ Define the complex-valued $M\times M$ overlap matrix
\begin{align}
{\bQ}^{ab}:=\Big(\frac{1}{N}\sum_{i\le N}x_{ik}^a\bar x_{i\ell}^b\Big)_{k,\ell\le M}=\frac{(\bx^a)^\intercal \bar \bx^b}{N}=({\bQ}^{ba})^\dagger.\label{overlap_HDiL}
\end{align}
Then the above formula \eqref{useful_form} applied thrice yields
\begin{widetext}
\begin{align}
\EE \mathcal{Z}(\bY)^u&=   \int dP_{X,N}(\{\bx\}_0^u)\prod_{a< b}^{0,u}\exp\frac{\lambda}{N}\sum_{i\le N}\|\bx_{i}^a\|^2\|\bx_{i}^b\|^2\nn
&\qquad\qquad\qquad\qquad\times \exp\frac{2\lambda}{N}\sum_{i<j}^{1,N}\frac{\langle \bx_{i}^a, \bar \bx_{j}^a\rangle+\langle\bar \bx_{i}^a, \bx_{j}^a\rangle}{2}\times\frac{\langle\bx_{i}^b, \bar \bx_{j}^b\rangle+\langle\bar \bx_{i}^b, \bx_{j}^b\rangle}{2}\nn
&\qquad\qquad\qquad\qquad\times \exp\frac{2\lambda}{N}\sum_{i<j}^{1,N}\frac{\langle\bx_i^a,\bar \bx_j^a\rangle-\langle\bar \bx_i^a,\bx_j^a\rangle}{2{\rm i}}\times\frac{\langle\bx_i^b,\bar \bx_j^b\rangle-\langle\bar \bx_i^b,\bx_j^b\rangle}{2{\rm i}}\nonumber\\
&=   \int dP_{X,N}(\{\bx\}_0^u)\prod_{a< b}^{0,u}\exp\frac{\lambda}{N}\sum_{i,j\le N}\Big(\Re \langle\bx_{i}^a, \bar \bx_{j}^a\rangle \Re \langle \bx_{i}^b, \bar \bx_{j}^b\rangle+\Im \langle\bx_i^a,\bar \bx_j^a\rangle\Im \langle\bx_i^b,\bar \bx_j^b\rangle\Big)\nn
&=   \int dP_{X,N}(\{\bx\}_0^u)\prod_{a< b}^{0,u}\exp\frac{\lambda}{N}\sum_{i,j\le N}\Re \big(\langle\bx_{i}^a, \bar \bx_{j}^a\rangle\langle\bar \bx_i^b, \bx_j^b\rangle\big)\nn
&=  \int dP_{X,N}(\{\bx\}_0^u) \prod_{a< b}^{0,u} \exp\lambda N\sum_{k,\ell\le M}\Big(\frac1N\sum_{i\le N}x_{ik}^a \bar x_{i\ell}^b \Big)\Big(\frac1N\sum_{j\le N}\bar x_{jk}^a x_{j\ell}^b\Big) \nn
&=   \int dP_{X,N}(\{\bx\}_0^u) \prod_{a< b}^{0,u} \exp\lambda N{\rm Tr}{\bQ}^{ab}({\bQ}^{ab})^\dagger.\label{complex_eq}
\end{align}
\end{widetext}
%

Until now the computation is standard. The novelty starts here. Let the singular value decomposition
\begin{align*}
\bQ^{ab}={\bA}^{ab}{\bsig}^{ab}{\bB}^{ab}.
\end{align*}
All matrices are of size $M\times M$; note that the overlaps $(\bQ^{ab})_{a<b}$ have rank equal to $n:=\min(N,M)$, so $({\bsig}^{ab})_{a<b}$ have $n$ non-zero entries on their diagonal. The dependence of the replicated system in $(\bx^a)_a$ is through the overlap matrices $(\bQ^{ab})_{a<b}$. Changing variables for $(\bQ^{ab})_{a<b}$ we have the completely generic change of density
\begin{align}
dP_{X,N}(\{\bx\}_0^u)&=  dP_{(Q),M}((\bQ^{ab})_{a<b}) \nn
&=  dP_{(A,B)|(\sigma),M}(({\bA}^{ab},\bB^{ab})_{a<b} \mid ({\bsig}^{ab})_{a<b}) \nn
&\quad\times dP_{(\sigma),M}(({\bsig}^{ab})_{a<b}) \label{changeDensity}
\end{align}
for a generic conditional j.p.d.f. $P_{(A,B)|(\sigma),M}$ of the singular vectors and j.p.d.f. of singular values $P_{(\sigma),M}$. With this, and because the density over singular vectors can be integrated right away, we now have
\begin{align}
&\EE \mathcal{Z}^u=   \int  dP_{(\sigma),M}(({\bsig}^{ab}))\prod_{a< b}^{0,u} \exp\frac \beta 2\lambda N{\rm Tr}(\bsig^{ab})^2.\label{replicatePart_bforeAnsatz}
\end{align}
This measure couples all matrices of singular values. It thus seems hopeless to go further without assuming some sort of simplification. We are now in position to move forward thanks to a novel type of decoupling assumption, which we think is the weakest (and most natural) possible assumption allowing to carry on computations from there.

The \emph{spectral replica symmetric ansatz} states that the replicated partition function is dominated by configurations such that the joint law $P_{(\sigma),M}(({\bsig}^{ab})_{a<b})$ factorizes as $N,M\to+\infty$ into a product of identical laws:
\begin{gather}
\text{Assumption (spectral replica symmetry):} \nn P_{(\sigma),M}(({\bsig}^{ab})_{a<b}) \to \prod_{a<b}^{0,u}\tilde p_{M}({\bsig}^{ab}).\label{RS_assump}
\end{gather}
The convergence $\to$ means that both the left and right hand sides weakly converge to the same asymptotic distribution as $N\to+\infty$. The j.p.d.f. $\tilde p_{M}({\bsig}^{ab})$, which is shared by assumption by the different overlap matrices/pairs of replica indices $a<b$, corresponds to the j.p.d.f. of the singular values of the overlap $N^{-1} (\bx^a)^\intercal \bar \bx^b$ (or equivalently $N^{-1} (\bx^b)^\dagger \bx^a$) between two i.i.d. matrices $\bx^a,\bx^b$ drawn from the prior $P_{X,N}$, and can thus in theory be obtained using random matrix theory; we will discuss further this point in the next section. Denoting $\bsig^{ab}=\bsig^{ab}(\bx^a,\bx^b)$ the singular values of the overlap,
\begin{align}
 \tilde p_M(\bsig)=\int dP_{X,N}(\bx^a) dP_{X,N}(\bx^b)\delta(\bsig^{ab}-\bsig). 
\end{align}
The law $\tilde p_M$ enforces $M-\min(N,M)$ singular values to be identically zero. We denote by $p_{n}$ the law of the remaining $n$ non-zero singular values only, that we continue to denote $\bsig^{ab}$ with a slight abuse of notation. Note that we did \emph{not} assume anything at the level of singular vectors which is an important feature. The decoupling is only assumed at the spectral level. So we have now that at leading exponential order and as $N$ gets large,
\begin{align}
\EE \mathcal{Z}^u &=   \int  \prod_{a< b}^{0,u} d\tilde p_M(\bsig^{ab}) \exp\frac \beta 2\lambda N{\rm Tr}(\bsig^{ab})^2\nn
&=   \int  \prod_{a< b}^{0,u} dp_n(\bsig^{ab}) \exp\frac \beta 2\lambda N{\rm Tr}(\bsig^{ab})^2\nn
&=\Big(\int dp_n(\bsig) \exp  \frac \beta 2\lambda N {\rm Tr}\bsig^2\Big)^{u(u+1)/2}.\label{integraltosolve}
\end{align}
Therefore the replica computation gives, using formula \eqref{replicaTrick}, that the expected free entropy limit is given by
\begin{align*}
\lim_{N\to+\infty}\frac1{2MN}\ln \int d\bsig \exp MN \Big(\frac{\ln p_n(\bsig)}{MN}+ \frac {\beta\lambda } {2M} {\rm Tr}\bsig^2 \Big)       
\end{align*}
which can be estimated by saddle-point.

\subsection{Replica symmetric formula}

Our computation shows that the order parameters are the (non-zero) singular values $\bsig$ of the overlap matrix $\bQ:=N^{-1}\bx^\dagger \tilde\bx$ between i.i.d. samples $\bx,\tilde \bx$ from the posterior distribution $P_{X\mid Y, N}$ (i.e., two conditionally independent replicas). Let us see how the scalar overlap $q$ defined by \eqref{q} can be deduced from it. A general Nishimori identity reads (see \cite{barbier2019overlap})
\begin{align}
  \mathbb{E}\langle g(\bx,\tilde \bx)\rangle  =\mathbb{E}\langle g(\bx,\bX)\rangle. \label{Nishi_2}
\end{align}
When two or more replicas appear inside a Gibbs bracket $\langle \,\cdot\,\rangle$ it has to be understood as the expectation with respect to the product Gibbs measure. From it, one can deduce the non-universal scalar overlap $q$. Indeed, the latter is equal to 
\begin{align}
q&:=\lim_{N\to+\infty}\frac1{N^2}\EE\big\langle \big|{\rm Tr}\bx\bX^\dagger\big|\big\rangle=\lim_{N\to+\infty}\frac1N\EE\big\langle\big|{\rm Tr} \bQ\big|\big \rangle\nn
&=\lim_{N\to+\infty}\frac1N\big|{\rm Tr} \EE\langle\bQ\rangle \big| =\frac1N\big|{\rm Tr} \bQ \big| + o_N(1).\label{q_2}  
\end{align}
The second equality follows from \eqref{Nishi_2}, while the third and last by concentration of the spectral moments of $\bQ$; this does not mean that $\bQ$ concentrates elementwise, only the moments $N^{-1}{\rm Tr}\bQ^k=N^{-1}{\rm Tr}\EE\langle \bQ\rangle^k+o_N(1)$ do. Finally, because $\EE\langle \bQ \rangle=\EE[\langle\bx\rangle\langle\bx\rangle^\dagger]$ is positive definite, its trace is also the sum of its singular values which, by the assumed self-averaging of the (moments of the) distribution of singular values, must be relatively close to ${\rm Tr} \bsig$ of singular values of $\bQ$ (which is not symmetric). Therefore, the mean of the singular values of the overlap matrix yield the scalar overlap $q$.
\begin{conjecture}[Replica symmetric formula for Hermitian dictionary learning]\label{conj:3}
Let $p_{n}$ be the j.p.d.f. of the $n:=\min(M,N)$ non-zero singular values (which are $O(1)$) of the matrix $N^{-1}\bx^\dagger \tilde \bx$, where ${\bm x},\tilde {\bm x}$ are i.i.d. $N\times M$ (not necessarily rotationally invariant) random matrices drawn from the prior $P_{X,N}$. 

The mutual information of model \eqref{model} verifies
%
\begin{align}
&\frac1{MN}I\Big(\bX;\sqrt{\frac{\lambda}{N}}\bX\bX^\dagger+\bZ\Big) \nn
&=-\frac{1}{2}\sup_{\bsig\in \mathbb{R}^n_{\ge 0}}\Big\{\frac{\ln p_{n}(\bsig)}{MN}+\frac{\beta\lambda}{2M} {\rm Tr}\bsig^2\Big\}\nn
&\quad+ \frac{\beta\lambda}{4} \frac{\mathbb{E}{\rm Tr}(\bX\bX^\dagger)^2}{MN^2}+\frac{1}{2}\sup_{\bsig\in\mathbb{R}^n_{\ge 0}}\frac{\ln p_{n}(\bsig)}{MN}+o_N(1).
\label{conj3:RS1}
\end{align}
Denote $\bsig^*$ the overlap singular values achieving the supremum in \eqref{conj3:RS1}. The overlap \eqref{q_2} is then
\begin{align}
 q=\frac1 N {\rm Tr}\bsig^* +o_N(1).
\end{align}
\end{conjecture}

\subsection{Validity of the spectral replica method in a simple scenario}\label{sec:simple-scenario}
Let us show that Result~\ref{conj:3} is correct in a simple (but non-trivial) setting. Let $N=M=n$. Consider the case where $\bX = \sqrt{N} \bO$ for an orthogonal ($\beta=1$) or unitary ($\beta=2$) matrix $\bO$. Then $\bX\bX^\dagger$ is $N$ times the identity and, therefore, the mutual information is null (the data $\bY$ being independent of $\bX$). The law $p_n$ in Result~\ref{conj:3} thus simply enforces all singular values to be one for any $\lambda$. Thus \eqref{conj3:RS1} becomes
\begin{align*}
 0 &=-\frac{\beta\lambda}{4N} {\rm Tr}\bsig^2 + \frac{\beta\lambda}{4} \frac{\mathbb{E}{\rm Tr}(\bX\bX^\dagger)^2}{N^3}
\end{align*}
with all $\sigma_i=1$, which is indeed true.

\subsection{Working with a density order parameter, and a weaker spectral replica symmetric ansatz}\label{sec:work_dens_param}

Let us show that the variational problem in Result~\ref{conj:3} can equivalently be expressed as an optimization over a density, similarly as in Section~\ref{sec:LDP}. For the rest of this section we consider that the prior $P_{X,N}$ defines a rotationally invariant ensemble of random matrices in the sense that matrices of left and right singular vectors of $\bX\sim P_{X,N}$ are Haar distributed. Another interest of the following derivation is that it will be based on a weaker form of the spectral replica symmetric ansatz \eqref{RS_assump}.

We start again from the averaged replicated partition function \eqref{replicatePart_bforeAnsatz}. Because the exponents are functions of the non-zero singular values only, we can simply think of $P_{(\sigma),M}(({\bsig}^{ab}))$  entering \eqref{replicatePart_bforeAnsatz} as the j.p.d.f. of these $n$ non-zero singular values. The empirical distribution $n^{-1}\sum_{k\le n}\delta((\sigma_{k}^{ab})^2 -x)$ of the non-zero squared singular values of the overlap $\bQ^{ab}$ is also the empirical distribution ${\hat{\rho}}^{ab}_n$ of the non-zero eigenvalues $(\lambda_i^{ab})_{i\le n}$ of the non-negative definite Hermitian matrix ${\bQ}^{ab}({\bQ}^{ab})^\dagger$:
\begin{align*}
{\hat{\rho}}^{ab}_n(x)&:=\frac1n\sum_{i\le n}\delta(\lambda_{i}^{ab} -x).
\end{align*}
The j.p.d.f. $P_{(\sigma),M}((\bsig^{ab})_{a<b})$ induces a joint probability density $\hat P_{n}[({\hat{\rho}}^{ab})_{a<b}]$ for the $u(u+1)/2$-dimensional random vector of empirical spectral measures $({\hat{\rho}}^{ab}_n)_{a<b}$. Then the integral \eqref{replicatePart_bforeAnsatz} can be re-expressed as
\begin{align}
\int D\hat P_{n}[({\hat{\rho}}^{ab})_{a<b}]\prod_{a<b}^{0,u}\exp \Big(\frac{\beta\lambda}2 NM \int d{\hat{\rho}}^{ab}(x)x \Big).\label{repPart_beforeLDP}
\end{align}
We \emph{assume} that an LDP at scale $M^2=\Theta(N^2)$ with rate functional $\tilde L$ holds for $({\hat{\rho}}^{ab}_n)_{a<b}$; this is quite generally correct for ESDs of (single) rotationally invariant random matrices \cite{hiai2000large,arous1997large,arous1998large}, which is the reason why we assume this invariance here. This LDP reads
\begin{align*}
 D\hat P_{n}[({\hat{\rho}}^{ab})_{a<b}]\sim \exp\big(-M^2 \tilde L[(\hat{\rho}^{ab})_{a<b}]\big) \prod_{a<b}^{0,u}D[{\hat{\rho}}^{ab}].
 \end{align*} 
The weaker form of the spectral replica symmetric ansatz \eqref{RS_assump} that we use here states that the rate functional $\tilde L$ is actually additive over the pairs of indices $a<b$ or, said differently, that the ESDs $({\hat{\rho}}^{ab}_n)_{a<b}$ asymptotically decouple with same laws dictated by a common rate function $L$:
\begin{gather}
\text{Assumption (spectral replica symmetry, weaker form):} \nn \tilde L[(\hat{\rho}^{ab})_{a<b}] \to \sum_{a<b}^{0,u}  L[\hat{\rho}^{ab}].\label{RS_assump_weak}
\end{gather}
This ansatz is indeed weaker than $\eqref{RS_assump}$ as decoupling is assumed for a much coarser statistics than the j.p.d.f. of the singular values, namely, decoupling of the ESDs over the replica pairs, which capture only the one-point marginals. The rate function $L$ is naturally taken to be the one corresponding to the ESD of ${\bQ}^{ab}({\bQ}^{ab})^\dagger$ where $\bx^a$ and $\bx^b$ are now simply two i.i.d. samples from the prior $P_{X,N}$, thus $L$ can be obtained, at least in certain cases as we will discuss in the next section. Therefore \eqref{repPart_beforeLDP} simplifies into
\begin{align*}
\Big(\int D[{\hat{\rho}}]\exp \Big(\frac{\beta\lambda}2 NM \int d{\hat{\rho}}(x)x-M^2L[\hat \rho] \Big)\Big)^{u(u+1)/2}.
\end{align*}
Plugging this expression in \eqref{replicaTrick} and evaluating the integral by saddle point over the density gives an expression in the case where the prior $P_{X,N}$ generates rotationally invariant matrices.
\begin{conjecture}[Replica symmetric formula for Hermitian dictionary learning with rotationally invariant prior]\label{conj:3_rotInv}
Consider the rotationally invariant random matrix ensemble defined by Hermitian matrices $\bT:=N^{-2}\bx^\dagger \tilde \bx\tilde \bx^\dagger\bx$ where ${\bm x},\tilde {\bm x}$ are i.i.d. $N\times M$ samples from the rotationally invariant prior $P_{X,N}$. Assume that the empirical spectral distribution of the non-zero eigenvalues of $\bT$ verifies a LDP at scale $M^2$ with rate functional $L$. The asymptotic mutual information of model \eqref{model} then verifies
\begin{align}
&\frac1{MN}I\Big(\bX;\sqrt{\frac{\lambda}{N}}\bX\bX^\dagger+\bZ\Big) \nn
&\to-\frac{1}{2}\sup_{\rho\in \mathcal{P}_{\ge 0}}\Big\{\frac{\beta\lambda}{2} \int d\rho(x) x-\alpha L[\rho]\Big\}\nn
&\quad+\frac{1}{2}\sup_{\rho\in \mathcal{P}_{\ge 0}}\Big\{-\alpha L[\rho]\Big\}+ \lim_{N\to+\infty}\frac{\beta\lambda}{4} \frac{\mathbb{E}{\rm Tr}(\bX\bX^\dagger)^2}{MN^2}.
\label{conj3:RS1_dens}
\end{align}

Denote $\rho^*$ the supremum in \eqref{conj3:RS1_dens}, which corresponds to the asymptotic density of non-zero eigenvalues of the (expected) overlap matrix times its conjugate transpose. Using a standard change of density we can access the asymptotic density $\gamma^* \in \mathcal{P}_{\ge 0}$ of singular values of the overlap using $\gamma^* (x)=2x \rho^*(x^2)$. Let $a:=\min(1,\alpha)$. The overlap \eqref{q_2} is then
\begin{align}
 q=\frac1{a} \int d \gamma^*(x)x=\frac1{a}\int d \rho^*(x)\sqrt x.
\end{align}
\end{conjecture}

In case the rate functional $L'$ for the LDP of the empirical density of non-zero singular values of the matrix $N^{-1}\bx^\dagger \tilde \bx$ is easier to obtain than $L$, then formula \eqref{conj3:RS1_dens} is simply changed by replacing $L$ by $L'$ as well as the expectation $\int d\rho(x)x$ by the second moment $\int d\rho(x)x^2$. The optimizer $\rho^*$ of the resulting formula then gives access to the scalar overlap thanks to $q=\int d\rho^*(x)x/a$.

Assume there exists an effective potential $V_T$ underlying this ensemble, in the sense that $\bT'\sim \exp (-\frac{\beta M}4{\rm Tr} V_T(\bT'))$ has same asymptotic spectral density as $\bT$. This effective potential $V_T$ can be deduced from the asymptotic spectral density using \eqref{Vfromrho}. Then by approximation of the rate functional $L$ by the usual form \eqref{rateFunc} for rotationally invariant ensembles, the asymptotic mutual information can be approximated by
\begin{align}
&\frac1{MN}I\Big(\bX;\sqrt{\frac{\lambda}{N}}\bX\bX^\dagger+\bZ\Big) \nn
&\approx-\frac{1}{2}\sup_{\rho\in \mathcal{P}_{\ge 0}}\Big\{\frac{\beta\lambda}{2} \int d\rho(x)x-\frac{\alpha\beta}4\int d\rho(x) V_T(x)\nn
&\quad+\frac{\alpha\beta}2\int d\rho(x)d \rho(y)\ln|x-y|\Big\}\nn
&\quad+\frac{1}{2}\sup_{\rho\in \mathcal{P}_{\ge 0}}\Big\{\frac{\alpha\beta}2\int d\rho(x)d \rho(y)\ln|x-y|\nn
&\quad-\frac{\alpha\beta}4\int d\rho(x) V_T(x)\Big\}+  \lim_{N\to+\infty}\frac{\beta\lambda}{4} \frac{\mathbb{E}{\rm Tr}(\bX\bX^\dagger)^2}{MN^2}.
\label{conj3:RS1_dens_2}
\end{align}

\subsection{A more explicit special case: the Ginibre signal}\label{sec:Ginibre}

Obtaining the j.p.d.f. $p_{n}$ of singular values entering Result~\ref{conj:3} from the knowledge of $P_{X,N}$ or the rate functional $L$ in Result~\ref{conj:3_rotInv} may be highly non-trivial, but this has the merit of being well defined ``standard'' random matrix theory problems. In certain cases these are known. E.g., for products of i.i.d. gaussian (Ginibre) matrices (or with an additional source \cite{forrester2016singular}). Starting with the j.p.d.f., in this case the law $p_n$ is related to the partition function of a two-matrix model, see next section, that can be integrated and yields a determinantal point process defined in terms of the Meijer G-function, see \cite{burda2010eigenvalues,ipsen2015products,akemann2013products,ipsen2013products,akemann2012universal,akemann2013singular}. Products of finitely but arbitrarily many matrices are considered in these references, but for us only the case of a product between two matrices is needed. There also exist results for products of truncated unitary matrices \cite{akemann2014universal,kieburg2016singular} and for more general product ensembles (but with less explicit formulas) \cite{ipsen2014weak}. We refer to \cite{akemann2015recent} for a review on the subject. See also \cite{liu2016bulk,fitzgerald2021fluctuations} for information about fluctuations and universality properties in such matrix product ensembles, or \cite{gudowska2003infinite,burda2010spectrum,burda2011eigenvalues,gotze2010asymptotic,o2011products} for results concerning their asymptotic density of eigenvalues and singular values. Using this body of work we can go further in explicitating Results~\ref{conj:3} and \ref{conj:3_rotInv} in these known cases. 

In this section we restrict ourselves to the special case where the signal $\bX$ is a Ginibre matrix, so that $\bX\bX^\dagger$ is Wishart.
We want to find the j.p.d.f. of the singular values of the matrix $N^{-1}\bx_0^\dagger \tilde \bx_0$ where ${\bm x}_0,\tilde {\bm x}_0$ are i.i.d. $N\times M$ Ginibre matrices with $O(1)$ entries, which is equivalent to find it for the matrix $\by_0^\dagger \tilde \by_0$ where ${\by}_0,\tilde {\by}_0$ are i.i.d. $N\times M$ standard Ginibre matrices with law $$P_{Y_0,N}(\by_0)\propto \exp{\rm Tr}\Big[-\frac{\beta N} 2 \by_0\by_0^\dagger\Big].$$ Recall $n:=\min(N,M)$. The steps leading to (2.11) of \cite{akemann2013singular} for the square case $M=N$, or those leading to (16) of \cite{akemann2013products} for the general rectangular case (where $M$ and $N$ do not necessarily match), imply that the j.p.d.f. $p_n$ of the $n$ non-zero singular values $\bsig$ of the matrix $\by_0^\dagger \tilde \by_0$ is
\begin{align}
p_{n}(\bsig)  &\propto  |\Delta_n(\bsig^2)|^\beta\big(\prod_{k\le n}\sigma_k\big)^{\beta(M-n+1)-1} R_n(\bsig)\nonumber\\
&\propto \exp MN\Big(\frac\beta{MN}\sum_{i<j}^{1,n}\ln |\sigma_i^2-\sigma_j^2|+\beta\frac{M-n}{MN}\Tr\ln \bsig \nn
&\quad+\frac1{MN}\ln R_n(\bsig)+o_n(1)\Big),\label{73}
\end{align}
where the function
\begin{align}
 R_n(\bsig)&:= \int_{\mathbb{R}_{\ge 0}^n} d\br\,  |\Delta_n(\br^2)|^\beta\exp{\rm Tr}\Big[-\frac{\beta N} 2\br^2\Big]\nonumber\\
 &\quad \times\int d\mu_n^{(\beta)}(\bU)\exp{\rm Tr}\Big[-\frac{\beta N} 2\bU^\dagger \bsig^2\bU\br^{-2}\Big]\nn
 &\quad\times\big(\prod_{k\le n}r_k\big)^{\beta(N-M-n+1)-1}.\label{74}
\end{align}
We recognize a two-matrix model. The spherical integral appears in the function $R_n(\bsig)$, that we re-express in a form appropriate for a saddle-point evaluation:
\begin{align*}
&\int d\br \exp MN\Big( \frac\beta{MN} \sum_{i<j}^{1,n}\ln|r_i^2-r_j^2|-\frac{\beta } {2M}{\rm Tr}\,\br^2\nn
&\quad+\beta\frac{N-M-n}{MN}\Tr\ln \br+I_n^{(\beta)}\Big(\bsig^2,\br^{-2},-\frac N n\Big)+o_n(1)\Big).
\end{align*}
Therefore we reach 
\begin{align*}
 &\frac {\ln R_n(\bsig)} {MN}= \sup_{\br\in\mathbb{R}_{\ge 0}^n}\Big\{ \frac\beta{MN} \sum_{i<j}^{1,n}\ln|r_i^2-r_j^2|-\frac{\beta} {2M}{\rm Tr}\,\br^2\nn
 &\quad+\beta\frac{N-M-n}{MN}\Tr\ln \br+I_n^{(\beta)}\Big(\bsig^2,\br^{-2},-\frac N n\Big)\Big\}+o_n(1).
\end{align*}
Thus we end up with 
\begin{align}
&\frac{\ln p_{n}(\bsig)}{MN} =   \beta \sup_{\br\in\mathbb{R}_{\ge 0}^n}\Big\{\sum_{i\neq j}^{1,n}\frac{\ln |\sigma_i^2-\sigma_j^2|}{2MN}+\frac{M-n}{2MN}\Tr\ln \bsig^2 \nn
&\quad+\sum_{i\neq j}^{1,n}\frac{\ln|r_{i}^2-r_{j}^2|}{2MN}-\frac{{\rm Tr}\,\br^2} {2M}+\frac{N-M-n}{2MN}\Tr\ln \br^2\nn
&\quad+\frac{n^2}{ \beta MN} I_n^{(\beta)}\Big(\bsig^2,\br^{-2},-\frac N n\Big)\Big\}+o_n(1).\label{pn_ginibre}
\end{align}
With this expression in hand we can write down a more explicit version of Result~\ref{conj:3} when the signal is Ginibre. By the same type of manipulations as those found in Section~\ref{sec:LDP} we can also indentify the rate functional $L$ starting from \eqref{pn_ginibre} and therefore express the result in terms of a density order parameter.
\begin{conjecture}[Replica symmetric formula for Hermitian dictionary learning with a Ginibre signal]\label{conj:Ginibre}

Let $n:=\min(N,M)$. When $P_{X,N}(\bX)\propto \exp\Tr[-(\beta /2)\bX\bX^\dagger]$ the mutual information of model \eqref{model} verifies 
\begin{align}
&\frac1{MN}I\Big(\bX;\sqrt{\frac{\lambda}{N}}\bX\bX^\dagger+\bZ\Big) \nn
&=-\frac{\beta}{2}\sup_{(\bsig,\br)\in \mathbb{R}^n_{\ge 0}\times \mathbb{R}^n_{\ge 0}}\Big\{\sum_{i\neq j}^{1,n}\frac{\ln |\sigma_i^2-\sigma_j^2|}{2MN}+\sum_{i\neq j}^{1,n}\frac{\ln|r_{i}^2-r_{j}^2|}{2MN} \nn
&\quad+\frac{M-n}{2MN} {\rm Tr} \ln \bsig^2 +\frac{N-M-n}{2MN} {\rm Tr} \ln \br^2\nn
&\quad+\frac{\lambda}{2M} {\rm Tr}\bsig^2-\frac1{2M} {{\rm Tr}\,\br^2}+\frac{n^2}{ \beta MN} I_n^{(\beta)}\Big(\bsig^2,\br^{-2},-\frac N n\Big)\Big\}\nn
&\quad+ \frac{\beta\lambda}{4} \frac{\mathbb{E}{\rm Tr}(\bX\bX^\dagger)^2}{MN^2}+\tau_N.
\label{conj3:RS1_Ginibre}
\end{align}
Constant $\tau_N$ fixes $I(\bX;\bZ)= 0$: 
$$
\tau_N:= \frac{\beta}2\sup_{(\bsig,\br)\in \mathbb{R}^n_{\ge 0}\times \mathbb{R}^n_{\ge 0}}g^{\rm{RS}}_n(\bsig,\br,\lambda=0)+o_N(1)
$$
for the replica symmetric potential function $g^{\rm{RS}}_n:\mathbb{R}^n_{\ge 0}\times \mathbb{R}^n_{\ge 0}\times \mathbb{R}_{\ge 0}\mapsto \mathbb{R}$ defined by the curly bracket $\{\cdots\}$ in the variational problem \eqref{conj3:RS1_Ginibre}.

Denote $\bsig^*$ the overlap singular values achieving the supremum in \eqref{conj3:RS1_Ginibre}. The overlap \eqref{q_2} is then given by $q=N^{-1} {\rm Tr}\bsig^* +o_N(1)$.

Introducing asymptotic densities $\rho$ and $\tilde \rho$ associated to the squared singular values $\bsig^2$ and $\br^2$, respectively, the conjecture can be re-expressed in the limit $N\to+\infty$ and $M/N \to \alpha$ as
\begin{widetext}
\begin{align}
&\frac1{MN}I\Big(\bX;\sqrt{\frac{\lambda}{N}}\bX\bX^\dagger+\bZ\Big)\to -\frac{\beta}{2}\sup_{(\rho,\tilde \rho)\in\mathcal{P}_{\ge 0}\times\mathcal{P}_{\ge 0}}\Big\{\min\Big(\frac\alpha2,\frac1{2\alpha}\Big) \Big(\int d\rho(x)\,d\rho(y)\ln |x-y|+\int d\tilde \rho(x)\,d\tilde \rho(y)\ln|x-y|\Big)\nn
&\qquad+\max\Big(0,\frac{1}2-\frac1{2\alpha}\Big)\int d\rho(x)\ln x  +\max\Big(\frac12-\alpha,-\frac12\Big)\int d\tilde \rho(x) \ln x +\min\Big(\frac12,\frac1{2\alpha}\Big)\Big(\lambda\int d\rho(x)x-\int d\tilde \rho(x)x\Big)\nn
&\qquad+\frac{1}{ \beta }\min\Big(\alpha,\frac1{\alpha}\Big) I^{(\beta)}\Big[\rho,\hat \rho[\tilde \rho],\frac{-1}{\min(1,\alpha)}\Big]\Big\}+ \frac{\beta\lambda}{4\alpha} \lim_{N\to+\infty}\frac{\mathbb{E}{\rm Tr}(\bX\bX^\dagger)^2}{N^3}+\tau.\label{conj3:RS2}
\end{align}
\end{widetext}
Using a standard change of density the density $\hat \rho[\tilde \rho]$ belonging to $\mathcal{P}_{\ge 0}$ can be expressed in terms of $\tilde \rho$ over which the optimization takes place:
\begin{align*}
\hat \rho[\tilde \rho](x)=\frac{\tilde \rho(1/x)}{x^{2}}.
\end{align*}
Constant $\tau$ fixes the contraint $I(\bX;\bZ)= 0$: $$\tau:= \frac{\beta}2\sup_{(\rho,\tilde \rho)\in\mathcal{P}_{\ge 0}\times\mathcal{P}_{\ge 0}}g^{\rm RS}[\rho,\tilde\rho,\lambda=0]$$
for the replica symmetric potential functional $g^{\rm RS}:\mathcal{P}_{\ge 0}\times\mathcal{P}_{\ge 0}\times\mathbb{R}_{\ge 0}\mapsto \mathbb{R}$ defined by the curly brackets $\{\cdots\}$ in \eqref{conj3:RS2}.

Denoting by $\rho^*$ the density of squared singular values of the overlap that achieves the supremum over $\rho$ in \eqref{conj3:RS2}, the scalar overlap reads $q=\int d\rho^*(x)\sqrt{x}/\min(1,\alpha)$.
\end{conjecture}

Due to the presence of the spherical integral, solving the above variational problem(s) is an important challenge. Because the Ginibre prior induces a rotationally invariant (Wishart) matrix $\bX\bX^\dagger$ this formula must match the simpler Result~\ref{conjectureHermitianRotInv_simple}. But, as already mentioned, the great advantage of Result~\ref{conj:Ginibre} (and more generically of Results~\ref{conj:3} and \ref{conj:3_rotInv}) is that their numerical solution would also give access to the scalar overlap $q$, which cannot be extracted from Result~\ref{conjectureHermitianRotInv_simple}. Solving the functional variational problems of Result~\ref{conj:Ginibre} and comparing it to Result~\ref{conjectureHermitianRotInv_simple}, whose solution can be numerically obtained thanks to Matytsin's formalism \cite{matytsin1994large}, is therefore one (rather challenging) route to check the validity of our spectral replica method. We do not know at the moment how to do that, and this is left for future investigations. 

But there is another alternative route to check the validity of our replica theory without actually solving the above variational problems. It goes as follows. Recall the integral appearing in \eqref{integraltosolve}. In the case of Ginibre signal it reads
\begin{align*}
 &\int d\bsig\, d\br \, |\Delta_n(\bsig^2)\Delta_n(\br^2)|^\beta\nn
 &\qquad\times\big(\prod_{k\le n}\sigma_k\big)^{\beta(M-n+1)-1}\big(\prod_{k\le n}r_k\big)^{\beta(N-M-n+1)-1}\nn 
 &\qquad\times \int d\mu_n^{(\beta)}(\bU)\exp \frac{\beta N} 2 {\rm Tr}\Big[\lambda  \bsig^2-\br^2-\bU^\dagger \bsig^2\bU\br^{-2}\Big].
\end{align*}
This is a two-matrix model. If one was able to actually compute this integral without using a saddle-point (which generates the variational formulation), the result could be compared to Result~\ref{conjectureHermitianRotInv_simple}. 
%
%
Fortunately, it can be computed exactly using the method of bi-orthogonal polynomials as used in \cite{akemann2013singular,akemann2013products,mehta1993method}, see also \cite{akemann2011oxford}. This promising strategy to test the validity of our spectral replica method is non-trivial and left for future work.

%% file: sections/DictionaryLearning.tex
\section{Dictionary learning}\label{sec:DiL_ST}

\subsection{The model}
Let the ground-truth dictionary ${\bS}=[S_{ik}]\in\mathbb{K}^{N\times K}$ be drawn from a centered distribution $\bS\sim P_{S,K}$, and the coefficients $\bT=[T_{jk}]\in\mathbb{K}^{M\times K}$ from $\bT\sim P_{T,K}$ centered also, where the entries of both $\bS$ and $\bT$ are typically $O(1)$. These two priors should induce symmetric j.p.d.f. of singular values for $\bS$ and $\bT$. We set $$N=\alpha K+o(K) \quad \mbox{and}\quad M=\gamma K+o(K)$$ with fixed $\alpha,\gamma>0$ as $K\to+\infty$. Consider having access to a data matrix $\bY=[Y_{ij}]\in\mathbb{K}^{N\times M}$ with entries generated according to 
\begin{align}
\bY=\sqrt{\frac\lambda N}\,\bS\bT^\dagger + {\bm Z},\label{DiL}
\end{align}
with a Ginibre noise matrix $\bZ\in\mathbb{K}^{N\times M}$ with law $$P_{Z,N}(\bZ)\propto \exp{\rm Tr}\Big[-\frac{\beta}{2} \bZ\bZ^\dagger\Big].$$
The scaling in $N$ of the matrix entries and eigenvalues are the same as in the positive definite case \eqref{model}. We assume this time that both priors are bi-orthogonal/unitary rotationally invariant, i.e., 
\begin{align}\label{bi-rotinv} 
\begin{cases}
dP_{S,K}(\bS)&=dP_{S,K}(\bO\bS\tilde \bO),\\
dP_{T,K}(\bT)&=dP_{T,K}(\bO\bT\tilde \bO)    
\end{cases}
\end{align}
for any orthogonal ($\beta=1$) or unitary ($\beta=2$) matrices $\bO$ and $\tilde \bO$. Rotational invariance of the prior was not needed in the Hermitian case of dictionary learning, but it seems required in the less symmetric present setting. We jointly denote $\bX:=(\bS,\bT)$ and $\bx:=(\bs,\bt)\in\mathbb{K}^{N\times K}\times \mathbb{K}^{M\times K}$. Let $dP_{X,K}(\bx):=dP_{S,K}(\bs)\,dP_{T,K}(\bt)$. The joint posterior is
\begin{align}
&dP_{X\mid Y,K}(\bx \mid \bY)=	\frac1{\mathcal{Z}(\bY)}dP_{X,K}(\bx)\nn
&\times\exp\frac\beta2 {\rm Tr}\Big[\sqrt{\frac{\lambda}{N}}\bY^\dagger \bs\bt^\dagger+\sqrt{\frac{\lambda}{N}}\bY \bt\bs^\dagger-\frac\lambda{N}\bs^\dagger\bs\bt^\dagger\bt  \Big].\label{posteriorST}
\end{align}
Note the invariance of the model under $({\bS},\bT)\to ({\bS}{\bm U},\bT{\bU})$ for any orthogonal/unitary ${\bU}$.

The object of interest is the average free entropy 
\begin{align*}
\EE f_N&:= \frac1{NM} \mathbb{E} \ln \mathcal{Z}(\bY).
\end{align*}
It is linked to the mutual information by
\begin{align*}
\frac{1}{MN}I\big(\bY;(\bS,\bT)\big) = - \EE f_N+ \frac{\beta \lambda}{2MN^2}\mathbb{E}{\rm Tr}\bS^\dagger \bS\bT^\dagger\bT.
\end{align*}

\subsection{Replica trick and spectral replica symmetry}

As before, working in the Bayes-optimal setting allows us to simply rename the ground truth $\bx^0=\bX$ which will play the same role as all other replicas $\bx^a=(\bs^a,\bt^a)\in \mathbb{K}^{N\times K}\times\mathbb{K}^{M\times K}$ of $\bx$. We set 
\begin{align*}
&\int dP_{X,K}(\{\bx\}_0^u)\,\cdots \nn
&= \int_{\mathbb{K}^{NK(u+1)}}\prod_{a=0}^u dP_{S,K}(\bs^a)\int_{\mathbb{K}^{MK(u+1)}}\prod_{a=0}^u dP_{T,K}(\bt^a)\,\cdots\,.	
\end{align*}
The replica trick \eqref{replicaTrick} requires computing the moments $\EE\mathcal{Z}^u$ of the partition function. As in Section~\ref{sec:hermitianDL} we consider first the more cumbersome complex case $\beta=2$. Model \eqref{DiL} is then equivalent to

\begin{align*}
\begin{cases}
\Re Y_{ij}\sim\mathcal{N}\big(\sqrt{\frac{\lambda}{N}}\frac{\langle\bS_{i},\bar \bT_{j}\rangle+\langle\bar \bS_{i},\bT_{j}\rangle}{2},\frac12\big) \\
\Im Y_{ij}\sim\mathcal{N}\big(\sqrt{\frac{\lambda}{N}}\frac{\langle\bS_{i},\bar \bT_{j}\rangle-\langle\bar \bS_{i}, \bT_{j}\rangle}{2{\rm i}}, \frac12\big)
\end{cases}
 (i, j) \in[N]\times [M].
\end{align*}

We introduce the $K\times K$ (a-priori non-Hermitian) overlap matrices and their SVD decompositions:
\begin{align*}
\bQ_s^{ab}&:=\Big(\frac1N\sum_{i\le N}{s_{ik}^a \bar s_{i\ell}^b}\Big)_{k,\ell\le K}=\frac{(\bs^a)^\intercal \bar\bs^b}N=\bA_s^{ab}\bsig_s^{ab}\bB_s^{ab}, \\ 
\bQ_t^{ab}&:=\Big(\frac1N\sum_{j\le M}{\bar t_{jk}^a t_{j\ell}^b}\Big)_{k,\ell\le K}=\frac{(\bt^a)^\dagger\bt^b}N=\bA_t^{ab}\bsig_t^{ab}\bB_t^{ab}. 
\end{align*}
The overlaps $(\bQ_s^{ab})_{a<b}$ are of rank $n_s:=\min(N,K)$, and thus $(\bsig_s^{ab})_{a<b}$ have $n_s$ non-zero entries on their diagonal, while $(\bQ_t^{ab})_{a<b}$ have rank $n_t:=\min(M,K)$ implying that $(\bsig_t^{ab})_{a<b}$ have $n_t$ non-zero entries on their diagonal. Now, we integrate $\bY$ which is, conditionally on $(\bS,\bT)$, a complex gaussian multivariate random variable, by using formula \eqref{useful_form} and obtain that
\begin{widetext}
\begin{align*}
\EE\mathcal{Z}^u&=\int dP_{X,K}(\{\bx\}_0^u) \,\mathbb{E}_{\bY|\bx^0} \prod_{a\le u}\exp2\sum_{i,j}^{N,M}\Big(\sqrt{\frac{\lambda}{N}}\Re  Y_{ij}\frac{\langle\bs_{i}^a, \bar \bt_{j}^a\rangle+\langle\bar \bs_{i}^a, \bt_{j}^a\rangle}{2}-\frac{\lambda}{2N}\Big(\frac{\langle\bs_{i}^a, \bar \bt_{j}^a\rangle+\langle\bar \bs_{i}^a, \bt_{j}^a\rangle}{2}\Big)^2\Big)\nn
&\qquad\qquad\qquad\qquad\qquad\qquad\times \exp2\sum_{i,j}^{N,M}\Big(\sqrt{\frac{\lambda}{N}}\Im  Y_{ij}\frac{\langle\bs_i^a,\bar \bt_j^a\rangle-\langle\bar \bs_i^a,\bt_j^a\rangle}{2{\rm i}}-\frac{\lambda}{2N}\Big(\frac{\langle\bs_i^a,\bar \bt_j^a\rangle-\langle\bar \bs_i^a,\bt_j^a\rangle}{2{\rm i}}\Big)^2\Big)\nn
&=\int dP_{X,K}(\{\bx\}_0^u) \prod_{a< b}^{0,u} \exp\frac{2\lambda}{N}\sum_{i,j}^{N,M}\Big(\frac{\langle\bs_{i}^a, \bar \bt_{j}^a\rangle+\langle\bar \bs_{i}^a, \bt_{j}^a\rangle}{2}\times\frac{\langle\bs_{i}^b, \bar \bt_{j}^b\rangle+\langle\bar \bs_{i}^b, \bt_{j}^b\rangle}{2} + \frac{\langle\bs_i^a,\bar \bt_j^a\rangle-\langle\bar \bs_i^a,\bt_j^a\rangle}{2{\rm i}}\times\frac{\langle\bs_i^b,\bar \bt_j^b\rangle-\langle\bar \bs_i^b,\bt_j^b\rangle}{2{\rm i}}\Big)\nn
&=   \int dP_{X,K}(\{\bx\}_0^u)\prod_{a< b}^{0,u}\exp\frac{2\lambda}{N}\sum_{i,j}^{N,M}\Big(\Re \langle\bs_{i}^a, \bar \bt_{j}^a\rangle \Re \langle\bs_{i}^b, \bar \bt_{j}^b\rangle+\Im \langle\bs_i^a,\bar \bt_j^a\rangle\Im \langle\bs_i^b,\bar \bt_j^b\rangle\Big)\nn
&=   \int dP_{X,K}(\{\bx\}_0^u)\prod_{a< b}^{0,u}\exp\frac{2\lambda}{N}\sum_{i,j}^{N,M}\Re \big(\langle\bs_{i}^a, \bar \bt_{j}^a\rangle\langle\bar \bs_i^b, \bt_j^b\rangle\big)\nn
&=  \int dP_{X,K}(\{\bx\}_0^u) \prod_{a< b}^{0,u} \exp\beta \lambda N\Re{\rm Tr}\bQ_s^{ab}(\bQ_t^{ab})^\dagger.
\end{align*}
\end{widetext}
The following product form of the prior measure $$dP_{X,K}(\{\bx\}_0^u)=\prod_{a=0}^u dP_{S,K}(\bs^a)\,dP_{T,K}(\bt^a)$$ induces a j.p.d.f. factorized over the two types of overlaps:
\begin{align}
&dP_{(Q_s,Q_t),K}((\bQ_s^{ab},\bQ_t^{ab})_{a<b})\nn
&\quad=dP_{(Q_s),K}((\bQ_s^{ab})_{a<b})\,dP_{(Q_t),K}((\bQ_t^{ab})_{a<b}).\label{productForm}    
\end{align}

At this stage we need one additional assumption when compared with the Hermitian case of Section~\ref{sec:hermitianDL}. For each pair $a<b$ of replica indices let i.i.d. Haar matrices $\bU^{ab}, \bV^{ab}\sim\mu_K^{(\beta)}$ independent of everything. We \emph{assume} the following equality in distribution in the large size limit \footnote{We note that this assumption is reminiscent of results in \cite{benaych2009rectangular} (see also \cite{belinschi2009regularization}) for the addition of two large random matrices with at least one being bi-unitary invariant.}, which is suggested by the combination of the independence \eqref{productForm} between the two types of overlaps together with the rotational invariance \eqref{bi-rotinv} of both priors (from which the overlap matrices $\bQ_s^{ab}$ and $\bQ_t^{ab}$ must inherit). 
%
\begin{gather*}
\text{Assumption (equality in law):}\ \text{For any pair $a<b$:}\\   \Tr \bQ_s^{ab}(\bQ_t^{ab})^\dagger \sim \Tr \big[\bU^{ab}\bQ_s^{ab}\bV^{ab}(\bQ_t^{ab})^\dagger\big]
\end{gather*}
as $M,N,K\to +\infty$ proportionally.
As a consequence, 
\begin{align*}
\EE\mathcal{Z}^u&\propto \int  dP_{(Q_s,Q_t),K}((\bQ_s^{ab},\bQ_t^{ab})_{a<b})\nn
&\quad\times\prod_{a< b}^{0,u} \exp\beta \lambda N\Re{\rm Tr}\bQ_s^{ab}(\bQ_t^{ab})^\dagger
\end{align*}
\begin{align}
&= \int  dP_{(Q_s),K}((\bQ_s^{ab})_{a<b})\,dP_{(Q_t),K}((\bQ_t^{ab})_{a<b})\nn
&\quad\times\int \prod_{a< b}^{0,u}d\mu_K^{(\beta)}(\bU^{ab})\,d\mu_K^{(\beta)}(\bV^{ab}) \nn
&\qquad\times\exp \beta\lambda N\Re{\rm Tr}\big[\bU^{ab}\bQ_s^{ab}\bV^{ab}(\bQ_t^{ab})^\dagger\big].\label{tobeused}
\end{align}
We now change variables for the SVD decompositions of the overlaps:
\begin{align*}
&dP_{(Q_s),K}((\bQ_s^{ab})_{a<b})dP_{(Q_t),K}((\bQ_t^{ab})_{a<b})\nn
&\quad=  dP_{(A_s,B_s)\mid(\sigma_s),K}((\bA_s^{ab},\bB_s^{ab})_{a<b} \mid (\bsig_s^{ab})_{a<b})\nn
&\qquad \times dP_{(A_t,B_t)\mid(\sigma_t),K}((\bA_t^{ab},\bB_t^{ab})_{a<b} \mid (\bsig_t^{ab})_{a<b})\nn
&\quad\quad\times dP_{(\sigma_s),K}((\bsig_s^{ab})_{a<b})\,dP_{(\sigma_t),K}((\bsig_t^{ab})_{a<b}).
\end{align*}
The spectral replica symmetric ansatz reads in this case:
\begin{align}
\text{Assumption (spectral replica symmetry):}  \nn
\begin{cases*}
	 P_{(\sigma_s),K}((\bsig_s^{ab})_{a<b}) \to \prod_{a<b}^{0,u} {\tilde p_{S,K}}(\bsig_s^{ab}),\\
	  P_{(\sigma_t),K}((\bsig_t^{ab})_{a<b}) \to \prod_{a<b}^{0,u} {\tilde p_{T,K}}(\bsig_t^{ab}),
\end{cases*}\label{spectralRS_dico}
\end{align}
where ${\tilde p_{S,K}}$ corresponds to the j.p.d.f. of singular values of a product $N^{-1}\bs^\dagger \tilde \bs$ between i.i.d. samples from the prior $P_{S,K}$, and similarly for ${\tilde p_{T,K}}$ with samples from $P_{T,K}$. Again, $\to$ means that both sides weakly converge to the same distribution as $N,M,K\to+\infty$. Thus
\begin{widetext}
\begin{align*}
\EE\mathcal{Z}^u&\propto \int \prod_{a<b}^{0,u}d\tilde p_{S,K}(\bsig_s^{ab})\,d\tilde p_{T,K}(\bsig_t^{ab})\nn
&\quad\times\int dP_{(A_s,B_s)\mid (\sigma_s),K}((\bA_s^{ab},\bB_s^{ab})_{a<b} \mid (\bsig_s^{ab})_{a<b})\,dP_{(A_t,B_t)\mid (\sigma_t),K}((\bA_t^{ab},\bB_t^{ab})_{a<b} \mid (\bsig_t^{ab})_{a<b})\nn
&\quad\quad\times\int \prod_{a< b}^{0,u}d\mu_K^{(\beta)}(\bU^{ab})\,d\mu_K^{(\beta)}(\bV^{ab}) \exp\beta \lambda N\Re{\rm Tr}\big[\bU^{ab}\bA_s^{ab}\bsig_s^{ab}\bB_s^{ab}\bV^{ab}(\bB_t^{ab})^\dagger\bsig^{ab}_t(\bA_t^{ab})^\dagger\big].
\end{align*}
\end{widetext}
The mechanism here is to absorb the left and right singular vectors into the Haar distributed matrices: we redefine $(\bA_t^{ab})^\dagger\bU^{ab} \bA_s^{ab}\to \bU^{ab}$ and $\bB_s^{ab}\bV^{ab}(\bB_t^{ab})^\dagger\to \bV^{ab}$ which remain independent and Haar distributed; these changes have unit Jacobian determinant. Thus, the distributions of singular vectors, which a-priori couple the different pairs of replicas, are integrated and decoupling of the integrals over the pairs of indices $a<b$ takes place. This yields
\begin{align*}
&\EE\mathcal{Z}(\bY)^u\propto \Big(\int d\tilde p_{S,K}(\bsig_s)\,d\tilde p_{T,K}(\bsig_t)\nn
&\quad\times d\mu_K^{(\beta)}(\bU)\, d\mu_K^{(\beta)}(\bV)\exp \beta\lambda N\Re{\rm Tr}\big[\bU\bsig_s\bV\bsig_t\big]\Big)^{u(u+1)/2}\nn
&=\Big(\int d\bsig_s\,d\bsig_t\exp  MN\Big[\frac{\ln {\tilde p_{S,K}}(\bsig_s)}{MN}\nn
&\quad+\frac{\ln {\tilde p_{T,K}}(\bsig_t)}{MN} +\frac{K^2}{MN}J_K^{(\beta)}(\bsig_s,\bsig_t,2\lambda)\Big] \Big)^{u(u+1)/2}.
\end{align*}
Saddle-point estimation and taking $u\to 0_+$ yields the following conjecture for the mutual information and, moreover, the non-universal scalar overlaps (see the justifications below \eqref{q_2}):
\begin{align}
q_s&:=\lim_{N\to+\infty}\frac1{N}\EE\big\langle \big|{\rm Tr}\bs\bS^\dagger\big|\big\rangle=\lim_{N\to+\infty}\frac1N\EE\big\langle\big|{\rm Tr} \bQ_s\big|\big \rangle\nn
&=\lim_{N\to+\infty}\frac1N\big|{\rm Tr} \EE\langle\bQ_s\rangle \big| =\frac1N\big|{\rm Tr} \bQ_s \big| + o_K(1).\label{q_s}  
\end{align}
Scalar overlap $q_t$ is defined similarly but replacing in \eqref{q_s} the matrices $(\bs,\bS,\bQ_s)$ by $(\bt,\bT,\bQ_t)$. Here the overlaps $$\bQ_s:=N^{-1}\bs^\dagger \tilde\bs \quad \mbox{and} \quad \bQ_t:=N^{-1}\bt^\dagger \tilde\bt$$ for two i.i.d. samples $\bx=(\bs,\bt)$ and $\tilde\bx=(\tilde \bs,\tilde \bt)$ from the joint posterior distribution $P_{X\mid Y, N}$ given by \eqref{posteriorST} (these should not be confused with the samples drawn from the priors appearing in the statement below).

\begin{conjecture}[Replica symmetric formula for dictionary learning] \label{conj:6}
Let ${\tilde p_{S,K}}$ by the j.p.d.f. of the $K$ singular values (which are $O(1)$) of $N^{-1}\bs^\dagger \tilde \bs$, where ${\bs},\tilde {\bs}$ are i.i.d. $N\times K$ random matrices drawn from prior $P_{S,K}$. Similarly, let ${\tilde p_{T,K}}$ be the j.p.d.f. of the $K$ singular values of $N^{-1}\bt^\dagger \tilde \bt$, where ${\bt},\tilde {\bt}$ are i.i.d. $M\times K$ matrices drawn from $P_{T,K}$.

The mutual information of model \eqref{DiL} verifies
\begin{align}
&\frac1{MN}I\Big((\bS,\bT);\sqrt{\frac{\lambda}{N}}\bS\bT^\dagger+\bZ\Big)\nn
&= -\frac{K^2}{2MN}\sup_{(\bsig_s,\bsig_t)\in \mathcal{S}_K}\Big\{ \frac{\ln {\tilde p_{S,K}}(\bsig_s)}{K^2}+\frac{\ln {\tilde p_{T,K}}(\bsig_t)}{K^2}\nn
&\quad +J_K^{(\beta)}(\bsig_s,\bsig_t,2\lambda)\Big\} + \frac{\beta\lambda}{2MN^2}\mathbb{E}{\rm Tr}\bS^\dagger \bS\bT^\dagger\bT+\tau_K.
\label{conj6}
\end{align}
The set $\mathcal{S}_K$ is defined as 
\begin{align*}
&\mathcal{S}_K:=\big\{(\bsig_s,\bsig_t)\in\mathbb{R}^K_{\ge 0}\times\mathbb{R}^K_{\ge 0}: \nn
&\quad\bsig_s \ \mbox{is rank} \ \min(N,K),\ \bsig_t \ \mbox{is rank} \ \min(M,K)\big\}.    
\end{align*}
Constant $\tau_K$ fixes $I((\bS,\bT);\bZ)= 0$, i.e., 
$$
\tau_K:= \frac{1}2\sup_{(\bsig_s,\bsig_t)\in\mathcal{S}_K}g^{\rm{RS}}_K(\bsig_s,\bsig_t,\lambda=0) + o_K(1),
$$
where $g^{\rm{RS}}_K:\mathbb{R}^K_{\ge 0}\times\mathbb{R}^K_{\ge 0}\times \mathbb{R}_{\ge 0} \mapsto \mathbb{R}$ is the replica symmetric potential function defined by the curly brackets $\{\cdots\}$ in the variational problem \eqref{conj6}. 

Denote $(\bsig_s^*,\bsig_t^*)$ the overlaps singular values achieving the supremum in \eqref{conj6}. The scalar overlaps are 
\begin{align}
 q_s=\frac1 N {\rm Tr}\bsig_s^* +o_K(1), \qquad q_t=\frac1 N {\rm Tr}\bsig_t^* +o_K(1).
\end{align}
\end{conjecture}

We can express the result using density order parameters, under an even weaker decoupling assumption than \eqref{spectralRS_dico}. The steps are similar to what is done in Section~\ref{sec:work_dens_param} so we will be more brief. Let the empirical distributions of singular values 
\begin{align*}
{\hat{\rho}}^{ab}_{s,K}(x)&:=\sum_{k\le K}\frac{\delta(\sigma_{s,i}^{ab} -x)}K, \ \ {\hat{\rho}}^{ab}_{t,K}(x):=\sum_{k\le K}\frac{\delta(\sigma_{t,i}^{ab} -x)}K.
\end{align*}
We assume that the random $({\hat{\rho}}^{ab}_{s,K})_{a<b}$ jointly verify a LDP at scale $K^2$:
\begin{align*}
 D\hat P_{s,K}[({\hat{\rho}}_s^{ab})_{a<b}]\sim \exp\big(-K^2 \tilde L_s[(\hat{\rho}^{ab}_s)_{a<b}]\big) \prod_{a<b}^{0,u}D[\hat{\rho}_s^{ab}],
\end{align*}
and similarly for $({\hat{\rho}}^{ab}_{t,K})_{a<b}$ with rate functional $\tilde L_t$. Then the weaker decoupling assumption reads
\begin{gather*}
\text{Assumption (spectral replica symmetry, weaker form):} \nn \tilde L_{s/t}[(\hat{\rho}^{ab}_{s/t})_{a<b}] \to \sum_{a<b}^{0,u}  L_{s/t}[\hat{\rho}_{s/t}^{ab}].
\label{RS_assump_weak-dico}
\end{gather*}
The rate functional $L_s$ is naturally taken to be the one corresponding to the empirical distribution of singular values of a product of two i.i.d. samples from the prior $P_{S,K}$, and similarly for $L_t$ but with samples from $P_{T,K}$. 

Start again from \eqref{tobeused}. Absorb the singular vectors of the overlaps in the Haar distributed matrices $\bU^{ab}$ and $\bV^{ab}$, so that the rectangular spherical integrals appearing in \eqref{tobeused} only depend on the singular values $(\bsig_s^{ab},\bsig_t^{ab})_{a<b}$. Only the measures $dP_{(\sigma_s),K}((\bsig_s^{ab})_{a<b})\,dP_{(\sigma_t),K}((\bsig_t^{ab})_{a<b})$ then remain. These induce j.p.d.f. over the above empirical distributions of singular values. As seen from \eqref{lim_RectSphInt}, asymptotically, the rectangular spherical depends only on the density of singular values. All these observations combined with the aforementioned LDPs and decoupling ansatz \eqref{RS_assump_weak-dico} allow us to simplify \eqref{tobeused} into
\begin{align*}
 &\EE\mathcal{Z}^u\propto \Big(\int D[\hat {\rho}_s]D[\hat {\rho}_t]\exp K^2  \Big(- L_s[\hat {\rho}_s]- L_t[\hat {\rho}_t]\nn 
 &\qquad\qquad+J^{(\beta)}[{\hat{\rho}}_{s},{\hat{\rho}}_{t},2\lambda]  +o_K(1)\Big)\Big)^{u(u+1)/2}.
\end{align*}
Saddle-point estimation and \eqref{replicaTrick} then yields the following variational formulation.

\begin{conjecture}[Replica symmetric formula for dictionary learning] \label{conj:6-2} Assume that the empirical distribution of singular values of $N^{-1}\bs^\dagger\tilde \bs$, where ${\bs},\tilde {\bs}$ are i.i.d. $N\times K$ random matrices drawn from prior $P_{S,K}$, verifies a LDP at scale $K^2$ with rate functional $L_s$. Similarly, assume that the empirical distribution of singular values of $N^{-1}\bt^\dagger\tilde \bt$, where ${\bt},\tilde {\bt}$ are i.i.d. $M\times K$ from prior $P_{S,K}$, verifies a LDP at scale $K^2$ with rate functional $L_t$.

The mutual information verifies, in the limit $K\to+\infty$ with $N/K\to \alpha$ and $M/K\to\gamma$,
\begin{align}
&\frac1{MN}I\Big((\bS,\bT);\sqrt{\frac{\lambda}{N}}\bS\bT^\dagger+\bZ\Big)\nn
&\to -\frac{1}{2\alpha\gamma}\sup_{(\rho_s,\rho_t)\in\mathcal{S}}\Big\{J^{(\beta)}[\rho_s,\rho_t,2\lambda]-L_s[\rho_s]- L_t[\rho_t]\Big\}\nonumber\\
&\quad+ \frac{\beta\lambda}{2} \lim_{K\to+\infty}\frac{\mathbb{E}{\rm Tr}\bS^\dagger \bS\bT^\dagger\bT}{MN^2}+\tau.\label{conj6:2}
\end{align}
The optimization is over probability densities with finite non-negative support belonging to
\begin{align*}
&\mathcal{S}:=\big\{(\rho_s,\rho_t)=((1-\min(1,\alpha))\delta_0+\min(1,\alpha)\tilde \rho_s,\nn
&\quad(1-\min(1,\gamma))\delta_0+\min(1,\gamma)\tilde \rho_t), (\tilde\rho_s,\tilde\rho_t)\in\mathcal{P}_{\ge 0}\times\mathcal{P}_{\ge 0}\big\}.
\end{align*}
Constant $\tau$ fixes the contraint $I((\bS,\bT);\bZ)= 0$, i.e., $$\tau:= \frac{1}{2\alpha\gamma}\sup_{(\rho_s,\rho_t)\in\mathcal{S}}g^{\rm RS}[\rho_s,\rho_t,\lambda=0],$$
for the replica symmetric potential functional $g^{\rm RS}:\mathcal{P}_{\ge 0}\times\mathcal{P}_{\ge 0}\times\mathbb{R}_{\ge 0}\mapsto \mathbb{R}$ defined by the curly brackets in \eqref{conj6:2}. The densities achieving the supremum in \eqref{conj6:2} are
\begin{align*}
(\rho_s^*,\rho_t^*)&=\big((1-\min(1,\alpha))\delta_0+\min(1,\alpha)\tilde \rho_s^*,\nn
&\quad(1-\min(1,\gamma))\delta_0+\min(1,\gamma)\tilde \rho_t^*\big)   .
\end{align*}
The overlaps are then given by
\begin{align*}
 q_s=\int d\rho_s^*(x)\,x, \qquad q_t=\int d\rho_t^*(x)\,x.
\end{align*}
\end{conjecture}

Note that in the more symmetric special case of $M=N$ and $P_{S,K}=P_{T,K}$ (which does \emph{not} correspond to the Hermitian dictionary learning problem \eqref{model} as $\bS$ and $\bT$ remain i.i.d.), by symmetry, the replica symmetric formula can be simplified as (setting $L_s=L_t=L$)
\begin{align*}
&\frac1{MN}I\Big((\bS,\bT);\sqrt{\frac{\lambda}{N}}\bS\bT^\dagger+\bZ\Big)\nn
&\to -\sup_{\rho\in\mathcal{S}_{\rm diag}}\Big\{  \frac{J^{(\beta)}[\rho,\rho,2\lambda]}{2\alpha^2}-\frac{L[\rho]}{\alpha^2}\Big\}\nn
&\qquad+ \frac{\beta\lambda}{2} \lim_{K\to+\infty}\frac{\mathbb{E}{\rm Tr}\bS^\dagger \bS\bT^\dagger\bT}{N^3}+\tau_{\rm diag}
\end{align*}
where $\mathcal{S}_{\rm diag}$ is the ``diagonal subset'' of $\mathcal{S}$ where additionally $\rho_s=\rho_t$ (and the definition of $\tau_{\rm diag}$ is modified from $\tau$ accordingly). The unique scalar overlap is in this case $q_s=q_t=\int d\rho^*(x)\,x$ where $\rho^*$ achieves the supremum in the variational problem for the mutual information.\\



%% file: sections/AppendixSpherical.tex
\section{Spherical integrals}\label{app:sphInt}

In this appendix we present the HCIZ formula in the Hermitian and general non-Hermitian matrix cases. For the derivations we refer to the very readable original papers by Itzykson and Zuber \cite{itzykson1980planar}, and Mehta \cite{mehta1993method}.

\subsection{Hermitian case}\label{App:mehta}

Consider Hermitian $M\times M$ matrices $\bA$, $\bB$. These are diagonalized by unitary matrices and have real eigenvalues. Recall the notation $\bA = \bU^A \boldsymbol{\lambda}^A (\bU^A)^\dagger$, $\bB = \bU^B \boldsymbol{\lambda}^B (\bU^B)^\dagger$. Recall the definition of the Vandermonde \eqref{Vander}.
The HCIZ formula reads
\begin{align}
&\int_{\mathcal{U}(M)} d\mu_M^{(2)}(\bU)\exp \gamma M  {\rm Tr} [\bA\bU^\dagger \bB \bU]\nn
&\quad=\frac{\prod_{k\leq M-1} k!}{(\gamma M)^{M(M-1)/2}}
\frac{{\rm det} [\exp\gamma M\lambda_{i}^A \lambda_{j}^B]}{\Delta_M(\bLam^A)\Delta_M(\bLam^B)},\label{hciz_integrated}
\end{align}
where $\mu_M^{(2)}$ is the normalized Haar measure over the group of unitary $M\times M$ matrices and $[\exp\gamma M \lambda_{i}^A \lambda_{j}^B]$ is the matrix $(\exp(\gamma M \lambda_{i}^A \lambda_{j}^B))_{i,j\le M}$.
Note that on the left hand side we can replace $\bA$, $\bB$ by $\boldsymbol{\lambda}^A$, $\boldsymbol{\lambda}^B$ since $\bU_A$, $\bU_B$ leave the Haar measure invariant. Note also that by permutation symmetry the ratio of determinants is positive and independent of the ordering of eigenvalues. In the limit $M\to+\infty$ the spherical integral can be described in terms of an hydrodynamical system (the complex Burgers equation) thanks to the work of Matytsin \cite{matytsin1994large} and proven in \cite{guionnet2002large}. See also \cite{bun2014instanton,menon2017complex}. This is not used in this paper but may be useful for future analyses.

\subsection{General non-Hermitian case}\label{App:generalmatrices}

Let $\bA$, $\bB$ be two general $M\times M$ matrices. Their singular value decomposition is 
$\bA = \bU^A \boldsymbol{\sigma}^A \bV^A$ and $\bB = \bU^B \boldsymbol{\sigma}^B \bV^B$ where 
$\bsig^A$, $\bsig^B$ are the diagonal matrices of non-negative singular values and $\bU^A$, 
$\bV^A$, $\bU^B$ and $\bV^B$ are unitary matrices. The spherical integral involves the modified Bessel function of first kind:
$$
I_0(x) := \int_0^\pi \frac{d\theta}{\pi} \exp (x\cos\theta).
$$ 
This function is positive, monotone increasing, $I_0(0) =1$, and grows as $\exp x$ at infinity. 
We have
\begin{align*}
&\int_{\mathcal{U}(M)\times \mathcal{U}(M)} d\mu_M^{(2)}(\bU) \,d\mu_M^{(2)}(\bV) \exp\gamma  M{\Re}{\rm Tr} [\bA \bU\bB \bV] \\
&\quad= \frac{2^{M(M-1)}(\prod_{k\leq M-1} k!)^2}{M!(M\gamma)^{M(M-1)}}
 \frac{{\det} [I_0(M\gamma \sigma_{i}^A \sigma_{j}^B)] }{\Delta_M((\boldsymbol{\sigma}^A)^2) \Delta_M((\boldsymbol{\sigma}^B)^2)}.
\label{IZSingVal}
\end{align*}
As before, on the left hand side we can replace $\bA$, $\bB$ by $\boldsymbol{\sigma}^A$, $\boldsymbol{\sigma}^B$, and the ratio of determinants is positive and invariant under permutations of singular values. This formula first obtained in \cite{schlittgen2003generalizations} and proven in \cite{ghaderipoor2008generalization} can be derived by the same methods used for the classical HCIZ formula based on the solution of the heat equation. Like for the standard spherical integral, there also exists an asymptotic $M\to+\infty$ representation of the rectangular spherical integral in terms of a complex hydrodynamical system \cite{forrester2016hydrodynamical,guionnet2021large}.

%% file: sections/AppendixMMSEderivative.tex
\section{Minimum mean-square error: a more explicit formula}\label{app:MMSEderivative}
By the I-MMSE relation \eqref{IMMSE}, we obtained in Result~\ref{conjectureHermitianRotInv_simple} that the MMSE is directly proportional to the derivative with respect to the signal-to-noise ratio $\lambda$ of the HCIZ integral. Its dependence in $\lambda$ is through the data-matrix eigenvalues $\bLam^Y$. From the HCIZ formula (see Appendix~\ref{app:sphInt}) we get in the complex case $\beta=2$, using the Jacobi formula 
$$
\frac{\partial}{\partial {X_{ij}}} \ln {\rm det} \bX = (\bX^{-1})_{ji},
$$
that the HCIZ derivative verifies
\begin{widetext}
\begin{align*}
&N^2\frac{d}{d\lambda} I_N^{(2)}(\bLam^S, \bLam^Y, \sqrt\lambda) 
 = \frac{N^2}{2\sqrt\lambda} \frac{d}{d\sqrt\lambda} I_N^{(2)}(\bLam, \sqrt\lambda\bLam^Y, 1)
\\ \nonumber &
\qquad=
\frac{1}{2\sqrt\lambda} \frac{d}{d\sqrt\lambda} \ln {\rm det}[\exp{N \lambda_c^S (\sqrt\lambda\lambda_d^Y)}] 
- 
\frac{1}{2\sqrt\lambda} \frac{d}{d\sqrt\lambda} \ln {\rm det}[(\sqrt\lambda \lambda^Y_c)^{d-1}]
\\ \nonumber &
\qquad= \frac{1}{2\sqrt\lambda}
\sum_{i, j\le N}
\frac{d \ln {\rm det}[\exp N{ \lambda_c^S (\sqrt\lambda\lambda_d^Y)}] }{d \exp N { \lambda_i^S (\sqrt\lambda\lambda_j^Y)}} 
\frac{d \exp N { \lambda_i^S (\sqrt\lambda\lambda_j^Y)}}{d \sqrt\lambda}- \frac{1}{2\sqrt\lambda}
\sum_{i, j\le N} \frac{d  \ln {\rm det}[(\sqrt\lambda\lambda_c^Y)^{d-1}]  }{d(\sqrt\lambda\lambda_i^Y)^{j-1}}
\frac{d  (\sqrt\lambda\lambda_i^Y)^{j-1}}{d \sqrt\lambda}
\nonumber \\ &
\qquad =
\frac{1}{2\sqrt\lambda}\sum_{i, j\le N}
([\exp N{ \lambda_c^S (\sqrt\lambda\lambda_d^Y)}]^{-1})_{ji}  N \lambda_i^S (\exp N{ \lambda_i^S (\sqrt\lambda\lambda_j^Y)})
\frac{d \sqrt\lambda\lambda_j^Y}{d \sqrt\lambda}\nn
&\qquad\qquad- \frac{1}{2\sqrt\lambda}
\sum_{i, j\le N} ([(\sqrt\lambda\lambda_c^Y)^{d-1}]^{-1})_{ji}  (j-1) (\sqrt\lambda\lambda_i^Y)^{j-2} 
\frac{d\sqrt\lambda\lambda_i^Y}{d \sqrt\lambda}.
\end{align*}
\end{widetext}
Introducing the $\bY$-eigenvectors $\bY  \boldsymbol{\psi}_i^Y = \lambda_i^Y \boldsymbol{\psi}_i^Y$, the Hellmann-Feynman theorem implies 
$$
\frac{d \lambda_i^Y}{d \sqrt\lambda} = (\boldsymbol{\psi}_i^Y)^\dagger \bS  \boldsymbol{\psi}_i^Y.
$$ 
Thus
\begin{align*}
\frac{d \sqrt\lambda\lambda_i^Y}{d \sqrt\lambda}  &=   (\boldsymbol{\psi}_i^Y)^\dagger(\sqrt\lambda \bS + \bxi)  \boldsymbol{\psi}_i^Y + \sqrt\lambda  (\boldsymbol{\psi}_i^Y)^\dagger \bS  \boldsymbol{\psi}_i^Y \\
&=  (\boldsymbol{\psi}_i^Y)^\dagger (2 \sqrt\lambda \bS + \bxi) \boldsymbol{\psi}_i^Y.
\end{align*}
Putting everything together in \eqref{MMSE_conj2} we find that the MMSE equals (when $\beta=2$)
\begin{widetext}
\begin{align*}
\frac{1}{N^2}\mathbb{E}\|\bS-\mathbb{E}[\bS\mid \bY]\|^2
&= 
 \frac{4}N \EE{\rm Tr} \bLam_S^2 
 \nonumber -\frac{2}{\sqrt\lambda N^2}\sum_{i, j\le N}
([\exp N{ \lambda_c^S (\sqrt\lambda\lambda_d^Y)}]^{-1})_{ji} N \lambda_i^S (\exp N{ \lambda_i^S (\sqrt\lambda\lambda_j^Y)})
(\boldsymbol{\psi}_j^Y)^\dagger (2 \sqrt\lambda \bS + \bxi) \boldsymbol{\psi}_j^Y\nn
&\qquad+ \frac{2}{\sqrt\lambda N^2}
\sum_{i, j\le N} ([(\sqrt\lambda\lambda_c^Y)^{d-1}]^{-1})_{ji}  (j-1)  (\sqrt\lambda\lambda_i^Y)^{j-2} 
(\boldsymbol{\psi}_i^Y)^\dagger (2 \sqrt\lambda \bS + \bxi) \boldsymbol{\psi}_i^Y+o_N(1).
\end{align*}
\end{widetext}
Besides the matrix inversions this formula also requires to compute eigenvectors of 
$\bY = \sqrt \lambda \bS + \bxi$; it may be more practical to use $(\boldsymbol{\psi}_i^Y)^\dagger  (2 \sqrt\lambda \bS + \bxi) \boldsymbol{\psi}_i^Y= \lambda_i^Y + \sqrt\lambda(\boldsymbol{\psi}_i^Y)^\dagger  \bS\boldsymbol{\psi}_i^Y$.

If one were to start instead from Result~\ref{conjectureHermitianRotInv} the computation would be similar, because at the stationary point, the total $\lambda$-derivative is computed by taking a partial derivative only with respect to the explicit $\lambda$ dependence. Indeed, the terms coming from the implicit dependence in the solution $\bLam^s$ of the fixed point equations do not contribute.

%% file: sections/mathematica.tex
\section{\textsc{Mathematica} codes}\label{app:mathematica}

\subsection{Small signal-to-noise expansion}
The two first functions provided below allowing to convert moments to free cumulants and vice-versa are taken from \cite{bryc2007computing}. 

This function gives the free cumulants as function of generic moments $(M_i)$ of a density. 
\begin{mmaCell}[functionlocal=y]{Code}
m[z_] := 1 + M1 z + M2 z^2 + M3 z^3 
           + M4 z^4 + M5 z^5  
           + M6 z^6 + M7 z^7 + M8 z^8; 

Simplify[Table[{k, (-(k - 1)^(-1)/k!)
          *D[m[z]^(1 - k), {z, k}] /. 
            {z -> 0}}, {k, 2, 8}]]
\end{mmaCell}
The next function gives the moment generating function expressed with the free cumulants and thus allows us to read the expression of the moments in terms of free cumulants. We force the values of the first two free cumulants to $k_1 = 0$ and $k_2 = 1$ (but this is not necessary). 
\begin{mmaCell}{Code}
k1 = 0;  k2 = 1;  m[z_] = 1; 

r[z_] := k1 + k2 z + k3 z^2 + k4 z^3 
         + k5 z^4 + k6 z^5  
         + k7 z^6 + k8 z^7 + k9 z^8; 

Do[rtmp[z_] = PolynomialMod[z r[z], 
                  z^(k + 1)]; 
      m[z_] = PolynomialMod[1 + 
      rtmp[z m[z]], z^(k + 1)], 
      {k, 0, 8}]; 

Collect[m[z], z]
\end{mmaCell}
We now express the asymptotic moments $m_i=\theta_i:=\lim_{N\to+\infty} N^{-1}{\rm Tr} \bS^i$ of the signal $\bS$ as a function of the free 
cumulants $(k_i)$ of its asymptotic spectral density thanks to the previous function. We focus on trace-less signals $m_1=0=k_1$ and with normalized variance $m_2=1=k_2$; the first condition does not change anything from the information-theoretic point of view as explained in Remark 1 below Result~\ref{conjectureHermitianRotInv_simple}, and the second condition simply amounts to a rescaling of $\lambda$ if not a-priori verified. 
\begin{mmaCell}{Code}
m3 = k3;
m4 = k4 + 2;
m5 = 5 k3 + k5;
m6 = 3 k3^2 + 6 k4 + k6 + 5;
m7 = 7 k3 k4 + 21 k3 + 7 k5 + k7;
m8 = 28 k3^2 + 8 k3 k5 + 4 k4^2 
     + 28 k4 + 8 k6 + k8 + 14;
\end{mmaCell}
We express the free cumulants $(c_i)$ of the data $\bY$ as a function of the free cumulants $(k_i)$ of the signal $\bS$. The $(c_i)$ are the $(\bar \phi_i)$ in the Zinn-Justin and Zuber expansion \cite{zinn2003some} (see also \cite{collins2003moments} for the same expansion in terms of trace-moments, also found in \cite{zinn2003some}). The Wigner matrix $\bxi$ only shifts the second free cumulant of $\sqrt\lambda \bS$ (which is $\lambda$) by $1$. Below, variable \texttt{snr} refers to $\lambda$.
\begin{mmaCell}{Code}
c2 = snr + 1; 
c3 = k3 snr^(3/2); 
c4 = k4 snr^2; 
c5 = k5 snr^(5/2); 
c6 = k6 snr^3; 
c7 = k7 snr^(7/2); 
c8 = k8 snr^4;
\end{mmaCell}
The terms $(F_n)$ in the Zinn-Justin and Zuber expansion of the spherical integral \cite{zinn2003some}, expressed with the free cumulants $(c_i)$ of the data matrix, and the moments $(m_i)$ of the signal $\bS$; the $(m_i)$ are the $(\theta_i)$ in the expansion found in \cite{zinn2003some}.
\begin{widetext}
\begin{mmaCell}{Code}
F2 = c2/2; 
F3 = c3 m3/3; 
F4 = c4 m4/4 - 1/2 (c2^2/2 + c4); 
F5 = c5 m5/5 - m3 (c2 c3 + c5); 
F6 = - 1/2 m3^2 (c2^3/3 + c2 c4 + c3^2 + c6) + 1/6 (2 c2^3 + 12 c2 c4 + 5 c3^2 + 7 c6) 
      - m4 (c2 c4 + c3^2/2 + c6) + (c6 m6)/6; 
F7 = - m3 m4 (c2^2 c3 + c2 c5 + 2 c3 c4 + c7) + m3 (5 c2^2 c3 + 7 c2 c5 
     + 8 c3 c4 + 4 c7) - m5 (c2 c5 + c3 c4 + c7) + c7 m7/7; 
F8 = - m3 m5 (c2^2 c4 + c2 c3^2 + c2 c6 + 2 c3 c5 + c4^2 + c8) + m3^2 (2 c2^4   
      + 16 c2^2 c4 + 20 c2 c3^2 + 16 c2 c6 + 24 c3 c5 + 11 c4^2 + 9 c8) 
      - 1/2 m4^2 (c2^4/4 + c2^2 c4 + 2 c2 c3^2 + c2 c6 + 2 c3 c5 + 3/2 c4^2 + c8) 
      + 1/2 m4 (c2^4 + 11 c2^2 c4 + 14 c2 c3^2 + 16 c2 c6 + 18 c3 c5 + 11 c4^2 + 9 c8) 
      - 3/8 (3 c2^4 + 24 c2^2 c4 + 24 c2 c3^2 + 24 c2 c6 + 24 c3 c5 + 15 c4^2 + 10 c8) 
      - m6 (c2 c6 + c3 c5 + c4^2/2 + c8) + c8 m8/8;
\end{mmaCell} 
\end{widetext}   
The expansion of the mutual information is, according to our Result~\ref{conjectureHermitianRotInv_simple}, given up to $O(\lambda^4)$ by
\begin{mmaCell}{Code}
MI8 = snr - F2 snr - F3 snr^(3/2) 
      - F4 snr^2 - F5 snr^(5/2)  
      - F6 snr^3 - F7 snr^(7/2) 
      - F8 snr S^4;

MutInfo = Collect[Simplify[MI8], snr];
\end{mmaCell} 
Only the first four order are reliable. This code gives the generic expansions \eqref{expansion_generic_freeCumul} and \eqref{expansion_generic_mom} in the case $m_1=0$ and $m_2=1$ (but this can be easily adapted using the code).

\subsection{Useful code to produce Figures~\ref{Fig1} and \ref{Fig2}}

This parts evaluates the spectral density of the data matrix $\bY$ by solving the transcendental equation \eqref{eq_for_Green}  for its Green function. The spectral density is then extracted from its imaginary part.
\begin{mmaCell}{Code}
snr = 1; step = 0.0005; zAndrho = {};  
init = I/5; bound = 4;

Do[zAndrho = Append[zAndrho, 
{z, Abs[Im[g /. 
FindRoot[SetAccuracy[z == Sqrt[3 snr] 
      * Coth[g Sqrt[3 snr]] + g, 30], 
      {g, init},
      WorkingPrecision->20]]] / Pi}], 
      {z, -bound, bound, step}];
\end{mmaCell}
As there may be multiple solutions depending on the initial point \texttt{init} for the search (that may need to be tuned), a sanity check is to check that the solution found is properly normalized:
\begin{mmaCell}{Code}
Print["Normalization = ", 
      Total[zAndrho[[All, 2]]] step];
\end{mmaCell} 
Now we find an interpolating function for the spectral density of $\bY$ from 
the previously equally spaced computed points, using Hermite polynomials. Plotting this interpolating function is what gives the asymptotic red curves in Figure~\ref{Fig1}:
\begin{mmaCell}{Code}
z = zAndrho[[All, 1]];  
rho = Chop[zAndrho[[All, 2]]];

rhoInterp = ListInterpolation[rho, 
 {Min[z], Max[z]}, Method->"Hermite"];
\end{mmaCell} 
We can now compute the asymptotic mutual information using \eqref{MI_uni_asympt}, based on the interpolation function, and compare the results to the Wigner case:
\begin{mmaCell}{Code}
MI = snr + Log[snr 12]/4 
   + 0.5 NIntegrate[rhoInterp[x] 
    * rhoInterp[y] Log[Abs[(x - y) 
    / (Exp[x Sqrt[12 snr]] 
    - Exp[y Sqrt[12 snr]])]], 
    {x, Min[z], Max[z]}, 
    {y, Min[z], Max[z]}];

Print[Abs[MI - 0.5 Log[1 + snr]]];
\end{mmaCell} 
Using these pieces of code and running them for various $\lambda$, one can obtain the pink dots in Figure~\ref{Fig2}. The finite size curves (blue and orange dots) are instead simply obtained by averaging the associated formulas over many large realizations of the model.

%% file: PRE_v1.bbl
\begin{thebibliography}{134}%
\makeatletter
\providecommand \@ifxundefined [1]{%
 \@ifx{#1\undefined}
}%
\providecommand \@ifnum [1]{%
 \ifnum #1\expandafter \@firstoftwo
 \else \expandafter \@secondoftwo
 \fi
}%
\providecommand \@ifx [1]{%
 \ifx #1\expandafter \@firstoftwo
 \else \expandafter \@secondoftwo
 \fi
}%
\providecommand \natexlab [1]{#1}%
\providecommand \enquote  [1]{``#1''}%
\providecommand \bibnamefont  [1]{#1}%
\providecommand \bibfnamefont [1]{#1}%
\providecommand \citenamefont [1]{#1}%
\providecommand \href@noop [0]{\@secondoftwo}%
\providecommand \href [0]{\begingroup \@sanitize@url \@href}%
\providecommand \@href[1]{\@@startlink{#1}\@@href}%
\providecommand \@@href[1]{\endgroup#1\@@endlink}%
\providecommand \@sanitize@url [0]{\catcode `\\12\catcode `\$12\catcode
  `\&12\catcode `\#12\catcode `\^12\catcode `\_12\catcode `\%12\relax}%
\providecommand \@@startlink[1]{}%
\providecommand \@@endlink[0]{}%
\providecommand \url  [0]{\begingroup\@sanitize@url \@url }%
\providecommand \@url [1]{\endgroup\@href {#1}{\urlprefix }}%
\providecommand \urlprefix  [0]{URL }%
\providecommand \Eprint [0]{\href }%
\providecommand \doibase [0]{https://doi.org/}%
\providecommand \selectlanguage [0]{\@gobble}%
\providecommand \bibinfo  [0]{\@secondoftwo}%
\providecommand \bibfield  [0]{\@secondoftwo}%
\providecommand \translation [1]{[#1]}%
\providecommand \BibitemOpen [0]{}%
\providecommand \bibitemStop [0]{}%
\providecommand \bibitemNoStop [0]{.\EOS\space}%
\providecommand \EOS [0]{\spacefactor3000\relax}%
\providecommand \BibitemShut  [1]{\csname bibitem#1\endcsname}%
\let\auto@bib@innerbib\@empty
\bibitem [{\citenamefont {Tao}(2012)}]{TaoRMT}%
  \BibitemOpen
  \bibfield  {author} {\bibinfo {author} {\bibfnamefont {T.}~\bibnamefont
  {Tao}},\ }\href@noop {} {\emph {\bibinfo {title} {Topics in random matrix
  theory}}},\ Vol.\ \bibinfo {volume} {132}\ (\bibinfo  {publisher} {American
  Mathematical Soc.},\ \bibinfo {year} {2012})\BibitemShut {NoStop}%
\bibitem [{\citenamefont {Anderson}\ \emph {et~al.}(2010)\citenamefont
  {Anderson}, \citenamefont {Guionnet},\ and\ \citenamefont
  {Zeitouni}}]{anderson2010introduction}%
  \BibitemOpen
  \bibfield  {author} {\bibinfo {author} {\bibfnamefont {G.~W.}\ \bibnamefont
  {Anderson}}, \bibinfo {author} {\bibfnamefont {A.}~\bibnamefont {Guionnet}},\
  and\ \bibinfo {author} {\bibfnamefont {O.}~\bibnamefont {Zeitouni}},\
  }\href@noop {} {\emph {\bibinfo {title} {An introduction to random
  matrices}}},\ \bibinfo {number} {118}\ (\bibinfo  {publisher} {Cambridge
  university press},\ \bibinfo {year} {2010})\BibitemShut {NoStop}%
\bibitem [{\citenamefont {Mehta}(2004)}]{mehta2004random}%
  \BibitemOpen
  \bibfield  {author} {\bibinfo {author} {\bibfnamefont {M.~L.}\ \bibnamefont
  {Mehta}},\ }\href@noop {} {\emph {\bibinfo {title} {Random matrices}}},\
  Vol.\ \bibinfo {volume} {142}\ (\bibinfo  {publisher} {Academic press},\
  \bibinfo {year} {2004})\BibitemShut {NoStop}%
\bibitem [{\citenamefont {Potters}\ and\ \citenamefont
  {Bouchaud}(2021)}]{bouchaudpotters}%
  \BibitemOpen
  \bibfield  {author} {\bibinfo {author} {\bibfnamefont {M.}~\bibnamefont
  {Potters}}\ and\ \bibinfo {author} {\bibfnamefont {J.-P.}\ \bibnamefont
  {Bouchaud}},\ }\href@noop {} {\emph {\bibinfo {title} {A first course in
  random matrix theory}}}\ (\bibinfo  {publisher} {Cambridge University
  Press},\ \bibinfo {year} {2021})\BibitemShut {NoStop}%
\bibitem [{\citenamefont {Livan}\ \emph {et~al.}(2018)\citenamefont {Livan},
  \citenamefont {Novaes},\ and\ \citenamefont {Vivo}}]{livan2018introduction}%
  \BibitemOpen
  \bibfield  {author} {\bibinfo {author} {\bibfnamefont {G.}~\bibnamefont
  {Livan}}, \bibinfo {author} {\bibfnamefont {M.}~\bibnamefont {Novaes}},\ and\
  \bibinfo {author} {\bibfnamefont {P.}~\bibnamefont {Vivo}},\ }\href@noop {}
  {\emph {\bibinfo {title} {Introduction to random matrices: theory and
  practice}}},\ Vol.~\bibinfo {volume} {26}\ (\bibinfo  {publisher}
  {Springer},\ \bibinfo {year} {2018})\BibitemShut {NoStop}%
\bibitem [{\citenamefont {Mairal}\ \emph {et~al.}(2009)\citenamefont {Mairal},
  \citenamefont {Bach}, \citenamefont {Ponce},\ and\ \citenamefont
  {Sapiro}}]{mairal2009online}%
  \BibitemOpen
  \bibfield  {author} {\bibinfo {author} {\bibfnamefont {J.}~\bibnamefont
  {Mairal}}, \bibinfo {author} {\bibfnamefont {F.}~\bibnamefont {Bach}},
  \bibinfo {author} {\bibfnamefont {J.}~\bibnamefont {Ponce}},\ and\ \bibinfo
  {author} {\bibfnamefont {G.}~\bibnamefont {Sapiro}},\ }\bibfield  {title}
  {\bibinfo {title} {Online dictionary learning for sparse coding},\ }in\
  \href@noop {} {\emph {\bibinfo {booktitle} {Proceedings of the 26th annual
  international conference on machine learning}}}\ (\bibinfo {year} {2009})\
  pp.\ \bibinfo {pages} {689--696}\BibitemShut {NoStop}%
\bibitem [{\citenamefont {{Tošić}}\ and\ \citenamefont
  {{Frossard}}(2011)}]{dictionaryLearning}%
  \BibitemOpen
  \bibfield  {author} {\bibinfo {author} {\bibfnamefont {I.}~\bibnamefont
  {{Tošić}}}\ and\ \bibinfo {author} {\bibfnamefont {P.}~\bibnamefont
  {{Frossard}}},\ }\bibfield  {title} {\bibinfo {title} {Dictionary learning},\
  }\href {https://doi.org/10.1109/MSP.2010.939537} {\bibfield  {journal}
  {\bibinfo  {journal} {IEEE Signal Processing Magazine}\ }\textbf {\bibinfo
  {volume} {28}},\ \bibinfo {pages} {27} (\bibinfo {year} {2011})}\BibitemShut
  {NoStop}%
\bibitem [{\citenamefont {Bengio}\ \emph {et~al.}(2013)\citenamefont {Bengio},
  \citenamefont {Courville},\ and\ \citenamefont
  {Vincent}}]{bengio2013representation}%
  \BibitemOpen
  \bibfield  {author} {\bibinfo {author} {\bibfnamefont {Y.}~\bibnamefont
  {Bengio}}, \bibinfo {author} {\bibfnamefont {A.}~\bibnamefont {Courville}},\
  and\ \bibinfo {author} {\bibfnamefont {P.}~\bibnamefont {Vincent}},\
  }\bibfield  {title} {\bibinfo {title} {Representation learning: A review and
  new perspectives},\ }\href@noop {} {\bibfield  {journal} {\bibinfo  {journal}
  {IEEE transactions on pattern analysis and machine intelligence}\ }\textbf
  {\bibinfo {volume} {35}},\ \bibinfo {pages} {1798} (\bibinfo {year}
  {2013})}\BibitemShut {NoStop}%
\bibitem [{\citenamefont {Olshausen}\ and\ \citenamefont
  {Field}(1996)}]{olshausen1996emergence}%
  \BibitemOpen
  \bibfield  {author} {\bibinfo {author} {\bibfnamefont {B.~A.}\ \bibnamefont
  {Olshausen}}\ and\ \bibinfo {author} {\bibfnamefont {D.~J.}\ \bibnamefont
  {Field}},\ }\bibfield  {title} {\bibinfo {title} {Emergence of simple-cell
  receptive field properties by learning a sparse code for natural images},\
  }\href@noop {} {\bibfield  {journal} {\bibinfo  {journal} {Nature}\ }\textbf
  {\bibinfo {volume} {381}},\ \bibinfo {pages} {607} (\bibinfo {year}
  {1996})}\BibitemShut {NoStop}%
\bibitem [{\citenamefont {Olshausen}\ and\ \citenamefont
  {Field}(1997)}]{olshausen1997sparse}%
  \BibitemOpen
  \bibfield  {author} {\bibinfo {author} {\bibfnamefont {B.~A.}\ \bibnamefont
  {Olshausen}}\ and\ \bibinfo {author} {\bibfnamefont {D.~J.}\ \bibnamefont
  {Field}},\ }\bibfield  {title} {\bibinfo {title} {Sparse coding with an
  overcomplete basis set: A strategy employed by v1?},\ }\href@noop {}
  {\bibfield  {journal} {\bibinfo  {journal} {Vision research}\ }\textbf
  {\bibinfo {volume} {37}},\ \bibinfo {pages} {3311} (\bibinfo {year}
  {1997})}\BibitemShut {NoStop}%
\bibitem [{\citenamefont {Kreutz-Delgado}\ \emph {et~al.}(2003)\citenamefont
  {Kreutz-Delgado}, \citenamefont {Murray}, \citenamefont {Rao}, \citenamefont
  {Engan}, \citenamefont {Lee},\ and\ \citenamefont
  {Sejnowski}}]{kreutz2003dictionary}%
  \BibitemOpen
  \bibfield  {author} {\bibinfo {author} {\bibfnamefont {K.}~\bibnamefont
  {Kreutz-Delgado}}, \bibinfo {author} {\bibfnamefont {J.~F.}\ \bibnamefont
  {Murray}}, \bibinfo {author} {\bibfnamefont {B.~D.}\ \bibnamefont {Rao}},
  \bibinfo {author} {\bibfnamefont {K.}~\bibnamefont {Engan}}, \bibinfo
  {author} {\bibfnamefont {T.-W.}\ \bibnamefont {Lee}},\ and\ \bibinfo {author}
  {\bibfnamefont {T.~J.}\ \bibnamefont {Sejnowski}},\ }\bibfield  {title}
  {\bibinfo {title} {Dictionary learning algorithms for sparse
  representation},\ }\href@noop {} {\bibfield  {journal} {\bibinfo  {journal}
  {Neural computation}\ }\textbf {\bibinfo {volume} {15}},\ \bibinfo {pages}
  {349} (\bibinfo {year} {2003})}\BibitemShut {NoStop}%
\bibitem [{\citenamefont {Cand{\`e}s}\ \emph {et~al.}(2011)\citenamefont
  {Cand{\`e}s}, \citenamefont {Li}, \citenamefont {Ma},\ and\ \citenamefont
  {Wright}}]{candes2011robust}%
  \BibitemOpen
  \bibfield  {author} {\bibinfo {author} {\bibfnamefont {E.~J.}\ \bibnamefont
  {Cand{\`e}s}}, \bibinfo {author} {\bibfnamefont {X.}~\bibnamefont {Li}},
  \bibinfo {author} {\bibfnamefont {Y.}~\bibnamefont {Ma}},\ and\ \bibinfo
  {author} {\bibfnamefont {J.}~\bibnamefont {Wright}},\ }\bibfield  {title}
  {\bibinfo {title} {Robust principal component analysis?},\ }\href@noop {}
  {\bibfield  {journal} {\bibinfo  {journal} {Journal of the ACM (JACM)}\
  }\textbf {\bibinfo {volume} {58}},\ \bibinfo {pages} {1} (\bibinfo {year}
  {2011})}\BibitemShut {NoStop}%
\bibitem [{\citenamefont {Perry}\ \emph {et~al.}(2018)\citenamefont {Perry},
  \citenamefont {Wein}, \citenamefont {Bandeira},\ and\ \citenamefont
  {Moitra}}]{perry2018optimality}%
  \BibitemOpen
  \bibfield  {author} {\bibinfo {author} {\bibfnamefont {A.}~\bibnamefont
  {Perry}}, \bibinfo {author} {\bibfnamefont {A.~S.}\ \bibnamefont {Wein}},
  \bibinfo {author} {\bibfnamefont {A.~S.}\ \bibnamefont {Bandeira}},\ and\
  \bibinfo {author} {\bibfnamefont {A.}~\bibnamefont {Moitra}},\ }\bibfield
  {title} {\bibinfo {title} {Optimality and sub-optimality of {PCA} {I}: Spiked
  random matrix models},\ }\href@noop {} {\bibfield  {journal} {\bibinfo
  {journal} {Annals of Statistics}\ }\textbf {\bibinfo {volume} {46}},\
  \bibinfo {pages} {2416} (\bibinfo {year} {2018})}\BibitemShut {NoStop}%
\bibitem [{\citenamefont {Hajek}\ \emph {et~al.}(2017)\citenamefont {Hajek},
  \citenamefont {Wu},\ and\ \citenamefont {Xu}}]{hajek2017information}%
  \BibitemOpen
  \bibfield  {author} {\bibinfo {author} {\bibfnamefont {B.}~\bibnamefont
  {Hajek}}, \bibinfo {author} {\bibfnamefont {Y.}~\bibnamefont {Wu}},\ and\
  \bibinfo {author} {\bibfnamefont {J.}~\bibnamefont {Xu}},\ }\bibfield
  {title} {\bibinfo {title} {Information limits for recovering a hidden
  community},\ }\href@noop {} {\bibfield  {journal} {\bibinfo  {journal} {IEEE
  Transactions on Information Theory}\ }\textbf {\bibinfo {volume} {63}},\
  \bibinfo {pages} {4729} (\bibinfo {year} {2017})}\BibitemShut {NoStop}%
\bibitem [{\citenamefont {Belouchrani}\ \emph {et~al.}(1997)\citenamefont
  {Belouchrani}, \citenamefont {Abed-Meraim}, \citenamefont {Cardoso},\ and\
  \citenamefont {Moulines}}]{belouchrani1997blind}%
  \BibitemOpen
  \bibfield  {author} {\bibinfo {author} {\bibfnamefont {A.}~\bibnamefont
  {Belouchrani}}, \bibinfo {author} {\bibfnamefont {K.}~\bibnamefont
  {Abed-Meraim}}, \bibinfo {author} {\bibfnamefont {J.-F.}\ \bibnamefont
  {Cardoso}},\ and\ \bibinfo {author} {\bibfnamefont {E.}~\bibnamefont
  {Moulines}},\ }\bibfield  {title} {\bibinfo {title} {A blind source
  separation technique using second-order statistics},\ }\href@noop {}
  {\bibfield  {journal} {\bibinfo  {journal} {IEEE Transactions on signal
  processing}\ }\textbf {\bibinfo {volume} {45}},\ \bibinfo {pages} {434}
  (\bibinfo {year} {1997})}\BibitemShut {NoStop}%
\bibitem [{\citenamefont {Cand{\`e}s}\ and\ \citenamefont
  {Recht}(2009)}]{candes2009exact}%
  \BibitemOpen
  \bibfield  {author} {\bibinfo {author} {\bibfnamefont {E.}~\bibnamefont
  {Cand{\`e}s}}\ and\ \bibinfo {author} {\bibfnamefont {B.}~\bibnamefont
  {Recht}},\ }\bibfield  {title} {\bibinfo {title} {Exact matrix completion via
  convex optimization},\ }\href@noop {} {\bibfield  {journal} {\bibinfo
  {journal} {Foundations of Computational mathematics}\ }\textbf {\bibinfo
  {volume} {9}},\ \bibinfo {pages} {717} (\bibinfo {year} {2009})}\BibitemShut
  {NoStop}%
\bibitem [{\citenamefont {Cand{\`e}s}\ and\ \citenamefont
  {Tao}(2010)}]{candes2010power}%
  \BibitemOpen
  \bibfield  {author} {\bibinfo {author} {\bibfnamefont {E.~J.}\ \bibnamefont
  {Cand{\`e}s}}\ and\ \bibinfo {author} {\bibfnamefont {T.}~\bibnamefont
  {Tao}},\ }\bibfield  {title} {\bibinfo {title} {The power of convex
  relaxation: Near-optimal matrix completion},\ }\href@noop {} {\bibfield
  {journal} {\bibinfo  {journal} {IEEE Transactions on Information Theory}\
  }\textbf {\bibinfo {volume} {56}},\ \bibinfo {pages} {2053} (\bibinfo {year}
  {2010})}\BibitemShut {NoStop}%
\bibitem [{\citenamefont {Abbe}(2017)}]{abbe2017community}%
  \BibitemOpen
  \bibfield  {author} {\bibinfo {author} {\bibfnamefont {E.}~\bibnamefont
  {Abbe}},\ }\bibfield  {title} {\bibinfo {title} {Community detection and
  stochastic block models: recent developments},\ }\href@noop {} {\bibfield
  {journal} {\bibinfo  {journal} {The Journal of Machine Learning Research}\
  }\textbf {\bibinfo {volume} {18}},\ \bibinfo {pages} {6446} (\bibinfo {year}
  {2017})}\BibitemShut {NoStop}%
\bibitem [{\citenamefont {{Lelarge}}\ and\ \citenamefont
  {{Miolane}}(2016)}]{2016arXiv161103888L}%
  \BibitemOpen
  \bibfield  {author} {\bibinfo {author} {\bibfnamefont {M.}~\bibnamefont
  {{Lelarge}}}\ and\ \bibinfo {author} {\bibfnamefont {L.}~\bibnamefont
  {{Miolane}}},\ }\bibfield  {title} {\bibinfo {title} {{Fundamental limits of
  symmetric low-rank matrix estimation}},\ }\href@noop {} {\bibfield  {journal}
  {\bibinfo  {journal} {ArXiv e-prints}\ } (\bibinfo {year} {2016})},\ \Eprint
  {https://arxiv.org/abs/1611.03888} {arXiv:1611.03888 [math.PR]} \BibitemShut
  {NoStop}%
\bibitem [{\citenamefont {Caltagirone}\ \emph {et~al.}(2017)\citenamefont
  {Caltagirone}, \citenamefont {Lelarge},\ and\ \citenamefont
  {Miolane}}]{caltagirone2017recovering}%
  \BibitemOpen
  \bibfield  {author} {\bibinfo {author} {\bibfnamefont {F.}~\bibnamefont
  {Caltagirone}}, \bibinfo {author} {\bibfnamefont {M.}~\bibnamefont
  {Lelarge}},\ and\ \bibinfo {author} {\bibfnamefont {L.}~\bibnamefont
  {Miolane}},\ }\bibfield  {title} {\bibinfo {title} {Recovering asymmetric
  communities in the stochastic block model},\ }\href@noop {} {\bibfield
  {journal} {\bibinfo  {journal} {IEEE Transactions on Network Science and
  Engineering}\ }\textbf {\bibinfo {volume} {5}},\ \bibinfo {pages} {237}
  (\bibinfo {year} {2017})}\BibitemShut {NoStop}%
\bibitem [{\citenamefont {Johnstone}(2001)}]{johnstone2001distribution}%
  \BibitemOpen
  \bibfield  {author} {\bibinfo {author} {\bibfnamefont {I.}~\bibnamefont
  {Johnstone}},\ }\bibfield  {title} {\bibinfo {title} {On the distribution of
  the largest eigenvalue in principal components analysis},\ }\href@noop {}
  {\bibfield  {journal} {\bibinfo  {journal} {The Annals of statistics}\
  }\textbf {\bibinfo {volume} {29}},\ \bibinfo {pages} {295} (\bibinfo {year}
  {2001})}\BibitemShut {NoStop}%
\bibitem [{\citenamefont {Johnstone}\ and\ \citenamefont
  {Lu}(2004)}]{johnstone2004sparse}%
  \BibitemOpen
  \bibfield  {author} {\bibinfo {author} {\bibfnamefont {I.}~\bibnamefont
  {Johnstone}}\ and\ \bibinfo {author} {\bibfnamefont {A.}~\bibnamefont {Lu}},\
  }\bibfield  {title} {\bibinfo {title} {Sparse principal components
  analysis},\ }\href@noop {} {\bibfield  {journal} {\bibinfo  {journal}
  {Unpublished manuscript}\ }\textbf {\bibinfo {volume} {7}} (\bibinfo {year}
  {2004})}\BibitemShut {NoStop}%
\bibitem [{\citenamefont {Zou}\ \emph {et~al.}(2006)\citenamefont {Zou},
  \citenamefont {Hastie},\ and\ \citenamefont {Tibshirani}}]{zou2006sparse}%
  \BibitemOpen
  \bibfield  {author} {\bibinfo {author} {\bibfnamefont {H.}~\bibnamefont
  {Zou}}, \bibinfo {author} {\bibfnamefont {T.}~\bibnamefont {Hastie}},\ and\
  \bibinfo {author} {\bibfnamefont {R.}~\bibnamefont {Tibshirani}},\ }\bibfield
   {title} {\bibinfo {title} {Sparse principal component analysis},\
  }\href@noop {} {\bibfield  {journal} {\bibinfo  {journal} {Journal of
  computational and graphical statistics}\ }\textbf {\bibinfo {volume} {15}},\
  \bibinfo {pages} {265} (\bibinfo {year} {2006})}\BibitemShut {NoStop}%
\bibitem [{\citenamefont {Johnstone}\ and\ \citenamefont
  {Lu}(2012)}]{johnstone2012consistency}%
  \BibitemOpen
  \bibfield  {author} {\bibinfo {author} {\bibfnamefont {I.}~\bibnamefont
  {Johnstone}}\ and\ \bibinfo {author} {\bibfnamefont {A.}~\bibnamefont {Lu}},\
  }\bibfield  {title} {\bibinfo {title} {On consistency and sparsity for
  principal components analysis in high dimensions},\ }\href@noop {} {\bibfield
   {journal} {\bibinfo  {journal} {Journal of the American Statistical
  Association}\ } (\bibinfo {year} {2012})}\BibitemShut {NoStop}%
\bibitem [{\citenamefont {Baik}\ \emph {et~al.}(2005)\citenamefont {Baik},
  \citenamefont {Arous},\ and\ \citenamefont {P{\'e}ch{\'e}}}]{baik2005phase}%
  \BibitemOpen
  \bibfield  {author} {\bibinfo {author} {\bibfnamefont {J.}~\bibnamefont
  {Baik}}, \bibinfo {author} {\bibfnamefont {G.~B.}\ \bibnamefont {Arous}},\
  and\ \bibinfo {author} {\bibfnamefont {S.}~\bibnamefont {P{\'e}ch{\'e}}},\
  }\bibfield  {title} {\bibinfo {title} {Phase transition of the largest
  eigenvalue for nonnull complex sample covariance matrices},\ }\href@noop {}
  {\bibfield  {journal} {\bibinfo  {journal} {Annals of Probability}\ ,\
  \bibinfo {pages} {1643}} (\bibinfo {year} {2005})}\BibitemShut {NoStop}%
\bibitem [{\citenamefont {Baik}\ and\ \citenamefont
  {Silverstein}(2006)}]{baik2006eigenvalues}%
  \BibitemOpen
  \bibfield  {author} {\bibinfo {author} {\bibfnamefont {J.}~\bibnamefont
  {Baik}}\ and\ \bibinfo {author} {\bibfnamefont {J.~W.}\ \bibnamefont
  {Silverstein}},\ }\bibfield  {title} {\bibinfo {title} {Eigenvalues of large
  sample covariance matrices of spiked population models},\ }\href@noop {}
  {\bibfield  {journal} {\bibinfo  {journal} {Journal of multivariate
  analysis}\ }\textbf {\bibinfo {volume} {97}},\ \bibinfo {pages} {1382}
  (\bibinfo {year} {2006})}\BibitemShut {NoStop}%
\bibitem [{\citenamefont {Korada}\ and\ \citenamefont
  {Macris}(2009)}]{koradamacris}%
  \BibitemOpen
  \bibfield  {author} {\bibinfo {author} {\bibfnamefont {S.~B.}\ \bibnamefont
  {Korada}}\ and\ \bibinfo {author} {\bibfnamefont {N.}~\bibnamefont
  {Macris}},\ }\bibfield  {title} {\bibinfo {title} {Exact solution of the
  gauge symmetric p-spin glass model on a complete graph},\ }\href@noop {}
  {\bibfield  {journal} {\bibinfo  {journal} {Journal of Statistical Physics}\
  }\textbf {\bibinfo {volume} {136}},\ \bibinfo {pages} {205} (\bibinfo {year}
  {2009})}\BibitemShut {NoStop}%
\bibitem [{\citenamefont {Lesieur}\ \emph {et~al.}(2017)\citenamefont
  {Lesieur}, \citenamefont {Krzakala},\ and\ \citenamefont
  {Zdeborov{\'a}}}]{lesieur2017constrained}%
  \BibitemOpen
  \bibfield  {author} {\bibinfo {author} {\bibfnamefont {T.}~\bibnamefont
  {Lesieur}}, \bibinfo {author} {\bibfnamefont {F.}~\bibnamefont {Krzakala}},\
  and\ \bibinfo {author} {\bibfnamefont {L.}~\bibnamefont {Zdeborov{\'a}}},\
  }\bibfield  {title} {\bibinfo {title} {Constrained low-rank matrix
  estimation: Phase transitions, approximate message passing and
  applications},\ }\href@noop {} {\bibfield  {journal} {\bibinfo  {journal}
  {Journal of Statistical Mechanics: Theory and Experiment}\ }\textbf {\bibinfo
  {volume} {2017}},\ \bibinfo {pages} {073403} (\bibinfo {year}
  {2017})}\BibitemShut {NoStop}%
\bibitem [{\citenamefont {Deshpande}\ \emph {et~al.}(2017)\citenamefont
  {Deshpande}, \citenamefont {Abbe},\ and\ \citenamefont
  {Montanari}}]{deshpande2017asymptotic}%
  \BibitemOpen
  \bibfield  {author} {\bibinfo {author} {\bibfnamefont {Y.}~\bibnamefont
  {Deshpande}}, \bibinfo {author} {\bibfnamefont {E.}~\bibnamefont {Abbe}},\
  and\ \bibinfo {author} {\bibfnamefont {A.}~\bibnamefont {Montanari}},\
  }\bibfield  {title} {\bibinfo {title} {Asymptotic mutual information for the
  balanced binary stochastic block model},\ }\href@noop {} {\bibfield
  {journal} {\bibinfo  {journal} {Information and Inference: A Journal of the
  IMA}\ }\textbf {\bibinfo {volume} {6}},\ \bibinfo {pages} {125} (\bibinfo
  {year} {2017})}\BibitemShut {NoStop}%
\bibitem [{\citenamefont {Miolane}(2017)}]{miolane2017fundamental}%
  \BibitemOpen
  \bibfield  {author} {\bibinfo {author} {\bibfnamefont {L.}~\bibnamefont
  {Miolane}},\ }\bibfield  {title} {\bibinfo {title} {Fundamental limits of
  low-rank matrix estimation: the non-symmetric case},\ }\href@noop {}
  {\bibfield  {journal} {\bibinfo  {journal} {arXiv preprint arXiv:1702.00473}\
  } (\bibinfo {year} {2017})}\BibitemShut {NoStop}%
\bibitem [{\citenamefont {{Lesieur}}\ \emph {et~al.}()\citenamefont
  {{Lesieur}}, \citenamefont {{Miolane}}, \citenamefont {{Lelarge}},
  \citenamefont {{Krzakala}},\ and\ \citenamefont
  {{Zdeborov{\'a}}}}]{2017arXiv170108010L}%
  \BibitemOpen
  \bibfield  {author} {\bibinfo {author} {\bibfnamefont {T.}~\bibnamefont
  {{Lesieur}}}, \bibinfo {author} {\bibfnamefont {L.}~\bibnamefont
  {{Miolane}}}, \bibinfo {author} {\bibfnamefont {M.}~\bibnamefont
  {{Lelarge}}}, \bibinfo {author} {\bibfnamefont {F.}~\bibnamefont
  {{Krzakala}}},\ and\ \bibinfo {author} {\bibfnamefont {L.}~\bibnamefont
  {{Zdeborov{\'a}}}},\ }\bibfield  {title} {\bibinfo {title} {{Statistical and
  computational phase transitions in spiked tensor estimation}},\ }in\
  \href@noop {} {\emph {\bibinfo {booktitle} {IEEE International Symposium on
  Information Theory (ISIT), 2017}}}\BibitemShut {NoStop}%
\bibitem [{\citenamefont {Barbier}\ \emph {et~al.}(2016)\citenamefont
  {Barbier}, \citenamefont {Dia}, \citenamefont {Macris}, \citenamefont
  {Krzakala}, \citenamefont {Lesieur},\ and\ \citenamefont
  {Zdeborov\'{a}}}]{XXT}%
  \BibitemOpen
  \bibfield  {author} {\bibinfo {author} {\bibfnamefont {J.}~\bibnamefont
  {Barbier}}, \bibinfo {author} {\bibfnamefont {M.}~\bibnamefont {Dia}},
  \bibinfo {author} {\bibfnamefont {N.}~\bibnamefont {Macris}}, \bibinfo
  {author} {\bibfnamefont {F.}~\bibnamefont {Krzakala}}, \bibinfo {author}
  {\bibfnamefont {T.}~\bibnamefont {Lesieur}},\ and\ \bibinfo {author}
  {\bibfnamefont {L.}~\bibnamefont {Zdeborov\'{a}}},\ }\bibfield  {title}
  {\bibinfo {title} {Mutual information for symmetric rank-one matrix
  estimation: A proof of the replica formula},\ }in\ \href@noop {} {\emph
  {\bibinfo {booktitle} {Advances in Neural Information Processing Systems
  (NIPS) 29}}}\ (\bibinfo {year} {2016})\ pp.\ \bibinfo {pages}
  {424--432}\BibitemShut {NoStop}%
\bibitem [{\citenamefont {Barbier}\ and\ \citenamefont
  {Macris}(2019{\natexlab{a}})}]{BarbierM17a}%
  \BibitemOpen
  \bibfield  {author} {\bibinfo {author} {\bibfnamefont {J.}~\bibnamefont
  {Barbier}}\ and\ \bibinfo {author} {\bibfnamefont {N.}~\bibnamefont
  {Macris}},\ }\bibfield  {title} {\bibinfo {title} {The adaptive interpolation
  method: a simple scheme to prove replica formulas in {B}ayesian inference},\
  }\href@noop {} {\bibfield  {journal} {\bibinfo  {journal} {Probability Theory
  and Related Fields}\ }\textbf {\bibinfo {volume} {174}},\ \bibinfo {pages}
  {1133} (\bibinfo {year} {2019}{\natexlab{a}})}\BibitemShut {NoStop}%
\bibitem [{\citenamefont {Barbier}\ and\ \citenamefont
  {Macris}(2019{\natexlab{b}})}]{BarbierMacris2019}%
  \BibitemOpen
  \bibfield  {author} {\bibinfo {author} {\bibfnamefont {J.}~\bibnamefont
  {Barbier}}\ and\ \bibinfo {author} {\bibfnamefont {N.}~\bibnamefont
  {Macris}},\ }\bibfield  {title} {\bibinfo {title} {The adaptive interpolation
  method for proving replica formulas. {A}pplications to the
  {C}urie{\textendash}{W}eiss and {W}igner spike models},\ }\href
  {https://doi.org/10.1088/1751-8121/ab2735} {\bibfield  {journal} {\bibinfo
  {journal} {Journal of Physics A: Mathematical and Theoretical}\ }\textbf
  {\bibinfo {volume} {52}},\ \bibinfo {pages} {294002} (\bibinfo {year}
  {2019}{\natexlab{b}})}\BibitemShut {NoStop}%
\bibitem [{\citenamefont {{Barbier}}\ \emph {et~al.}(2018)\citenamefont
  {{Barbier}}, \citenamefont {{Dia}}, \citenamefont {{Macris}}, \citenamefont
  {{Krzakala}},\ and\ \citenamefont {{Zdeborov\'a}}}]{2018arXiv181202537B}%
  \BibitemOpen
  \bibfield  {author} {\bibinfo {author} {\bibfnamefont {J.}~\bibnamefont
  {{Barbier}}}, \bibinfo {author} {\bibfnamefont {M.}~\bibnamefont {{Dia}}},
  \bibinfo {author} {\bibfnamefont {N.}~\bibnamefont {{Macris}}}, \bibinfo
  {author} {\bibfnamefont {F.}~\bibnamefont {{Krzakala}}},\ and\ \bibinfo
  {author} {\bibfnamefont {L.}~\bibnamefont {{Zdeborov\'a}}},\ }\bibfield
  {title} {\bibinfo {title} {{Rank-one matrix estimation: analysis of
  algorithmic and information theoretic limits by the spatial coupling
  method}},\ }\href@noop {} {\bibfield  {journal} {\bibinfo  {journal} {arXiv
  e-prints}\ ,\ \bibinfo {pages} {arXiv:1812.02537}} (\bibinfo {year}
  {2018})},\ \Eprint {https://arxiv.org/abs/1812.02537} {arXiv:1812.02537
  [cs.IT]} \BibitemShut {NoStop}%
\bibitem [{\citenamefont {{Barbier}}\ \emph {et~al.}(2017)\citenamefont
  {{Barbier}}, \citenamefont {{Macris}},\ and\ \citenamefont
  {{Miolane}}}]{2017arXiv170910368B}%
  \BibitemOpen
  \bibfield  {author} {\bibinfo {author} {\bibfnamefont {J.}~\bibnamefont
  {{Barbier}}}, \bibinfo {author} {\bibfnamefont {N.}~\bibnamefont
  {{Macris}}},\ and\ \bibinfo {author} {\bibfnamefont {L.}~\bibnamefont
  {{Miolane}}},\ }\bibfield  {title} {\bibinfo {title} {{The Layered Structure
  of Tensor Estimation and its Mutual Information}},\ }in\ \href@noop {} {\emph
  {\bibinfo {booktitle} {47th Annual Allerton Conference on Communication,
  Control, and Computing (Allerton)}}}\ (\bibinfo {year} {2017})\ \Eprint
  {https://arxiv.org/abs/1709.10368} {arXiv:1709.10368} \BibitemShut {NoStop}%
\bibitem [{\citenamefont {Alaoui}\ \emph {et~al.}(2017)\citenamefont {Alaoui},
  \citenamefont {Krzakala},\ and\ \citenamefont {Jordan}}]{alaoui2017finite}%
  \BibitemOpen
  \bibfield  {author} {\bibinfo {author} {\bibfnamefont {A.~E.}\ \bibnamefont
  {Alaoui}}, \bibinfo {author} {\bibfnamefont {F.}~\bibnamefont {Krzakala}},\
  and\ \bibinfo {author} {\bibfnamefont {M.~I.}\ \bibnamefont {Jordan}},\
  }\bibfield  {title} {\bibinfo {title} {Finite size corrections and likelihood
  ratio fluctuations in the spiked {W}igner model},\ }\href@noop {} {\bibfield
  {journal} {\bibinfo  {journal} {arXiv preprint arXiv:1710.02903}\ } (\bibinfo
  {year} {2017})}\BibitemShut {NoStop}%
\bibitem [{\citenamefont {El~Alaoui}\ and\ \citenamefont
  {Krzakala}(2018)}]{el2018estimation}%
  \BibitemOpen
  \bibfield  {author} {\bibinfo {author} {\bibfnamefont {A.}~\bibnamefont
  {El~Alaoui}}\ and\ \bibinfo {author} {\bibfnamefont {F.}~\bibnamefont
  {Krzakala}},\ }\bibfield  {title} {\bibinfo {title} {Estimation in the spiked
  {W}igner model: a short proof of the replica formula},\ }in\ \href@noop {}
  {\emph {\bibinfo {booktitle} {2018 IEEE International Symposium on
  Information Theory (ISIT)}}}\ (\bibinfo {organization} {IEEE},\ \bibinfo
  {year} {2018})\ pp.\ \bibinfo {pages} {1874--1878}\BibitemShut {NoStop}%
\bibitem [{\citenamefont {Mourrat}(2018)}]{mourrat2018hamilton}%
  \BibitemOpen
  \bibfield  {author} {\bibinfo {author} {\bibfnamefont {J.-C.}\ \bibnamefont
  {Mourrat}},\ }\bibfield  {title} {\bibinfo {title} {{H}amilton-{J}acobi
  equations for mean-field disordered systems},\ }\href@noop {} {\bibfield
  {journal} {\bibinfo  {journal} {arXiv:1811.01432}\ } (\bibinfo {year}
  {2018})}\BibitemShut {NoStop}%
\bibitem [{\citenamefont {Mourrat}(2019)}]{mourrat2019hamilton}%
  \BibitemOpen
  \bibfield  {author} {\bibinfo {author} {\bibfnamefont {J.-C.}\ \bibnamefont
  {Mourrat}},\ }\bibfield  {title} {\bibinfo {title} {{H}amilton-{J}acobi
  equations for finite-rank matrix inference},\ }\href@noop {} {\bibfield
  {journal} {\bibinfo  {journal} {arXiv:1904.05294}\ } (\bibinfo {year}
  {2019})}\BibitemShut {NoStop}%
\bibitem [{\citenamefont {Barbier}\ \emph {et~al.}(2020)\citenamefont
  {Barbier}, \citenamefont {Macris},\ and\ \citenamefont
  {Rush}}]{barbier2020all}%
  \BibitemOpen
  \bibfield  {author} {\bibinfo {author} {\bibfnamefont {J.}~\bibnamefont
  {Barbier}}, \bibinfo {author} {\bibfnamefont {N.}~\bibnamefont {Macris}},\
  and\ \bibinfo {author} {\bibfnamefont {C.}~\bibnamefont {Rush}},\ }\bibfield
  {title} {\bibinfo {title} {All-or-nothing statistical and computational phase
  transitions in sparse spiked matrix estimation},\ }in\ \href@noop {} {\emph
  {\bibinfo {booktitle} {Advances in Neural Information Processing Systems}}}\
  (\bibinfo {year} {2020})\BibitemShut {NoStop}%
\bibitem [{\citenamefont {Reeves}(2020)}]{reeves2020information}%
  \BibitemOpen
  \bibfield  {author} {\bibinfo {author} {\bibfnamefont {G.}~\bibnamefont
  {Reeves}},\ }\bibfield  {title} {\bibinfo {title} {Information-theoretic
  limits for the matrix tensor product},\ }\href@noop {} {\bibfield  {journal}
  {\bibinfo  {journal} {IEEE Journal on Selected Areas in Information Theory}\
  } (\bibinfo {year} {2020})}\BibitemShut {NoStop}%
\bibitem [{\citenamefont {Alberici}\ \emph
  {et~al.}(2021{\natexlab{a}})\citenamefont {Alberici}, \citenamefont
  {Camilli}, \citenamefont {Contucci},\ and\ \citenamefont
  {Mingione}}]{alberici2021multi}%
  \BibitemOpen
  \bibfield  {author} {\bibinfo {author} {\bibfnamefont {D.}~\bibnamefont
  {Alberici}}, \bibinfo {author} {\bibfnamefont {F.}~\bibnamefont {Camilli}},
  \bibinfo {author} {\bibfnamefont {P.}~\bibnamefont {Contucci}},\ and\
  \bibinfo {author} {\bibfnamefont {E.}~\bibnamefont {Mingione}},\ }\bibfield
  {title} {\bibinfo {title} {The multi-species mean-field spin-glass on the
  {N}ishimori line},\ }\href@noop {} {\bibfield  {journal} {\bibinfo  {journal}
  {Journal of Statistical Physics}\ }\textbf {\bibinfo {volume} {182}},\
  \bibinfo {pages} {1} (\bibinfo {year} {2021}{\natexlab{a}})}\BibitemShut
  {NoStop}%
\bibitem [{\citenamefont {Camilli}\ \emph {et~al.}(2021)\citenamefont
  {Camilli}, \citenamefont {Contucci},\ and\ \citenamefont
  {Mingione}}]{camilli2021inference}%
  \BibitemOpen
  \bibfield  {author} {\bibinfo {author} {\bibfnamefont {F.}~\bibnamefont
  {Camilli}}, \bibinfo {author} {\bibfnamefont {P.}~\bibnamefont {Contucci}},\
  and\ \bibinfo {author} {\bibfnamefont {E.}~\bibnamefont {Mingione}},\
  }\bibfield  {title} {\bibinfo {title} {An inference problem in a mismatched
  setting as a spin-glass model with mattis interaction},\ }\href@noop {}
  {\bibfield  {journal} {\bibinfo  {journal} {arXiv preprint arXiv:2107.11689}\
  } (\bibinfo {year} {2021})}\BibitemShut {NoStop}%
\bibitem [{\citenamefont {Pourkamali}\ and\ \citenamefont
  {Macris}(2021)}]{pourkamali2021mismatched}%
  \BibitemOpen
  \bibfield  {author} {\bibinfo {author} {\bibfnamefont {F.}~\bibnamefont
  {Pourkamali}}\ and\ \bibinfo {author} {\bibfnamefont {N.}~\bibnamefont
  {Macris}},\ }\bibfield  {title} {\bibinfo {title} {Mismatched estimation of
  rank-one symmetric matrices under gaussian noise},\ }\href@noop {} {\bibfield
   {journal} {\bibinfo  {journal} {arXiv preprint arXiv:2107.08927}\ }
  (\bibinfo {year} {2021})}\BibitemShut {NoStop}%
\bibitem [{\citenamefont {Alberici}\ \emph
  {et~al.}(2021{\natexlab{b}})\citenamefont {Alberici}, \citenamefont
  {Camilli}, \citenamefont {Contucci},\ and\ \citenamefont
  {Mingione}}]{alberici2021solution}%
  \BibitemOpen
  \bibfield  {author} {\bibinfo {author} {\bibfnamefont {D.}~\bibnamefont
  {Alberici}}, \bibinfo {author} {\bibfnamefont {F.}~\bibnamefont {Camilli}},
  \bibinfo {author} {\bibfnamefont {P.}~\bibnamefont {Contucci}},\ and\
  \bibinfo {author} {\bibfnamefont {E.}~\bibnamefont {Mingione}},\ }\bibfield
  {title} {\bibinfo {title} {The solution of the deep boltzmann machine on the
  {N}ishimori line},\ }\href@noop {} {\bibfield  {journal} {\bibinfo  {journal}
  {Communications in Mathematical Physics}\ ,\ \bibinfo {pages} {1}} (\bibinfo
  {year} {2021}{\natexlab{b}})}\BibitemShut {NoStop}%
\bibitem [{\citenamefont {Mannelli}\ \emph
  {et~al.}(2019{\natexlab{a}})\citenamefont {Mannelli}, \citenamefont {Biroli},
  \citenamefont {Cammarota}, \citenamefont {Krzakala},\ and\ \citenamefont
  {Zdeborov{\'a}}}]{mannelli2019afraid}%
  \BibitemOpen
  \bibfield  {author} {\bibinfo {author} {\bibfnamefont {S.~S.}\ \bibnamefont
  {Mannelli}}, \bibinfo {author} {\bibfnamefont {G.}~\bibnamefont {Biroli}},
  \bibinfo {author} {\bibfnamefont {C.}~\bibnamefont {Cammarota}}, \bibinfo
  {author} {\bibfnamefont {F.}~\bibnamefont {Krzakala}},\ and\ \bibinfo
  {author} {\bibfnamefont {L.}~\bibnamefont {Zdeborov{\'a}}},\ }\bibfield
  {title} {\bibinfo {title} {Who is afraid of big bad minima? analysis of
  gradient-flow in spiked matrix-tensor models},\ }in\ \href@noop {} {\emph
  {\bibinfo {booktitle} {Advances in Neural Information Processing Systems}}}\
  (\bibinfo {year} {2019})\ pp.\ \bibinfo {pages} {8676--8686}\BibitemShut
  {NoStop}%
\bibitem [{\citenamefont {Mannelli}\ \emph
  {et~al.}(2019{\natexlab{b}})\citenamefont {Mannelli}, \citenamefont
  {Krzakala}, \citenamefont {Urbani},\ and\ \citenamefont
  {Zdeborova}}]{mannelli2019passed}%
  \BibitemOpen
  \bibfield  {author} {\bibinfo {author} {\bibfnamefont {S.~S.}\ \bibnamefont
  {Mannelli}}, \bibinfo {author} {\bibfnamefont {F.}~\bibnamefont {Krzakala}},
  \bibinfo {author} {\bibfnamefont {P.}~\bibnamefont {Urbani}},\ and\ \bibinfo
  {author} {\bibfnamefont {L.}~\bibnamefont {Zdeborova}},\ }\bibfield  {title}
  {\bibinfo {title} {Passed \& spurious: Descent algorithms and local minima in
  spiked matrix-tensor models},\ }in\ \href@noop {} {\emph {\bibinfo
  {booktitle} {international conference on machine learning}}}\ (\bibinfo
  {organization} {PMLR},\ \bibinfo {year} {2019})\ pp.\ \bibinfo {pages}
  {4333--4342}\BibitemShut {NoStop}%
\bibitem [{\citenamefont {Liu}\ \emph {et~al.}(2021)\citenamefont {Liu},
  \citenamefont {Li}, \citenamefont {Wei}, \citenamefont {Zhou},\ and\
  \citenamefont {Zhao}}]{liu2021noisy}%
  \BibitemOpen
  \bibfield  {author} {\bibinfo {author} {\bibfnamefont {T.}~\bibnamefont
  {Liu}}, \bibinfo {author} {\bibfnamefont {Y.}~\bibnamefont {Li}}, \bibinfo
  {author} {\bibfnamefont {S.}~\bibnamefont {Wei}}, \bibinfo {author}
  {\bibfnamefont {E.}~\bibnamefont {Zhou}},\ and\ \bibinfo {author}
  {\bibfnamefont {T.}~\bibnamefont {Zhao}},\ }\bibfield  {title} {\bibinfo
  {title} {Noisy gradient descent converges to flat minima for nonconvex matrix
  factorization},\ }\href@noop {} {\bibfield  {journal} {\bibinfo  {journal}
  {arXiv preprint arXiv:2102.12430}\ } (\bibinfo {year} {2021})}\BibitemShut
  {NoStop}%
\bibitem [{\citenamefont {Maillard}\ \emph {et~al.}(2019)\citenamefont
  {Maillard}, \citenamefont {Foini}, \citenamefont {Castellanos}, \citenamefont
  {Krzakala}, \citenamefont {M{\'e}zard},\ and\ \citenamefont
  {Zdeborov{\'a}}}]{maillard2019high}%
  \BibitemOpen
  \bibfield  {author} {\bibinfo {author} {\bibfnamefont {A.}~\bibnamefont
  {Maillard}}, \bibinfo {author} {\bibfnamefont {L.}~\bibnamefont {Foini}},
  \bibinfo {author} {\bibfnamefont {A.~L.}\ \bibnamefont {Castellanos}},
  \bibinfo {author} {\bibfnamefont {F.}~\bibnamefont {Krzakala}}, \bibinfo
  {author} {\bibfnamefont {M.}~\bibnamefont {M{\'e}zard}},\ and\ \bibinfo
  {author} {\bibfnamefont {L.}~\bibnamefont {Zdeborov{\'a}}},\ }\bibfield
  {title} {\bibinfo {title} {High-temperature expansions and message passing
  algorithms},\ }\href@noop {} {\bibfield  {journal} {\bibinfo  {journal}
  {Journal of Statistical Mechanics: Theory and Experiment}\ }\textbf {\bibinfo
  {volume} {2019}},\ \bibinfo {pages} {113301} (\bibinfo {year}
  {2019})}\BibitemShut {NoStop}%
\bibitem [{\citenamefont {Bodin}\ and\ \citenamefont
  {Macris}(2021)}]{bodin2021rank}%
  \BibitemOpen
  \bibfield  {author} {\bibinfo {author} {\bibfnamefont {A.}~\bibnamefont
  {Bodin}}\ and\ \bibinfo {author} {\bibfnamefont {N.}~\bibnamefont {Macris}},\
  }\bibfield  {title} {\bibinfo {title} {Rank-one matrix estimation: analytic
  time evolution of gradient descent dynamics},\ }\href@noop {} {\bibfield
  {journal} {\bibinfo  {journal} {arXiv preprint arXiv:2105.12257}\ } (\bibinfo
  {year} {2021})}\BibitemShut {NoStop}%
\bibitem [{\citenamefont {Sakata}\ and\ \citenamefont
  {Kabashima}(2013{\natexlab{a}})}]{SK13EPL}%
  \BibitemOpen
  \bibfield  {author} {\bibinfo {author} {\bibfnamefont {A.}~\bibnamefont
  {Sakata}}\ and\ \bibinfo {author} {\bibfnamefont {Y.}~\bibnamefont
  {Kabashima}},\ }\bibfield  {title} {\bibinfo {title} {Statistical mechanics
  of dictionary learning},\ }\href
  {https://doi.org/10.1209/0295-5075/103/28008} {\bibfield  {journal} {\bibinfo
   {journal} {Europhysics Letters}\ }\textbf {\bibinfo {volume} {103}},\
  \bibinfo {pages} {28008} (\bibinfo {year} {2013}{\natexlab{a}})}\BibitemShut
  {NoStop}%
\bibitem [{\citenamefont {Sakata}\ and\ \citenamefont
  {Kabashima}(2013{\natexlab{b}})}]{SK13ISIT}%
  \BibitemOpen
  \bibfield  {author} {\bibinfo {author} {\bibfnamefont {A.}~\bibnamefont
  {Sakata}}\ and\ \bibinfo {author} {\bibfnamefont {Y.}~\bibnamefont
  {Kabashima}},\ }\bibfield  {title} {\bibinfo {title} {Sample complexity of
  {B}ayesian optimal dictionary learning},\ }\href@noop {} {\bibfield
  {journal} {\bibinfo  {journal} {2013 IEEE International Symposium on
  Information Theory}\ ,\ \bibinfo {pages} {669 }} (\bibinfo {year}
  {2013}{\natexlab{b}})}\BibitemShut {NoStop}%
\bibitem [{\citenamefont {Krzakala}\ \emph {et~al.}(2013)\citenamefont
  {Krzakala}, \citenamefont {M{\'e}zard},\ and\ \citenamefont
  {Zdeborov{\'a}}}]{FML2013}%
  \BibitemOpen
  \bibfield  {author} {\bibinfo {author} {\bibfnamefont {F.}~\bibnamefont
  {Krzakala}}, \bibinfo {author} {\bibfnamefont {M.}~\bibnamefont
  {M{\'e}zard}},\ and\ \bibinfo {author} {\bibfnamefont {L.}~\bibnamefont
  {Zdeborov{\'a}}},\ }\bibfield  {title} {\bibinfo {title} {Phase diagram and
  approximate message passing for blind calibration and dictionary learning},\
  }\href@noop {} {\bibfield  {journal} {\bibinfo  {journal} {2013 IEEE
  International Symposium on Information Theory}\ ,\ \bibinfo {pages} {659 }}
  (\bibinfo {year} {2013})}\BibitemShut {NoStop}%
\bibitem [{\citenamefont {Kabashima}\ \emph {et~al.}(2016)\citenamefont
  {Kabashima}, \citenamefont {Krzakala}, \citenamefont {M{\'e}zard},
  \citenamefont {Sakata},\ and\ \citenamefont
  {Zdeborov{\'a}}}]{kabashima2016phase}%
  \BibitemOpen
  \bibfield  {author} {\bibinfo {author} {\bibfnamefont {Y.}~\bibnamefont
  {Kabashima}}, \bibinfo {author} {\bibfnamefont {F.}~\bibnamefont {Krzakala}},
  \bibinfo {author} {\bibfnamefont {M.}~\bibnamefont {M{\'e}zard}}, \bibinfo
  {author} {\bibfnamefont {A.}~\bibnamefont {Sakata}},\ and\ \bibinfo {author}
  {\bibfnamefont {L.}~\bibnamefont {Zdeborov{\'a}}},\ }\bibfield  {title}
  {\bibinfo {title} {Phase transitions and sample complexity in {B}ayes-optimal
  matrix factorization},\ }\href@noop {} {\bibfield  {journal} {\bibinfo
  {journal} {IEEE Transactions on information theory}\ }\textbf {\bibinfo
  {volume} {62}},\ \bibinfo {pages} {4228} (\bibinfo {year}
  {2016})}\BibitemShut {NoStop}%
\bibitem [{\citenamefont {Schmidt}(2018)}]{thesis_schmidt}%
  \BibitemOpen
  \bibfield  {author} {\bibinfo {author} {\bibfnamefont {H.~C.}\ \bibnamefont
  {Schmidt}},\ }\href {http://www.theses.fr/2018SACLS366} {\bibinfo {title}
  {Statistical physics of sparse and dense models in optimization and
  inference}} (\bibinfo {year} {2018})\BibitemShut {NoStop}%
\bibitem [{\citenamefont {Maillard}\ \emph {et~al.}(2021)\citenamefont
  {Maillard}, \citenamefont {Krzakala}, \citenamefont {M{\'e}zard},\ and\
  \citenamefont {Zdeborov{\'a}}}]{maillard2021perturbative}%
  \BibitemOpen
  \bibfield  {author} {\bibinfo {author} {\bibfnamefont {A.}~\bibnamefont
  {Maillard}}, \bibinfo {author} {\bibfnamefont {F.}~\bibnamefont {Krzakala}},
  \bibinfo {author} {\bibfnamefont {M.}~\bibnamefont {M{\'e}zard}},\ and\
  \bibinfo {author} {\bibfnamefont {L.}~\bibnamefont {Zdeborov{\'a}}},\
  }\href@noop {} {\bibinfo {title} {Perturbative construction of mean-field
  equations in extensive-rank matrix factorization and denoising}} (\bibinfo
  {year} {2021}),\ \Eprint {https://arxiv.org/abs/2110.08775} {arXiv:2110.08775
  [cond-mat.dis-nn]} \BibitemShut {NoStop}%
\bibitem [{\citenamefont {Parker}\ \emph
  {et~al.}(2014{\natexlab{a}})\citenamefont {Parker}, \citenamefont
  {Schniter},\ and\ \citenamefont {Cevher}}]{PSCvI}%
  \BibitemOpen
  \bibfield  {author} {\bibinfo {author} {\bibfnamefont {J.~T.}\ \bibnamefont
  {Parker}}, \bibinfo {author} {\bibfnamefont {P.}~\bibnamefont {Schniter}},\
  and\ \bibinfo {author} {\bibfnamefont {V.}~\bibnamefont {Cevher}},\
  }\bibfield  {title} {\bibinfo {title} {Bilinear generalized approximate
  message passing - part i: Derivation},\ }\href@noop {} {\bibfield  {journal}
  {\bibinfo  {journal} {IEEE Transactions on Signal Processing}\ }\textbf
  {\bibinfo {volume} {62}},\ \bibinfo {pages} {5839} (\bibinfo {year}
  {2014}{\natexlab{a}})}\BibitemShut {NoStop}%
\bibitem [{\citenamefont {Parker}\ \emph
  {et~al.}(2014{\natexlab{b}})\citenamefont {Parker}, \citenamefont
  {Schniter},\ and\ \citenamefont {Cevher}}]{PSCvII}%
  \BibitemOpen
  \bibfield  {author} {\bibinfo {author} {\bibfnamefont {J.~T.}\ \bibnamefont
  {Parker}}, \bibinfo {author} {\bibfnamefont {P.}~\bibnamefont {Schniter}},\
  and\ \bibinfo {author} {\bibfnamefont {V.}~\bibnamefont {Cevher}},\
  }\bibfield  {title} {\bibinfo {title} {Bilinear generalized approximate
  message passing - part ii: Applications},\ }\href@noop {} {\bibfield
  {journal} {\bibinfo  {journal} {IEEE Transactions on Signal Processing}\
  }\textbf {\bibinfo {volume} {62}},\ \bibinfo {pages} {5854} (\bibinfo {year}
  {2014}{\natexlab{b}})}\BibitemShut {NoStop}%
\bibitem [{\citenamefont {Bun}\ \emph {et~al.}(2016)\citenamefont {Bun},
  \citenamefont {Allez}, \citenamefont {Bouchaud},\ and\ \citenamefont
  {Potters}}]{bun2016rotational}%
  \BibitemOpen
  \bibfield  {author} {\bibinfo {author} {\bibfnamefont {J.}~\bibnamefont
  {Bun}}, \bibinfo {author} {\bibfnamefont {R.}~\bibnamefont {Allez}}, \bibinfo
  {author} {\bibfnamefont {J.-P.}\ \bibnamefont {Bouchaud}},\ and\ \bibinfo
  {author} {\bibfnamefont {M.}~\bibnamefont {Potters}},\ }\bibfield  {title}
  {\bibinfo {title} {Rotational invariant estimator for general noisy
  matrices},\ }\href@noop {} {\bibfield  {journal} {\bibinfo  {journal} {IEEE
  Transactions on Information Theory}\ }\textbf {\bibinfo {volume} {62}},\
  \bibinfo {pages} {7475} (\bibinfo {year} {2016})}\BibitemShut {NoStop}%
\bibitem [{Note1()}]{Note1}%
  \BibitemOpen
  \bibinfo {note} {It is now known that the estimator proposed in paper \cite
  {bun2016rotational} is Bayes-optimal in certain settings, see \cite
  {maillard2021perturbative}.}\BibitemShut {Stop}%
\bibitem [{\citenamefont {Harish-Chandra}(1957)}]{harish1957differential}%
  \BibitemOpen
  \bibfield  {author} {\bibinfo {author} {\bibnamefont {Harish-Chandra}},\
  }\bibfield  {title} {\bibinfo {title} {Differential operators on a semisimple
  lie algebra},\ }\href@noop {} {\bibfield  {journal} {\bibinfo  {journal}
  {American Journal of Mathematics}\ ,\ \bibinfo {pages} {87}} (\bibinfo {year}
  {1957})}\BibitemShut {NoStop}%
\bibitem [{\citenamefont {Itzykson}\ and\ \citenamefont
  {Zuber}(1980)}]{itzykson1980planar}%
  \BibitemOpen
  \bibfield  {author} {\bibinfo {author} {\bibfnamefont {C.}~\bibnamefont
  {Itzykson}}\ and\ \bibinfo {author} {\bibfnamefont {J.-B.}\ \bibnamefont
  {Zuber}},\ }\bibfield  {title} {\bibinfo {title} {The planar approximation.
  ii},\ }\href@noop {} {\bibfield  {journal} {\bibinfo  {journal} {Journal of
  Mathematical Physics}\ }\textbf {\bibinfo {volume} {21}},\ \bibinfo {pages}
  {411} (\bibinfo {year} {1980})}\BibitemShut {NoStop}%
\bibitem [{\citenamefont {Bleher}\ and\ \citenamefont
  {Its}(2001)}]{bleher2001random}%
  \BibitemOpen
  \bibfield  {author} {\bibinfo {author} {\bibfnamefont {P.}~\bibnamefont
  {Bleher}}\ and\ \bibinfo {author} {\bibfnamefont {A.}~\bibnamefont {Its}},\
  }\href@noop {} {\emph {\bibinfo {title} {Random matrix models and their
  applications}}},\ Vol.~\bibinfo {volume} {40}\ (\bibinfo  {publisher}
  {Cambridge university press},\ \bibinfo {year} {2001})\BibitemShut {NoStop}%
\bibitem [{\citenamefont {Br{\'e}zin}\ and\ \citenamefont
  {Wadia}(1993)}]{brezin1993large}%
  \BibitemOpen
  \bibfield  {author} {\bibinfo {author} {\bibfnamefont {E.}~\bibnamefont
  {Br{\'e}zin}}\ and\ \bibinfo {author} {\bibfnamefont {S.~R.}\ \bibnamefont
  {Wadia}},\ }\href@noop {} {\emph {\bibinfo {title} {The large {N} expansion
  in quantum field theory and statistical physics}}}\ (\bibinfo  {publisher}
  {World scientific},\ \bibinfo {year} {1993})\BibitemShut {NoStop}%
\bibitem [{\citenamefont {Mehta}(1993)}]{mehta1993method}%
  \BibitemOpen
  \bibfield  {author} {\bibinfo {author} {\bibfnamefont {M.~L.}\ \bibnamefont
  {Mehta}},\ }\bibfield  {title} {\bibinfo {title} {A method of integration
  over matrix variables},\ }in\ \href@noop {} {\emph {\bibinfo {booktitle} {The
  large {N} Expansion In Quantum Field Theory And Statistical Physics: From
  Spin Systems to 2-Dimensional Gravity}}}\ (\bibinfo  {publisher} {World
  Scientific},\ \bibinfo {year} {1993})\ pp.\ \bibinfo {pages}
  {616--629}\BibitemShut {NoStop}%
\bibitem [{\citenamefont {Chadha}\ \emph {et~al.}(1981)\citenamefont {Chadha},
  \citenamefont {Mahoux},\ and\ \citenamefont {Mehta}}]{chadha1981method}%
  \BibitemOpen
  \bibfield  {author} {\bibinfo {author} {\bibfnamefont {S.}~\bibnamefont
  {Chadha}}, \bibinfo {author} {\bibfnamefont {G.}~\bibnamefont {Mahoux}},\
  and\ \bibinfo {author} {\bibfnamefont {M.~L.}\ \bibnamefont {Mehta}},\
  }\bibfield  {title} {\bibinfo {title} {A method of integration over matrix
  variables: 2},\ }\href@noop {} {\bibfield  {journal} {\bibinfo  {journal}
  {Journal of Physics A: Mathematical and General}\ }\textbf {\bibinfo {volume}
  {14}},\ \bibinfo {pages} {579} (\bibinfo {year} {1981})}\BibitemShut
  {NoStop}%
\bibitem [{\citenamefont {Kazakov}(2000)}]{kazakov2000solvable}%
  \BibitemOpen
  \bibfield  {author} {\bibinfo {author} {\bibfnamefont {V.~A.}\ \bibnamefont
  {Kazakov}},\ }\bibfield  {title} {\bibinfo {title} {Solvable matrix models},\
  }\href@noop {} {\bibfield  {journal} {\bibinfo  {journal} {arXiv preprint
  hep-th/0003064}\ } (\bibinfo {year} {2000})}\BibitemShut {NoStop}%
\bibitem [{\citenamefont {Kazakov}\ and\ \citenamefont
  {Migdal}(1993)}]{kazakov1993induced}%
  \BibitemOpen
  \bibfield  {author} {\bibinfo {author} {\bibfnamefont {V.}~\bibnamefont
  {Kazakov}}\ and\ \bibinfo {author} {\bibfnamefont {A.}~\bibnamefont
  {Migdal}},\ }\bibfield  {title} {\bibinfo {title} {Induced gauge theory at
  large {N}},\ }\href@noop {} {\bibfield  {journal} {\bibinfo  {journal}
  {Nuclear Physics B}\ }\textbf {\bibinfo {volume} {397}},\ \bibinfo {pages}
  {214} (\bibinfo {year} {1993})}\BibitemShut {NoStop}%
\bibitem [{\citenamefont {Kazakov}(1993)}]{kazakov1993d}%
  \BibitemOpen
  \bibfield  {author} {\bibinfo {author} {\bibfnamefont {V.}~\bibnamefont
  {Kazakov}},\ }\bibfield  {title} {\bibinfo {title} {D-dimensional induced
  gauge theory as a solvable matrix model},\ }\href@noop {} {\bibfield
  {journal} {\bibinfo  {journal} {Nuclear Physics B-Proceedings Supplements}\
  }\textbf {\bibinfo {volume} {30}},\ \bibinfo {pages} {149} (\bibinfo {year}
  {1993})}\BibitemShut {NoStop}%
\bibitem [{\citenamefont {Eynard}\ \emph {et~al.}(2015)\citenamefont {Eynard},
  \citenamefont {Kimura},\ and\ \citenamefont {Ribault}}]{eynard2015random}%
  \BibitemOpen
  \bibfield  {author} {\bibinfo {author} {\bibfnamefont {B.}~\bibnamefont
  {Eynard}}, \bibinfo {author} {\bibfnamefont {T.}~\bibnamefont {Kimura}},\
  and\ \bibinfo {author} {\bibfnamefont {S.}~\bibnamefont {Ribault}},\
  }\bibfield  {title} {\bibinfo {title} {Random matrices},\ }\href@noop {}
  {\bibfield  {journal} {\bibinfo  {journal} {arXiv preprint arXiv:1510.04430}\
  } (\bibinfo {year} {2015})}\BibitemShut {NoStop}%
\bibitem [{\citenamefont {Zinn-Justin}(2000)}]{zinn2000dilute}%
  \BibitemOpen
  \bibfield  {author} {\bibinfo {author} {\bibfnamefont {P.}~\bibnamefont
  {Zinn-Justin}},\ }\bibfield  {title} {\bibinfo {title} {The dilute potts
  model on random surfaces},\ }\href@noop {} {\bibfield  {journal} {\bibinfo
  {journal} {Journal of Statistical Physics}\ }\textbf {\bibinfo {volume}
  {98}},\ \bibinfo {pages} {245} (\bibinfo {year} {2000})}\BibitemShut
  {NoStop}%
\bibitem [{\citenamefont {Stephanov}\ \emph {et~al.}(2005)\citenamefont
  {Stephanov}, \citenamefont {Verbaarschot},\ and\ \citenamefont
  {Wettig}}]{stephanov2005random}%
  \BibitemOpen
  \bibfield  {author} {\bibinfo {author} {\bibfnamefont {M.}~\bibnamefont
  {Stephanov}}, \bibinfo {author} {\bibfnamefont {J.}~\bibnamefont
  {Verbaarschot}},\ and\ \bibinfo {author} {\bibfnamefont {T.}~\bibnamefont
  {Wettig}},\ }\bibfield  {title} {\bibinfo {title} {Random matrices},\
  }\href@noop {} {\bibfield  {journal} {\bibinfo  {journal} {arXiv preprint
  hep-ph/0509286}\ } (\bibinfo {year} {2005})}\BibitemShut {NoStop}%
\bibitem [{\citenamefont {Zvonkin}(1997)}]{zvonkin1997matrix}%
  \BibitemOpen
  \bibfield  {author} {\bibinfo {author} {\bibfnamefont {A.}~\bibnamefont
  {Zvonkin}},\ }\bibfield  {title} {\bibinfo {title} {Matrix integrals and map
  enumeration: an accessible introduction},\ }\href@noop {} {\bibfield
  {journal} {\bibinfo  {journal} {Mathematical and Computer Modelling}\
  }\textbf {\bibinfo {volume} {26}},\ \bibinfo {pages} {281} (\bibinfo {year}
  {1997})}\BibitemShut {NoStop}%
\bibitem [{\citenamefont {Matytsin}\ and\ \citenamefont
  {Zaugg}(1997)}]{matytsin1997kosterlitz}%
  \BibitemOpen
  \bibfield  {author} {\bibinfo {author} {\bibfnamefont {A.}~\bibnamefont
  {Matytsin}}\ and\ \bibinfo {author} {\bibfnamefont {P.}~\bibnamefont
  {Zaugg}},\ }\bibfield  {title} {\bibinfo {title} {Kosterlitz-{T}houless phase
  transitions on discretized random surfaces},\ }\href@noop {} {\bibfield
  {journal} {\bibinfo  {journal} {Nuclear Physics B}\ }\textbf {\bibinfo
  {volume} {497}},\ \bibinfo {pages} {658} (\bibinfo {year}
  {1997})}\BibitemShut {NoStop}%
\bibitem [{\citenamefont {Barbier}(2019)}]{barbier2019overlap}%
  \BibitemOpen
  \bibfield  {author} {\bibinfo {author} {\bibfnamefont {J.}~\bibnamefont
  {Barbier}},\ }\bibfield  {title} {\bibinfo {title} {Overlap matrix
  concentration in optimal {B}ayesian inference},\ }\href@noop {} {\bibfield
  {journal} {\bibinfo  {journal} {Information and Inference: A Journal of the
  IMA}\ } (\bibinfo {year} {2019})}\BibitemShut {NoStop}%
\bibitem [{\citenamefont {Guionnet}(2004)}]{guionnet2004first}%
  \BibitemOpen
  \bibfield  {author} {\bibinfo {author} {\bibfnamefont {A.}~\bibnamefont
  {Guionnet}},\ }\bibfield  {title} {\bibinfo {title} {First order asymptotics
  of matrix integrals; a rigorous approach towards the understanding of matrix
  models},\ }\href@noop {} {\bibfield  {journal} {\bibinfo  {journal}
  {Communications in mathematical physics}\ }\textbf {\bibinfo {volume}
  {244}},\ \bibinfo {pages} {527} (\bibinfo {year} {2004})}\BibitemShut
  {NoStop}%
\bibitem [{\citenamefont {M{\'e}zard}\ \emph {et~al.}(1987)\citenamefont
  {M{\'e}zard}, \citenamefont {Parisi},\ and\ \citenamefont
  {Virasoro}}]{MezardParisi87b}%
  \BibitemOpen
  \bibfield  {author} {\bibinfo {author} {\bibfnamefont {M.}~\bibnamefont
  {M{\'e}zard}}, \bibinfo {author} {\bibfnamefont {G.}~\bibnamefont {Parisi}},\
  and\ \bibinfo {author} {\bibfnamefont {M.~A.}\ \bibnamefont {Virasoro}},\
  }\href@noop {} {\emph {\bibinfo {title} {Spin-Glass Theory and Beyond}}},\
  \bibinfo {series} {Lecture Notes in Physics}, Vol.~\bibinfo {volume} {9}\
  (\bibinfo  {publisher} {World Scientific},\ \bibinfo {address} {Singapore},\
  \bibinfo {year} {1987})\BibitemShut {NoStop}%
\bibitem [{\citenamefont {M\'ezard}\ and\ \citenamefont
  {Montanari}(2009)}]{MezardMontanari09}%
  \BibitemOpen
  \bibfield  {author} {\bibinfo {author} {\bibfnamefont {M.}~\bibnamefont
  {M\'ezard}}\ and\ \bibinfo {author} {\bibfnamefont {A.}~\bibnamefont
  {Montanari}},\ }\href@noop {} {\emph {\bibinfo {title} {Information, Physics
  and Computation}}}\ (\bibinfo  {publisher} {Oxford Press},\ \bibinfo {year}
  {2009})\BibitemShut {NoStop}%
\bibitem [{\citenamefont {Johansson}(2001)}]{johansson2001universality}%
  \BibitemOpen
  \bibfield  {author} {\bibinfo {author} {\bibfnamefont {K.~J.}\ \bibnamefont
  {Johansson}},\ }\bibfield  {title} {\bibinfo {title} {Universality of the
  local spacing distribution in certain ensembles of hermitian {W}igner
  matrices},\ }\href@noop {} {\bibfield  {journal} {\bibinfo  {journal}
  {Communications in Mathematical Physics}\ }\textbf {\bibinfo {volume}
  {215}},\ \bibinfo {pages} {683} (\bibinfo {year} {2001})}\BibitemShut
  {NoStop}%
\bibitem [{\citenamefont {Rosenzweig}\ and\ \citenamefont
  {Porter}(1960)}]{rosenzweig1960repulsion}%
  \BibitemOpen
  \bibfield  {author} {\bibinfo {author} {\bibfnamefont {N.}~\bibnamefont
  {Rosenzweig}}\ and\ \bibinfo {author} {\bibfnamefont {C.~E.}\ \bibnamefont
  {Porter}},\ }\bibfield  {title} {\bibinfo {title} {" repulsion of energy
  levels" in complex atomic spectra},\ }\href@noop {} {\bibfield  {journal}
  {\bibinfo  {journal} {Physical Review}\ }\textbf {\bibinfo {volume} {120}},\
  \bibinfo {pages} {1698} (\bibinfo {year} {1960})}\BibitemShut {NoStop}%
\bibitem [{\citenamefont {Br{\'e}zin}\ and\ \citenamefont
  {Hikami}(1996)}]{brezin1996correlations}%
  \BibitemOpen
  \bibfield  {author} {\bibinfo {author} {\bibfnamefont {E.}~\bibnamefont
  {Br{\'e}zin}}\ and\ \bibinfo {author} {\bibfnamefont {S.}~\bibnamefont
  {Hikami}},\ }\bibfield  {title} {\bibinfo {title} {Correlations of nearby
  levels induced by a random potential},\ }\href@noop {} {\bibfield  {journal}
  {\bibinfo  {journal} {Nuclear Physics B}\ }\textbf {\bibinfo {volume}
  {479}},\ \bibinfo {pages} {697} (\bibinfo {year} {1996})}\BibitemShut
  {NoStop}%
\bibitem [{\citenamefont {Biroli}\ and\ \citenamefont
  {Tarzia}(2021)}]{biroli2021levy}%
  \BibitemOpen
  \bibfield  {author} {\bibinfo {author} {\bibfnamefont {G.}~\bibnamefont
  {Biroli}}\ and\ \bibinfo {author} {\bibfnamefont {M.}~\bibnamefont
  {Tarzia}},\ }\bibfield  {title} {\bibinfo {title}
  {L{\'e}vy-{R}osenzweig-{P}orter random matrix ensemble},\ }\href@noop {}
  {\bibfield  {journal} {\bibinfo  {journal} {Physical Review B}\ }\textbf
  {\bibinfo {volume} {103}},\ \bibinfo {pages} {104205} (\bibinfo {year}
  {2021})}\BibitemShut {NoStop}%
\bibitem [{\citenamefont {Kravtsov}\ \emph {et~al.}(2015)\citenamefont
  {Kravtsov}, \citenamefont {Khaymovich}, \citenamefont {Cuevas},\ and\
  \citenamefont {Amini}}]{kravtsov2015random}%
  \BibitemOpen
  \bibfield  {author} {\bibinfo {author} {\bibfnamefont {V.}~\bibnamefont
  {Kravtsov}}, \bibinfo {author} {\bibfnamefont {I.}~\bibnamefont
  {Khaymovich}}, \bibinfo {author} {\bibfnamefont {E.}~\bibnamefont {Cuevas}},\
  and\ \bibinfo {author} {\bibfnamefont {M.}~\bibnamefont {Amini}},\ }\bibfield
   {title} {\bibinfo {title} {A random matrix model with localization and
  ergodic transitions},\ }\href@noop {} {\bibfield  {journal} {\bibinfo
  {journal} {New Journal of Physics}\ }\textbf {\bibinfo {volume} {17}},\
  \bibinfo {pages} {122002} (\bibinfo {year} {2015})}\BibitemShut {NoStop}%
\bibitem [{\citenamefont {Facoetti}\ \emph {et~al.}(2016)\citenamefont
  {Facoetti}, \citenamefont {Vivo},\ and\ \citenamefont
  {Biroli}}]{facoetti2016non}%
  \BibitemOpen
  \bibfield  {author} {\bibinfo {author} {\bibfnamefont {D.}~\bibnamefont
  {Facoetti}}, \bibinfo {author} {\bibfnamefont {P.}~\bibnamefont {Vivo}},\
  and\ \bibinfo {author} {\bibfnamefont {G.}~\bibnamefont {Biroli}},\
  }\bibfield  {title} {\bibinfo {title} {From non-ergodic eigenvectors to local
  resolvent statistics and back: A random matrix perspective},\ }\href@noop {}
  {\bibfield  {journal} {\bibinfo  {journal} {EPL (Europhysics Letters)}\
  }\textbf {\bibinfo {volume} {115}},\ \bibinfo {pages} {47003} (\bibinfo
  {year} {2016})}\BibitemShut {NoStop}%
\bibitem [{Note2()}]{Note2}%
  \BibitemOpen
  \bibinfo {note} {We require symmetry except in the trivial case
  $p_{S,N}({\protect \bm {\lambda }}^S)=\delta ({\protect \bm {\lambda
  }}^S-{\protect \bm {\lambda }}^S_0)$ for some fixed ${\protect \bm {\lambda
  }}^S_0$.}\BibitemShut {Stop}%
\bibitem [{\citenamefont {Guionnet}\ and\ \citenamefont
  {Zeitouni}(2002)}]{guionnet2002large}%
  \BibitemOpen
  \bibfield  {author} {\bibinfo {author} {\bibfnamefont {A.}~\bibnamefont
  {Guionnet}}\ and\ \bibinfo {author} {\bibfnamefont {O.}~\bibnamefont
  {Zeitouni}},\ }\bibfield  {title} {\bibinfo {title} {Large deviations
  asymptotics for spherical integrals},\ }\href
  {https://www.sciencedirect.com/science/article/pii/S0022123601938339}
  {\bibfield  {journal} {\bibinfo  {journal} {Journal of functional analysis}\
  }\textbf {\bibinfo {volume} {188}},\ \bibinfo {pages} {461} (\bibinfo {year}
  {2002})}\BibitemShut {NoStop}%
\bibitem [{\citenamefont {Matytsin}(1994)}]{matytsin1994large}%
  \BibitemOpen
  \bibfield  {author} {\bibinfo {author} {\bibfnamefont {A.}~\bibnamefont
  {Matytsin}},\ }\bibfield  {title} {\bibinfo {title} {On the large-n limit of
  the {I}tzykson-{Z}uber integral},\ }\href@noop {} {\bibfield  {journal}
  {\bibinfo  {journal} {Nuclear Physics B}\ }\textbf {\bibinfo {volume}
  {411}},\ \bibinfo {pages} {805} (\bibinfo {year} {1994})}\BibitemShut
  {NoStop}%
\bibitem [{\citenamefont {Mingo}\ and\ \citenamefont
  {Speicher}(2017)}]{mingo2017free}%
  \BibitemOpen
  \bibfield  {author} {\bibinfo {author} {\bibfnamefont {J.~A.}\ \bibnamefont
  {Mingo}}\ and\ \bibinfo {author} {\bibfnamefont {R.}~\bibnamefont
  {Speicher}},\ }\href@noop {} {\emph {\bibinfo {title} {Free probability and
  random matrices}}},\ Vol.~\bibinfo {volume} {35}\ (\bibinfo  {publisher}
  {Springer},\ \bibinfo {year} {2017})\BibitemShut {NoStop}%
\bibitem [{\citenamefont {Pielaszkiewicz}\ and\ \citenamefont
  {Singull}(2015)}]{pielaszkiewicz2015closed}%
  \BibitemOpen
  \bibfield  {author} {\bibinfo {author} {\bibfnamefont {J.}~\bibnamefont
  {Pielaszkiewicz}}\ and\ \bibinfo {author} {\bibfnamefont {M.}~\bibnamefont
  {Singull}},\ }\href@noop {} {\bibinfo {title} {Closed form of the asymptotic
  spectral distribution of random matrices using free independence}} (\bibinfo
  {year} {2015})\BibitemShut {NoStop}%
\bibitem [{\citenamefont {Guo}\ \emph {et~al.}(2005)\citenamefont {Guo},
  \citenamefont {Shamai},\ and\ \citenamefont {Verd{\'{u}}}}]{GuoShamaiVerdu}%
  \BibitemOpen
  \bibfield  {author} {\bibinfo {author} {\bibfnamefont {D.}~\bibnamefont
  {Guo}}, \bibinfo {author} {\bibfnamefont {S.}~\bibnamefont {Shamai}},\ and\
  \bibinfo {author} {\bibfnamefont {S.}~\bibnamefont {Verd{\'{u}}}},\
  }\bibfield  {title} {\bibinfo {title} {Mutual information and minimum
  mean-square error in gaussian channels},\ }\href@noop {} {\bibfield
  {journal} {\bibinfo  {journal} {IEEE Trans. on Inf. Theory}\ }\textbf
  {\bibinfo {volume} {51}} (\bibinfo {year} {2005})}\BibitemShut {NoStop}%
\bibitem [{Note3()}]{Note3}%
  \BibitemOpen
  \bibinfo {note} {The factor $4$ that differs from the $2$ in the usual I-MMSE
  relation \cite {GuoShamaiVerdu} comes from the fact that the Wigner matrix to
  denoise has only a fraction $N(N+1)/(2N^2)= 1/2+O(1/N)$ of independent
  entries. The $O(1/N)$ correction comes from the diagonal terms in matrix
  ${\protect \bm {S}}$ for which the signal-to-noise ratio is different than
  the one of the off-diagonal entries. The complex noise case of the I-MMSE
  relation is discussed in Section V.D of \cite {GuoShamaiVerdu}.}\BibitemShut
  {Stop}%
\bibitem [{\citenamefont {Fischmann}\ \emph {et~al.}(2012)\citenamefont
  {Fischmann}, \citenamefont {Bruzda}, \citenamefont {Khoruzhenko},
  \citenamefont {Sommers},\ and\ \citenamefont
  {{\.Z}yczkowski}}]{fischmann2012induced}%
  \BibitemOpen
  \bibfield  {author} {\bibinfo {author} {\bibfnamefont {J.}~\bibnamefont
  {Fischmann}}, \bibinfo {author} {\bibfnamefont {W.}~\bibnamefont {Bruzda}},
  \bibinfo {author} {\bibfnamefont {B.~A.}\ \bibnamefont {Khoruzhenko}},
  \bibinfo {author} {\bibfnamefont {H.-J.}\ \bibnamefont {Sommers}},\ and\
  \bibinfo {author} {\bibfnamefont {K.}~\bibnamefont {{\.Z}yczkowski}},\
  }\bibfield  {title} {\bibinfo {title} {Induced {G}inibre ensemble of random
  matrices and quantum operations},\ }\href@noop {} {\bibfield  {journal}
  {\bibinfo  {journal} {Journal of Physics A: Mathematical and Theoretical}\
  }\textbf {\bibinfo {volume} {45}},\ \bibinfo {pages} {075203} (\bibinfo
  {year} {2012})}\BibitemShut {NoStop}%
\bibitem [{\citenamefont {Akemann}\ \emph
  {et~al.}(2013{\natexlab{a}})\citenamefont {Akemann}, \citenamefont {Ipsen},\
  and\ \citenamefont {Kieburg}}]{akemann2013products}%
  \BibitemOpen
  \bibfield  {author} {\bibinfo {author} {\bibfnamefont {G.}~\bibnamefont
  {Akemann}}, \bibinfo {author} {\bibfnamefont {J.~R.}\ \bibnamefont {Ipsen}},\
  and\ \bibinfo {author} {\bibfnamefont {M.}~\bibnamefont {Kieburg}},\
  }\bibfield  {title} {\bibinfo {title} {Products of rectangular random
  matrices: singular values and progressive scattering},\ }\href@noop {}
  {\bibfield  {journal} {\bibinfo  {journal} {Physical Review E}\ }\textbf
  {\bibinfo {volume} {88}},\ \bibinfo {pages} {052118} (\bibinfo {year}
  {2013}{\natexlab{a}})}\BibitemShut {NoStop}%
\bibitem [{\citenamefont {Ipsen}\ and\ \citenamefont
  {Kieburg}(2014)}]{ipsen2014weak}%
  \BibitemOpen
  \bibfield  {author} {\bibinfo {author} {\bibfnamefont {J.~R.}\ \bibnamefont
  {Ipsen}}\ and\ \bibinfo {author} {\bibfnamefont {M.}~\bibnamefont
  {Kieburg}},\ }\bibfield  {title} {\bibinfo {title} {Weak commutation
  relations and eigenvalue statistics for products of rectangular random
  matrices},\ }\href@noop {} {\bibfield  {journal} {\bibinfo  {journal}
  {Physical Review E}\ }\textbf {\bibinfo {volume} {89}},\ \bibinfo {pages}
  {032106} (\bibinfo {year} {2014})}\BibitemShut {NoStop}%
\bibitem [{\citenamefont {Hiai}\ and\ \citenamefont
  {Petz}(2000)}]{hiai2000large}%
  \BibitemOpen
  \bibfield  {author} {\bibinfo {author} {\bibfnamefont {F.}~\bibnamefont
  {Hiai}}\ and\ \bibinfo {author} {\bibfnamefont {D.}~\bibnamefont {Petz}},\
  }\bibfield  {title} {\bibinfo {title} {A large deviation theorem for the
  empirical eigenvalue distribution of random unitary matrices},\ }in\
  \href@noop {} {\emph {\bibinfo {booktitle} {Annales de l'Institut Henri
  Poincare (B) Probability and Statistics}}},\ Vol.~\bibinfo {volume} {36}\
  (\bibinfo {organization} {Elsevier},\ \bibinfo {year} {2000})\ pp.\ \bibinfo
  {pages} {71--85}\BibitemShut {NoStop}%
\bibitem [{\citenamefont {Arous}\ and\ \citenamefont
  {Guionnet}(1997)}]{arous1997large}%
  \BibitemOpen
  \bibfield  {author} {\bibinfo {author} {\bibfnamefont {G.~B.}\ \bibnamefont
  {Arous}}\ and\ \bibinfo {author} {\bibfnamefont {A.}~\bibnamefont
  {Guionnet}},\ }\bibfield  {title} {\bibinfo {title} {Large deviations for
  wigner's law and voiculescu's non-commutative entropy},\ }\href@noop {}
  {\bibfield  {journal} {\bibinfo  {journal} {Probability theory and related
  fields}\ }\textbf {\bibinfo {volume} {108}},\ \bibinfo {pages} {517}
  (\bibinfo {year} {1997})}\BibitemShut {NoStop}%
\bibitem [{\citenamefont {Arous}\ and\ \citenamefont
  {Zeitouni}(1998)}]{arous1998large}%
  \BibitemOpen
  \bibfield  {author} {\bibinfo {author} {\bibfnamefont {G.~B.}\ \bibnamefont
  {Arous}}\ and\ \bibinfo {author} {\bibfnamefont {O.}~\bibnamefont
  {Zeitouni}},\ }\bibfield  {title} {\bibinfo {title} {Large deviations from
  the circular law},\ }\href@noop {} {\bibfield  {journal} {\bibinfo  {journal}
  {ESAIM: Probability and Statistics}\ }\textbf {\bibinfo {volume} {2}},\
  \bibinfo {pages} {123} (\bibinfo {year} {1998})}\BibitemShut {NoStop}%
\bibitem [{\citenamefont {Fyodorov}\ and\ \citenamefont
  {Williams}(2007)}]{fyodorov2007replica}%
  \BibitemOpen
  \bibfield  {author} {\bibinfo {author} {\bibfnamefont {Y.~V.}\ \bibnamefont
  {Fyodorov}}\ and\ \bibinfo {author} {\bibfnamefont {I.}~\bibnamefont
  {Williams}},\ }\bibfield  {title} {\bibinfo {title} {Replica symmetry
  breaking condition exposed by random matrix calculation of landscape
  complexity},\ }\href@noop {} {\bibfield  {journal} {\bibinfo  {journal}
  {Journal of Statistical Physics}\ }\textbf {\bibinfo {volume} {129}},\
  \bibinfo {pages} {1081} (\bibinfo {year} {2007})}\BibitemShut {NoStop}%
\bibitem [{\citenamefont {Barbier}\ and\ \citenamefont
  {Panchenko}(2020)}]{barbier2020strong}%
  \BibitemOpen
  \bibfield  {author} {\bibinfo {author} {\bibfnamefont {J.}~\bibnamefont
  {Barbier}}\ and\ \bibinfo {author} {\bibfnamefont {D.}~\bibnamefont
  {Panchenko}},\ }\bibfield  {title} {\bibinfo {title} {Strong replica symmetry
  in high-dimensional optimal {B}ayesian inference},\ }\href@noop {} {\bibfield
   {journal} {\bibinfo  {journal} {arXiv preprint arXiv:2005.03115}\ }
  (\bibinfo {year} {2020})}\BibitemShut {NoStop}%
\bibitem [{\citenamefont {Leake}\ \emph {et~al.}(2021)\citenamefont {Leake},
  \citenamefont {McSwiggen},\ and\ \citenamefont
  {Vishnoi}}]{leake2021sampling}%
  \BibitemOpen
  \bibfield  {author} {\bibinfo {author} {\bibfnamefont {J.}~\bibnamefont
  {Leake}}, \bibinfo {author} {\bibfnamefont {C.}~\bibnamefont {McSwiggen}},\
  and\ \bibinfo {author} {\bibfnamefont {N.~K.}\ \bibnamefont {Vishnoi}},\
  }\bibfield  {title} {\bibinfo {title} {Sampling matrices from
  {H}arish-{C}handra--{I}tzykson--{Z}uber densities with applications to
  quantum inference and differential privacy},\ }in\ \href@noop {} {\emph
  {\bibinfo {booktitle} {Proceedings of the 53rd Annual ACM SIGACT Symposium on
  Theory of Computing}}}\ (\bibinfo {year} {2021})\ pp.\ \bibinfo {pages}
  {1384--1397}\BibitemShut {NoStop}%
\bibitem [{\citenamefont {Bun}\ \emph {et~al.}(2014)\citenamefont {Bun},
  \citenamefont {Bouchaud}, \citenamefont {Majumdar},\ and\ \citenamefont
  {Potters}}]{bun2014instanton}%
  \BibitemOpen
  \bibfield  {author} {\bibinfo {author} {\bibfnamefont {J.}~\bibnamefont
  {Bun}}, \bibinfo {author} {\bibfnamefont {J.-P.}\ \bibnamefont {Bouchaud}},
  \bibinfo {author} {\bibfnamefont {S.~N.}\ \bibnamefont {Majumdar}},\ and\
  \bibinfo {author} {\bibfnamefont {M.}~\bibnamefont {Potters}},\ }\bibfield
  {title} {\bibinfo {title} {Instanton approach to large {N}
  {H}arish-{C}handra-{I}tzykson-{Z}uber integrals},\ }\href@noop {} {\bibfield
  {journal} {\bibinfo  {journal} {Physical review letters}\ }\textbf {\bibinfo
  {volume} {113}},\ \bibinfo {pages} {070201} (\bibinfo {year}
  {2014})}\BibitemShut {NoStop}%
\bibitem [{\citenamefont {Zuber}(2008)}]{zuber2008large}%
  \BibitemOpen
  \bibfield  {author} {\bibinfo {author} {\bibfnamefont {J.-B.}\ \bibnamefont
  {Zuber}},\ }\bibfield  {title} {\bibinfo {title} {The large-n limit of matrix
  integrals over the orthogonal group},\ }\href@noop {} {\bibfield  {journal}
  {\bibinfo  {journal} {Journal of Physics A: Mathematical and Theoretical}\
  }\textbf {\bibinfo {volume} {41}},\ \bibinfo {pages} {382001} (\bibinfo
  {year} {2008})}\BibitemShut {NoStop}%
\bibitem [{\citenamefont {Potters}\ and\ \citenamefont
  {Bouchaud}(2020)}]{potters2020first}%
  \BibitemOpen
  \bibfield  {author} {\bibinfo {author} {\bibfnamefont {M.}~\bibnamefont
  {Potters}}\ and\ \bibinfo {author} {\bibfnamefont {J.-P.}\ \bibnamefont
  {Bouchaud}},\ }\href@noop {} {\emph {\bibinfo {title} {A First Course in
  Random Matrix Theory: For Physicists, Engineers and Data Scientists}}}\
  (\bibinfo  {publisher} {Cambridge University Press},\ \bibinfo {year}
  {2020})\BibitemShut {NoStop}%
\bibitem [{\citenamefont {Collins}(2003)}]{collins2003moments}%
  \BibitemOpen
  \bibfield  {author} {\bibinfo {author} {\bibfnamefont {B.}~\bibnamefont
  {Collins}},\ }\bibfield  {title} {\bibinfo {title} {Moments and cumulants of
  polynomial random variables on unitarygroups, the {I}tzykson-{Z}uber
  integral, and free probability},\ }\href@noop {} {\bibfield  {journal}
  {\bibinfo  {journal} {International Mathematics Research Notices}\ }\textbf
  {\bibinfo {volume} {2003}},\ \bibinfo {pages} {953} (\bibinfo {year}
  {2003})}\BibitemShut {NoStop}%
\bibitem [{\citenamefont {Zinn-Justin}\ and\ \citenamefont
  {Zuber}(2003)}]{zinn2003some}%
  \BibitemOpen
  \bibfield  {author} {\bibinfo {author} {\bibfnamefont {P.}~\bibnamefont
  {Zinn-Justin}}\ and\ \bibinfo {author} {\bibfnamefont {J.-B.}\ \bibnamefont
  {Zuber}},\ }\bibfield  {title} {\bibinfo {title} {On some integrals over the
  {U}({N}) unitary group and their large {N} limit},\ }\href@noop {} {\bibfield
   {journal} {\bibinfo  {journal} {Journal of Physics A: Mathematical and
  General}\ }\textbf {\bibinfo {volume} {36}},\ \bibinfo {pages} {3173}
  (\bibinfo {year} {2003})}\BibitemShut {NoStop}%
\bibitem [{\citenamefont {Edelman}\ and\ \citenamefont
  {Rao}(2005)}]{edelman2005random}%
  \BibitemOpen
  \bibfield  {author} {\bibinfo {author} {\bibfnamefont {A.}~\bibnamefont
  {Edelman}}\ and\ \bibinfo {author} {\bibfnamefont {N.~R.}\ \bibnamefont
  {Rao}},\ }\bibfield  {title} {\bibinfo {title} {Random matrix theory},\
  }\href@noop {} {\bibfield  {journal} {\bibinfo  {journal} {Acta numerica}\
  }\textbf {\bibinfo {volume} {14}},\ \bibinfo {pages} {233} (\bibinfo {year}
  {2005})}\BibitemShut {NoStop}%
\bibitem [{\citenamefont {Schlittgen}\ and\ \citenamefont
  {Wettig}(2003)}]{schlittgen2003generalizations}%
  \BibitemOpen
  \bibfield  {author} {\bibinfo {author} {\bibfnamefont {B.}~\bibnamefont
  {Schlittgen}}\ and\ \bibinfo {author} {\bibfnamefont {T.}~\bibnamefont
  {Wettig}},\ }\bibfield  {title} {\bibinfo {title} {Generalizations of some
  integrals over the unitary group},\ }\href@noop {} {\bibfield  {journal}
  {\bibinfo  {journal} {Journal of Physics A: Mathematical and General}\
  }\textbf {\bibinfo {volume} {36}},\ \bibinfo {pages} {3195} (\bibinfo {year}
  {2003})}\BibitemShut {NoStop}%
\bibitem [{\citenamefont {Ghaderipoor}\ and\ \citenamefont
  {Tellambura}(2008)}]{ghaderipoor2008generalization}%
  \BibitemOpen
  \bibfield  {author} {\bibinfo {author} {\bibfnamefont {A.}~\bibnamefont
  {Ghaderipoor}}\ and\ \bibinfo {author} {\bibfnamefont {C.}~\bibnamefont
  {Tellambura}},\ }\bibfield  {title} {\bibinfo {title} {Generalization of some
  integrals over unitary matrices by character expansion of groups},\
  }\href@noop {} {\bibfield  {journal} {\bibinfo  {journal} {Journal of
  mathematical physics}\ }\textbf {\bibinfo {volume} {49}},\ \bibinfo {pages}
  {073519} (\bibinfo {year} {2008})}\BibitemShut {NoStop}%
\bibitem [{\citenamefont {Forrester}\ and\ \citenamefont
  {Grela}(2016)}]{forrester2016hydrodynamical}%
  \BibitemOpen
  \bibfield  {author} {\bibinfo {author} {\bibfnamefont {P.~J.}\ \bibnamefont
  {Forrester}}\ and\ \bibinfo {author} {\bibfnamefont {J.}~\bibnamefont
  {Grela}},\ }\bibfield  {title} {\bibinfo {title} {Hydrodynamical spectral
  evolution for random matrices},\ }\href@noop {} {\bibfield  {journal}
  {\bibinfo  {journal} {Journal of Physics A: Mathematical and Theoretical}\
  }\textbf {\bibinfo {volume} {49}},\ \bibinfo {pages} {085203} (\bibinfo
  {year} {2016})}\BibitemShut {NoStop}%
\bibitem [{\citenamefont {Guionnet}\ and\ \citenamefont
  {Huang}(2021)}]{guionnet2021large}%
  \BibitemOpen
  \bibfield  {author} {\bibinfo {author} {\bibfnamefont {A.}~\bibnamefont
  {Guionnet}}\ and\ \bibinfo {author} {\bibfnamefont {J.}~\bibnamefont
  {Huang}},\ }\bibfield  {title} {\bibinfo {title} {Large deviations
  asymptotics of rectangular spherical integral},\ }\href@noop {} {\bibfield
  {journal} {\bibinfo  {journal} {arXiv preprint arXiv:2106.07146}\ } (\bibinfo
  {year} {2021})}\BibitemShut {NoStop}%
\bibitem [{\citenamefont {Tao}\ and\ \citenamefont {Vu}(2012)}]{tao2012random}%
  \BibitemOpen
  \bibfield  {author} {\bibinfo {author} {\bibfnamefont {T.}~\bibnamefont
  {Tao}}\ and\ \bibinfo {author} {\bibfnamefont {V.}~\bibnamefont {Vu}},\
  }\bibfield  {title} {\bibinfo {title} {Random covariance matrices:
  Universality of local statistics of eigenvalues},\ }\href@noop {} {\bibfield
  {journal} {\bibinfo  {journal} {The Annals of Probability}\ }\textbf
  {\bibinfo {volume} {40}},\ \bibinfo {pages} {1285} (\bibinfo {year}
  {2012})}\BibitemShut {NoStop}%
\bibitem [{\citenamefont {Forrester}\ and\ \citenamefont
  {Liu}(2016)}]{forrester2016singular}%
  \BibitemOpen
  \bibfield  {author} {\bibinfo {author} {\bibfnamefont {P.~J.}\ \bibnamefont
  {Forrester}}\ and\ \bibinfo {author} {\bibfnamefont {D.-Z.}\ \bibnamefont
  {Liu}},\ }\bibfield  {title} {\bibinfo {title} {Singular values for products
  of complex {G}inibre matrices with a source: hard edge limit and phase
  transition},\ }\href@noop {} {\bibfield  {journal} {\bibinfo  {journal}
  {Communications in Mathematical Physics}\ }\textbf {\bibinfo {volume}
  {344}},\ \bibinfo {pages} {333} (\bibinfo {year} {2016})}\BibitemShut
  {NoStop}%
\bibitem [{\citenamefont {Burda}\ \emph
  {et~al.}(2010{\natexlab{a}})\citenamefont {Burda}, \citenamefont {Jarosz},
  \citenamefont {Livan}, \citenamefont {Nowak},\ and\ \citenamefont
  {Swiech}}]{burda2010eigenvalues}%
  \BibitemOpen
  \bibfield  {author} {\bibinfo {author} {\bibfnamefont {Z.}~\bibnamefont
  {Burda}}, \bibinfo {author} {\bibfnamefont {A.}~\bibnamefont {Jarosz}},
  \bibinfo {author} {\bibfnamefont {G.}~\bibnamefont {Livan}}, \bibinfo
  {author} {\bibfnamefont {M.~A.}\ \bibnamefont {Nowak}},\ and\ \bibinfo
  {author} {\bibfnamefont {A.}~\bibnamefont {Swiech}},\ }\bibfield  {title}
  {\bibinfo {title} {Eigenvalues and singular values of products of rectangular
  gaussian random matrices},\ }\href@noop {} {\bibfield  {journal} {\bibinfo
  {journal} {Physical Review E}\ }\textbf {\bibinfo {volume} {82}},\ \bibinfo
  {pages} {061114} (\bibinfo {year} {2010}{\natexlab{a}})}\BibitemShut
  {NoStop}%
\bibitem [{\citenamefont {Ipsen}(2015)}]{ipsen2015products}%
  \BibitemOpen
  \bibfield  {author} {\bibinfo {author} {\bibfnamefont {J.~R.}\ \bibnamefont
  {Ipsen}},\ }\bibfield  {title} {\bibinfo {title} {Products of independent
  gaussian random matrices},\ }\href@noop {} {\bibfield  {journal} {\bibinfo
  {journal} {arXiv preprint arXiv:1510.06128}\ } (\bibinfo {year}
  {2015})}\BibitemShut {NoStop}%
\bibitem [{\citenamefont {Ipsen}(2013)}]{ipsen2013products}%
  \BibitemOpen
  \bibfield  {author} {\bibinfo {author} {\bibfnamefont {J.~R.}\ \bibnamefont
  {Ipsen}},\ }\bibfield  {title} {\bibinfo {title} {Products of independent
  quaternion {G}inibre matrices and their correlation functions},\ }\href@noop
  {} {\bibfield  {journal} {\bibinfo  {journal} {Journal of Physics A:
  Mathematical and Theoretical}\ }\textbf {\bibinfo {volume} {46}},\ \bibinfo
  {pages} {265201} (\bibinfo {year} {2013})}\BibitemShut {NoStop}%
\bibitem [{\citenamefont {Akemann}\ and\ \citenamefont
  {Burda}(2012)}]{akemann2012universal}%
  \BibitemOpen
  \bibfield  {author} {\bibinfo {author} {\bibfnamefont {G.}~\bibnamefont
  {Akemann}}\ and\ \bibinfo {author} {\bibfnamefont {Z.}~\bibnamefont
  {Burda}},\ }\bibfield  {title} {\bibinfo {title} {Universal microscopic
  correlation functions for products of independent {G}inibre matrices},\
  }\href@noop {} {\bibfield  {journal} {\bibinfo  {journal} {Journal of Physics
  A: Mathematical and Theoretical}\ }\textbf {\bibinfo {volume} {45}},\
  \bibinfo {pages} {465201} (\bibinfo {year} {2012})}\BibitemShut {NoStop}%
\bibitem [{\citenamefont {Akemann}\ \emph
  {et~al.}(2013{\natexlab{b}})\citenamefont {Akemann}, \citenamefont
  {Kieburg},\ and\ \citenamefont {Wei}}]{akemann2013singular}%
  \BibitemOpen
  \bibfield  {author} {\bibinfo {author} {\bibfnamefont {G.}~\bibnamefont
  {Akemann}}, \bibinfo {author} {\bibfnamefont {M.}~\bibnamefont {Kieburg}},\
  and\ \bibinfo {author} {\bibfnamefont {L.}~\bibnamefont {Wei}},\ }\bibfield
  {title} {\bibinfo {title} {Singular value correlation functions for products
  of {W}ishart random matrices},\ }\href@noop {} {\bibfield  {journal}
  {\bibinfo  {journal} {Journal of Physics A: Mathematical and Theoretical}\
  }\textbf {\bibinfo {volume} {46}},\ \bibinfo {pages} {275205} (\bibinfo
  {year} {2013}{\natexlab{b}})}\BibitemShut {NoStop}%
\bibitem [{\citenamefont {Akemann}\ \emph {et~al.}(2014)\citenamefont
  {Akemann}, \citenamefont {Burda}, \citenamefont {Kieburg},\ and\
  \citenamefont {Nagao}}]{akemann2014universal}%
  \BibitemOpen
  \bibfield  {author} {\bibinfo {author} {\bibfnamefont {G.}~\bibnamefont
  {Akemann}}, \bibinfo {author} {\bibfnamefont {Z.}~\bibnamefont {Burda}},
  \bibinfo {author} {\bibfnamefont {M.}~\bibnamefont {Kieburg}},\ and\ \bibinfo
  {author} {\bibfnamefont {T.}~\bibnamefont {Nagao}},\ }\bibfield  {title}
  {\bibinfo {title} {Universal microscopic correlation functions for products
  of truncated unitary matrices},\ }\href@noop {} {\bibfield  {journal}
  {\bibinfo  {journal} {Journal of Physics A: Mathematical and Theoretical}\
  }\textbf {\bibinfo {volume} {47}},\ \bibinfo {pages} {255202} (\bibinfo
  {year} {2014})}\BibitemShut {NoStop}%
\bibitem [{\citenamefont {Kieburg}\ \emph {et~al.}(2016)\citenamefont
  {Kieburg}, \citenamefont {Kuijlaars},\ and\ \citenamefont
  {Stivigny}}]{kieburg2016singular}%
  \BibitemOpen
  \bibfield  {author} {\bibinfo {author} {\bibfnamefont {M.}~\bibnamefont
  {Kieburg}}, \bibinfo {author} {\bibfnamefont {A.~B.}\ \bibnamefont
  {Kuijlaars}},\ and\ \bibinfo {author} {\bibfnamefont {D.}~\bibnamefont
  {Stivigny}},\ }\bibfield  {title} {\bibinfo {title} {Singular value
  statistics of matrix products with truncated unitary matrices},\ }\href@noop
  {} {\bibfield  {journal} {\bibinfo  {journal} {International Mathematics
  Research Notices}\ }\textbf {\bibinfo {volume} {2016}},\ \bibinfo {pages}
  {3392} (\bibinfo {year} {2016})}\BibitemShut {NoStop}%
\bibitem [{\citenamefont {Akemann}\ and\ \citenamefont
  {Ipsen}(2015)}]{akemann2015recent}%
  \BibitemOpen
  \bibfield  {author} {\bibinfo {author} {\bibfnamefont {G.}~\bibnamefont
  {Akemann}}\ and\ \bibinfo {author} {\bibfnamefont {J.~R.}\ \bibnamefont
  {Ipsen}},\ }\bibfield  {title} {\bibinfo {title} {Recent exact and asymptotic
  results for products of independent random matrices},\ }\href@noop {}
  {\bibfield  {journal} {\bibinfo  {journal} {arXiv preprint arXiv:1502.01667}\
  } (\bibinfo {year} {2015})}\BibitemShut {NoStop}%
\bibitem [{\citenamefont {Liu}\ \emph {et~al.}(2016)\citenamefont {Liu},
  \citenamefont {Wang},\ and\ \citenamefont {Zhang}}]{liu2016bulk}%
  \BibitemOpen
  \bibfield  {author} {\bibinfo {author} {\bibfnamefont {D.-Z.}\ \bibnamefont
  {Liu}}, \bibinfo {author} {\bibfnamefont {D.}~\bibnamefont {Wang}},\ and\
  \bibinfo {author} {\bibfnamefont {L.}~\bibnamefont {Zhang}},\ }\bibfield
  {title} {\bibinfo {title} {Bulk and soft-edge universality for singular
  values of products of {G}inibre random matrices},\ }in\ \href@noop {} {\emph
  {\bibinfo {booktitle} {Annales de l'Institut Henri Poincar{\'e},
  Probabilit{\'e}s et Statistiques}}},\ Vol.~\bibinfo {volume} {52}\ (\bibinfo
  {organization} {Institut Henri Poincar{\'e}},\ \bibinfo {year} {2016})\ pp.\
  \bibinfo {pages} {1734--1762}\BibitemShut {NoStop}%
\bibitem [{\citenamefont {FitzGerald}\ and\ \citenamefont
  {Simm}(2021)}]{fitzgerald2021fluctuations}%
  \BibitemOpen
  \bibfield  {author} {\bibinfo {author} {\bibfnamefont {W.}~\bibnamefont
  {FitzGerald}}\ and\ \bibinfo {author} {\bibfnamefont {N.}~\bibnamefont
  {Simm}},\ }\bibfield  {title} {\bibinfo {title} {Fluctuations and
  correlations for products of real asymmetric random matrices},\ }\href@noop
  {} {\bibfield  {journal} {\bibinfo  {journal} {arXiv preprint
  arXiv:2109.00322}\ } (\bibinfo {year} {2021})}\BibitemShut {NoStop}%
\bibitem [{\citenamefont {Gudowska-Nowak}\ \emph {et~al.}(2003)\citenamefont
  {Gudowska-Nowak}, \citenamefont {Janik}, \citenamefont {Jurkiewicz},\ and\
  \citenamefont {Nowak}}]{gudowska2003infinite}%
  \BibitemOpen
  \bibfield  {author} {\bibinfo {author} {\bibfnamefont {E.}~\bibnamefont
  {Gudowska-Nowak}}, \bibinfo {author} {\bibfnamefont {R.~A.}\ \bibnamefont
  {Janik}}, \bibinfo {author} {\bibfnamefont {J.}~\bibnamefont {Jurkiewicz}},\
  and\ \bibinfo {author} {\bibfnamefont {M.~A.}\ \bibnamefont {Nowak}},\
  }\bibfield  {title} {\bibinfo {title} {Infinite products of large random
  matrices and matrix-valued diffusion},\ }\href@noop {} {\bibfield  {journal}
  {\bibinfo  {journal} {Nuclear Physics B}\ }\textbf {\bibinfo {volume}
  {670}},\ \bibinfo {pages} {479} (\bibinfo {year} {2003})}\BibitemShut
  {NoStop}%
\bibitem [{\citenamefont {Burda}\ \emph
  {et~al.}(2010{\natexlab{b}})\citenamefont {Burda}, \citenamefont {Janik},\
  and\ \citenamefont {Waclaw}}]{burda2010spectrum}%
  \BibitemOpen
  \bibfield  {author} {\bibinfo {author} {\bibfnamefont {Z.}~\bibnamefont
  {Burda}}, \bibinfo {author} {\bibfnamefont {R.~A.}\ \bibnamefont {Janik}},\
  and\ \bibinfo {author} {\bibfnamefont {B.}~\bibnamefont {Waclaw}},\
  }\bibfield  {title} {\bibinfo {title} {Spectrum of the product of independent
  random gaussian matrices},\ }\href@noop {} {\bibfield  {journal} {\bibinfo
  {journal} {Physical Review E}\ }\textbf {\bibinfo {volume} {81}},\ \bibinfo
  {pages} {041132} (\bibinfo {year} {2010}{\natexlab{b}})}\BibitemShut
  {NoStop}%
\bibitem [{\citenamefont {Burda}\ \emph
  {et~al.}(2010{\natexlab{c}})\citenamefont {Burda}, \citenamefont {Jarosz},
  \citenamefont {Livan}, \citenamefont {Nowak},\ and\ \citenamefont
  {Swiech}}]{burda2011eigenvalues}%
  \BibitemOpen
  \bibfield  {author} {\bibinfo {author} {\bibfnamefont {Z.}~\bibnamefont
  {Burda}}, \bibinfo {author} {\bibfnamefont {A.}~\bibnamefont {Jarosz}},
  \bibinfo {author} {\bibfnamefont {G.}~\bibnamefont {Livan}}, \bibinfo
  {author} {\bibfnamefont {M.~A.}\ \bibnamefont {Nowak}},\ and\ \bibinfo
  {author} {\bibfnamefont {A.}~\bibnamefont {Swiech}},\ }\bibfield  {title}
  {\bibinfo {title} {Eigenvalues and singular values of products of rectangular
  gaussian random matrices},\ }\href@noop {} {\bibfield  {journal} {\bibinfo
  {journal} {Physical Review E}\ }\textbf {\bibinfo {volume} {82}},\ \bibinfo
  {pages} {061114} (\bibinfo {year} {2010}{\natexlab{c}})}\BibitemShut
  {NoStop}%
\bibitem [{\citenamefont {G{\"o}tze}\ and\ \citenamefont
  {Tikhomirov}(2010)}]{gotze2010asymptotic}%
  \BibitemOpen
  \bibfield  {author} {\bibinfo {author} {\bibfnamefont {F.}~\bibnamefont
  {G{\"o}tze}}\ and\ \bibinfo {author} {\bibfnamefont {A.}~\bibnamefont
  {Tikhomirov}},\ }\bibfield  {title} {\bibinfo {title} {On the asymptotic
  spectrum of products of independent random matrices},\ }\href@noop {}
  {\bibfield  {journal} {\bibinfo  {journal} {arXiv preprint arXiv:1012.2710}\
  } (\bibinfo {year} {2010})}\BibitemShut {NoStop}%
\bibitem [{\citenamefont {O'Rourke}\ and\ \citenamefont
  {Soshnikov}(2011)}]{o2011products}%
  \BibitemOpen
  \bibfield  {author} {\bibinfo {author} {\bibfnamefont {S.}~\bibnamefont
  {O'Rourke}}\ and\ \bibinfo {author} {\bibfnamefont {A.}~\bibnamefont
  {Soshnikov}},\ }\bibfield  {title} {\bibinfo {title} {Products of independent
  non-hermitian random matrices},\ }\href@noop {} {\bibfield  {journal}
  {\bibinfo  {journal} {Electronic Journal of Probability}\ }\textbf {\bibinfo
  {volume} {16}},\ \bibinfo {pages} {2219} (\bibinfo {year}
  {2011})}\BibitemShut {NoStop}%
\bibitem [{\citenamefont {Akemann}\ \emph {et~al.}(2011)\citenamefont
  {Akemann}, \citenamefont {Baik},\ and\ \citenamefont
  {Di~Francesco}}]{akemann2011oxford}%
  \BibitemOpen
  \bibfield  {author} {\bibinfo {author} {\bibfnamefont {G.}~\bibnamefont
  {Akemann}}, \bibinfo {author} {\bibfnamefont {J.}~\bibnamefont {Baik}},\ and\
  \bibinfo {author} {\bibfnamefont {P.}~\bibnamefont {Di~Francesco}},\
  }\href@noop {} {\emph {\bibinfo {title} {The Oxford handbook of random matrix
  theory}}}\ (\bibinfo  {publisher} {Oxford University Press},\ \bibinfo {year}
  {2011})\BibitemShut {NoStop}%
\bibitem [{Note4()}]{Note4}%
  \BibitemOpen
  \bibinfo {note} {We note that this assumption is reminiscent of results in
  \cite {benaych2009rectangular} (see also \cite {belinschi2009regularization})
  for the addition of two large random matrices with at least one being
  bi-unitary invariant.}\BibitemShut {Stop}%
\bibitem [{\citenamefont {Menon}(2017)}]{menon2017complex}%
  \BibitemOpen
  \bibfield  {author} {\bibinfo {author} {\bibfnamefont {G.}~\bibnamefont
  {Menon}},\ }\href@noop {} {\bibinfo {title} {The complex {B}urgers equation,
  the {H}{C}{I}{Z} integral and the {C}alogero-{M}oser system}} (\bibinfo
  {year} {2017})\BibitemShut {NoStop}%
\bibitem [{\citenamefont {Bryc}(2007)}]{bryc2007computing}%
  \BibitemOpen
  \bibfield  {author} {\bibinfo {author} {\bibfnamefont {W.}~\bibnamefont
  {Bryc}},\ }\bibfield  {title} {\bibinfo {title} {Computing moments of free
  additive convolution of measures},\ }\href@noop {} {\bibfield  {journal}
  {\bibinfo  {journal} {Applied mathematics and computation}\ }\textbf
  {\bibinfo {volume} {194}},\ \bibinfo {pages} {561} (\bibinfo {year}
  {2007})}\BibitemShut {NoStop}%
\bibitem [{\citenamefont {Benaych-Georges}(2009)}]{benaych2009rectangular}%
  \BibitemOpen
  \bibfield  {author} {\bibinfo {author} {\bibfnamefont {F.}~\bibnamefont
  {Benaych-Georges}},\ }\bibfield  {title} {\bibinfo {title} {Rectangular
  random matrices, related convolution},\ }\href@noop {} {\bibfield  {journal}
  {\bibinfo  {journal} {Probability Theory and Related Fields}\ }\textbf
  {\bibinfo {volume} {144}},\ \bibinfo {pages} {471} (\bibinfo {year}
  {2009})}\BibitemShut {NoStop}%
\bibitem [{\citenamefont {Belinschi}\ \emph {et~al.}(2009)\citenamefont
  {Belinschi}, \citenamefont {Benaych-Georges},\ and\ \citenamefont
  {Guionnet}}]{belinschi2009regularization}%
  \BibitemOpen
  \bibfield  {author} {\bibinfo {author} {\bibfnamefont {S.~T.}\ \bibnamefont
  {Belinschi}}, \bibinfo {author} {\bibfnamefont {F.}~\bibnamefont
  {Benaych-Georges}},\ and\ \bibinfo {author} {\bibfnamefont {A.}~\bibnamefont
  {Guionnet}},\ }\bibfield  {title} {\bibinfo {title} {Regularization by free
  additive convolution, square and rectangular cases},\ }\href@noop {}
  {\bibfield  {journal} {\bibinfo  {journal} {Complex Analysis and Operator
  Theory}\ }\textbf {\bibinfo {volume} {3}},\ \bibinfo {pages} {611} (\bibinfo
  {year} {2009})}\BibitemShut {NoStop}%
\end{thebibliography}%
